%% file: main.tex
\documentclass[11pt]{article}

\input{preamble}

\title{\LARGE \bf System-Level Performance and Communication
Tradeoff in Networked Control with Predictions}
\author{Yifei~Wu,
        Jing~Yu,
        Tongxin~Li\thanks{
     Wu and Li are with School of Data Science, The Chinese University of Hong Kong, Shenzhen, Guangdong, China~(e-mails: {\tt yifeiwu1@link.cuhk.edu.cn, litongxin@cuhk.edu.cn}). Yu is with the Department of Electrical and Computer Engineering, University of Washington, Seattle, WA, USA.(e-mail:{\tt jing5@uw.edu}).}
        }

\date{\vspace*{-1cm}}

\begin{document}
\maketitle

\begin{abstract}
   Distributed control of large-scale systems is challenging due to the need for scalable and localized communication and computation. In this work, we introduce a \textsc{Pred}ictive \textsc{S}ystem-\textsc{L}evel \textsc{S}ynthesis (\psls) framework that designs controllers by jointly integrating communication constraints and local disturbance predictions into an affine feedback structure.  Rather than focusing on the worst-case uncertainty, \psls leverages both current state feedback and future system disturbance predictions to achieve distributed control of networked systems. In particular, \psls enables a unified system synthesis of the optimal $\kappa$-localized controller, therefore outperforms approaches with post hoc communication truncation, as was commonly seen in the literature.  
The \psls framework can be naturally decomposed into spatial and temporal components for efficient and parallelizable computation across the network, yielding a regret upper bound that explicitly depends on the prediction error and communication range. Our regret analysis not only reveals a
non-monotonic trade-off between control performance and communication range when prediction errors are present, but also  guides the identification of an optimal size for local communication neighborhoods, thereby enabling the co-design of controller and its underlying communication topology.
\end{abstract}
{\keywords{} Networked control systems, Predictive control, System level synthesis.}
\addtocontents{toc}{\protect\setcounter{tocdepth}{0}}
\section{Introduction}

Modern large-scale and networked systems, such as power networks~\cite{matni2017scalable,yu2023online,shin2023near,10681576} and building temperature control~\cite{lian2021system,xu2024stability}, pose significant challenges for traditional control design. These applications have motivated the development of control synthesis at the system level, as known as system-level synthesis (SLS) as a new framework that shifts the design focus from crafting an individual controller to designing the entire closed loop system response. This framework has been shown to enable the systematic incorporation of structural constraints such as locality and sparsity while delivering scalable and robust performance in distributed control architectures~\cite{anderson2019system,han2020localized,chen2024robust}.

Despite its proven potential, the standard SLS framework typically assumes static or worst-case uncertainty models, thus overlooking the value of predictive information that is available in many modern applications. In the context of networked control systems, short term forecasts generated by artificial intelligence tools or human inputs can be exploited to anticipate disturbances and adapt control actions in advance. The benefits of prediction in control have long been recognized in the literature on model predictive control~\cite{bemporad1999control,camacho2007constrained}, and regret-optimal control~\cite{yu2020power,goel2022power}. However, a unified SLS framework that fully integrates future disturbance predictions remains absent.

Incorporating prediction into distributed control scenarios, however, introduces new challenges. First, constructing system-level responses compatible with predictive control is nontrivial. The classic predictive synthesis problem typically considers only causal closed-loop mappings, which are not directly generalizable to incorporate prediction. Thus, the problem of designing controllers that not only meet general structural communication constraints such as locality and sparsity but also seamlessly integrate predictions in the system-level remains unexplored.
Second,
in distributed control systems, a prevalent approach involves designing an optimal controller and then truncating its gains to fit the communication constraints, as explored in recent studies ~\cite{zhang2023optimal,shin2023near,xu2024stability}. While this truncation method simplifies implementation by adapting precomputed gains to limited communication topologies, e.g., $\distance$-hop networks and sparse connections, it often sacrifices optimality and the flexibility to incorporate general communication constraints. Achieving optimal control requires a unified approach that simultaneously optimizes the controller and accounts for communication constraints, rather than relying on post hoc truncation. Consequently, the control parameters themselves need to be optimized in light of the communication constraints, and a unified predictive and distributed control framework thus becomes essential.

\begin{figure}
    \centering
\includegraphics[width=1\linewidth]{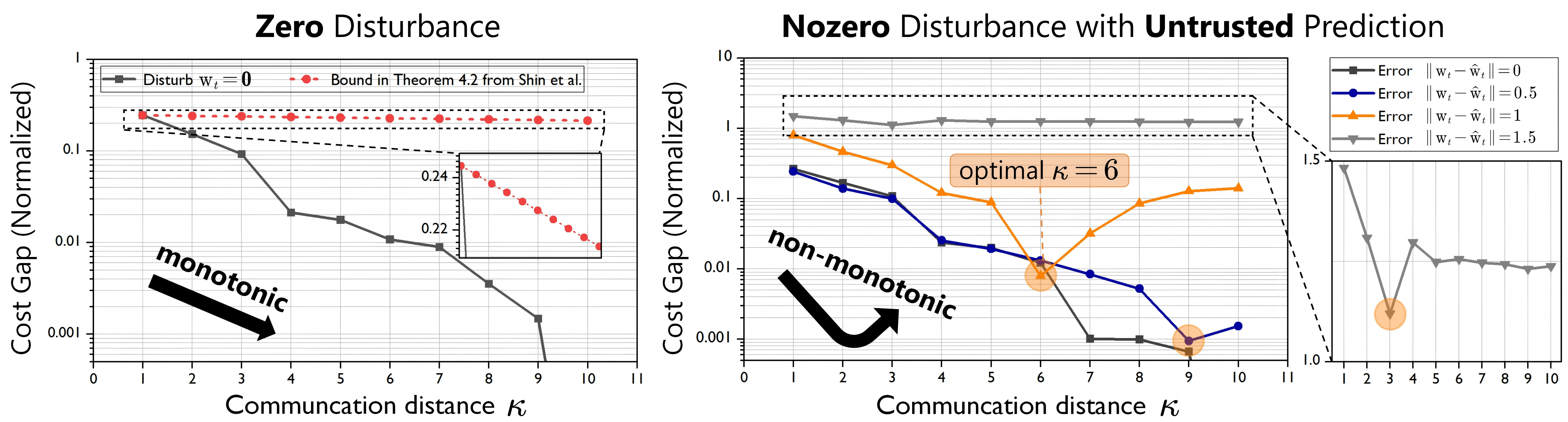}
    \caption{Impact of prediction errors on the \textbf{communication-performance trade-off}. Both graphs plot the normalized cost gap between truncated controllers (defined precisely in Section~\ref{sec:case_studies}) and optimal non-causal gain versus communication range $\distance$. \textsc{\textbf{Left}} (\textbf{Zero} disturbance): The cost gap decreases monotonically for $\mathrm{w}_t=\mathbf{0}$, compared to a bound in Theorem 4.2 from Shin et al.~\cite{shin2023near} (plotted by omitting constants); \textsc{\textbf{Right}} (\textbf{Nonzero} Disturbance with Untrusted Predictions): The cost gap varies non-monotonically with prediction errors {($\|\mathrm{w}_t-\widehat{\mathrm{w}}_t\|=0,0.5,1,1.5$)}. The initial state is set as $\|\mathrm{x}_0\|=1$.}
    \label{fig:enter-label}
\end{figure}

More importantly, the predictions available in practice are often imperfect, and these inaccuracies can fundamentally alter the trade-off between decentralization and performance. For instance, Shin et al.~\cite{shin2023near} and Xu et al.~\cite{xu2024stability} explored an idealized scenario with zero prediction error, revealing a \textit{monotonic} trade-off where the performance gap between the truncated controller and the centralized optimal controller decays exponentially with increasing communication range. In contrast, with realistic prediction errors, the trade-off becomes \textit{non-monotonic} as illustrated in Figure~\ref{fig:enter-label}. In such settings, excessive decentralization leaves each local controller with insufficient information, while overly centralized schemes incur substantial communication and computational burdens. Moreover, prediction errors can propagate between neighboring nodes, exacerbating performance degradation. As a result, there exists an optimal balance in the degree of controller decentralization where the adverse effects of prediction error are minimized while still benefiting from adequate local data exchange. This insight highlights a critical and practical challenge that must be addressed to enable effective predictive localized control:

\begin{center}
    \textit{How can we formulate a unified System-Level Synthesis (SLS) framework that leverages imperfect predictions to optimize controller gains subject to communication constraints? Moreover, can we characterize the non-monotonic trade-off between decentralization and performance?}
\end{center}

To address these challenges, we introduce a Predictive SLS (termed \psls) framework that integrates predictive information into the synthesis of system-level responses and controllers. This framework is designed to provide performance guarantees under bounded prediction error and specified degrees of decentralization. By explicitly accounting for the impact of prediction inaccuracies and local information exchange, \psls enables a co-design of the controller and the communication topology. The resulting framework offers a systematic approach to achieve robust performance in distributed control settings where prediction error is unavoidable and balancing local and global data exchange is crucial. 
Our main results are three-fold:

\begin{enumerate}
    \item  
\textbf{The \psls framework.} We propose the \texttt{Pred}ictive \texttt{S}ystem-\texttt{L}evel \texttt{S}ynthesis (\psls) framework of the form $\mathbf{u} = \mathbf{Kx} + \widehat{\mathbf{L}}\widehat{\mathbf{w}}$
(see~\Cref{eq:noncausal_form} for more details), where $\mathbf{u}$, $\mathbf{x}$, and $\widehat{\mathbf{w}}$ represent action, state, and disturbance predictions respectively; $\mathbf{K}$ and $\widehat{\mathbf{L}}$ are proper and improper transfer matrices (Theorem~\ref{theorem constraints}). This enables optimal controller synthesis via a convex quadratic program over system responses, matching the offline optimal policy for arbitrary disturbances when predictions are error-free (Proposition~\ref{proposition:1}). The framework guarantees stability and performance for networked systems under localized constraints, without requiring open-loop stability or restrictive spatial exponential decaying (SED) assumptions on system dynamics. This is achieved through a novel synthesis of an equivalence principle and constraint transformations (\Cref{proposition:2}).

\item \textbf{Localized control with  predictions.} We incorporate general communication constraints into \psls, and propose a scalable decomposition of the controller synthesis and implementation. When the communication constraints are column-decomposable, spatial and temporal separability is used to split the distributed \psls into agent-wise and temporally independent sub-problems.  
     To ensure computational tractability, we introduce a finite-horizon approximation in Section~\ref{subsec:finite_time_approx} using finite impulse response (FIR) mappings. 
     This is made possible by leveraging the exponential decay property of the closed-loop mappings, which holds under broad classes of topologies and even for open-loop unstable systems, as established in Theorem~\ref{theorem:temporal} and~\ref{theorem:spatial}. Unlike the results in~\cite{zhang2023optimal}, which require the system to be open-loop stable, and approaches that do not integrate predictions~\cite{shin2023near}, \psls leverages system-level parameterization to directly incorporate any communication constraints that can be expressed as convex constraints into the controller synthesis. By unifying $\kappa$-hop communication with untrusted predictions, we derive regret bounds that explicitly quantify the trade-off between prediction accuracy ($\varepsilon$) and communication range ($\distance$). 
   \item \textbf{Co-design of control and communication.} While general optimization-based co-design of controllers and topologies remains open for networked control with predictions, our analysis on  $\distance$-hop information exchange patterns  provides foundational insights:  
   1. \textit{Communication-Performance Trade-offs}: We establish that the parameter $\distance$ (communication range for each local controller) can be tuned based on the trade-off implicated by the regret bounds. 
   This bridges \textbf{SLS theory} with recent \textbf{regret-based} approaches in control systems~\cite{zhang2021regret,goel2023regret}.  
   2. \textit{Pathway for Co-Design}: Our numerical experiments (e.g., chain-graph-induced graph instance and other graph instances in~\cite{wu2025}) highlight how communication range impacts system performance and motivating future work on communication structure-aware control design. 
\end{enumerate}

\begin{table*}[t]
\scriptsize
\centering
\caption{Comparison with Related Distributed Controllers.}
\label{tab:comparison}
\begin{tabular}{lccccc}
\toprule
 & \textbf{Zhang et al. \cite{zhang2023optimal}} & \textbf{Shin et al. \cite{shin2023near}} & \textbf{Yu et al. \cite{yu2023online}} & \textbf{Xu et al. \cite{xu2024stability}} & \textbf{\psls (Ours)} \\
\midrule
\textbf{Framework} & DAC & Linear LQR & SLS & MPC & \textbf{Predictive SLS}  \\ 
\textbf{Prediction integration}     & No  & No  & No & Yes &\textbf{Yes} \\
\textbf{Open-loop stability} & Yes & No Req. & No Req. & No Req. & \textbf{No Req.} \\
\textbf{Topology}$^1$            & SED$^2$  & 
Undirected$^{3}$  & No Req. &
Undirected &  \textbf{No Req.}\\
\textbf{Perturbation type} & Gaussian$^{4}$ & --- & Bounded Arbitrary & Bounded Arbitrary & \textbf{Bounded Arbitrary} \\
\textbf{Optimality guarantee}     &  Causal  & Causal  & --- &  Non-causal &\textbf{Non-causal} \\
\textbf{Trade-off}              & Monotonic & Monotonic & --- & Monotonic & \textbf{Non-monotonic}$^{5}$ \\
\bottomrule\\
\multicolumn{6}{c}{\footnotesize  $^{1}$: 
system dynamics topology requirement. $^{2}$: 
spatially-exponential decay \cite{zhang2023optimal}. $^{3}$: 
undirected communication graph. } \\ \multicolumn{6}{c}{\footnotesize  $^{4}$: zero-mean i.i.d. Gaussian disturbances. 
 $^{5}$:  non-monotonic communication-performance trade-off (Theorem \ref{theorem:regret}). }
 \\
\end{tabular}
\end{table*}

\subsection*{Related Work}

There are many lines of work that is related to our approach both in networked control theory and SLS theory. Below we compare \psls with closely related works and summarize key differences in Table~\ref{tab:comparison}.

\textbf{Distributed control for networked dynamical systems.} 
Compared to disturbance-action control (DAC) controllers~\cite{agarwal2019online,martin2024guarantees}, particularly the distributed controller in~\cite{zhang2023optimal}, our work offers a more general approach to optimal control in networked linear systems. Zhang et al.~\cite{zhang2023optimal} explore the optimal control of network linear quadratic regulators (LQR) with spatially-exponentially decaying (SED) structures, where Theorem 1 in~\cite{zhang2023optimal} demonstrates that optimal closed-loop matrices (CLMs) are SED with respect to the distance $d(i,j)$, assuming open-loop stability and SED system matrices $(A,B,Q,R)$. Our results, as stated in Theorem 5.2, generalize this finding by showing that for stabilizable $(A,B,Q,R)$ and any subsystem $i$, the structural difference between distributed CLMs and optimal CLMs is bounded by an exponential decay with respect to the maximum distance $\distance$, without ensuring SED compliance. This extends applicability beyond the restrictive conditions of open-loop stability and SED matrices required by~\cite{zhang2023optimal}. From the perspective of closed-loop gain, our framework relaxes the SED requirement, yet when $(A,B,Q,R)$ are SED and open-loop stable, our results coincide with theirs as a special case, guaranteeing an exponentially decaying difference between optimal and localized CLMs. Considering cost, Theorem 2 in \cite{zhang2023optimal} identifies a gap between the truncated gain $K_{\text{trunc}}$ and the optimal causal gain $K^\star$ when $K^\star$ is SED, whereas our work analyzes the difference between an optimal non-causal policy and our \psls approach, parameterized by communication hops $\kappa$ and prediction error $\varepsilon$. When predictions are omitted and our CLMs exhibit SED, our results similarly yield an exponential decay, aligning with \cite{zhang2023optimal} while providing a broader and more flexible framework.

Xu and Qu \cite{xu2024stability} explored distributed truncated predictive control within a model predictive control (MPC) framework for linear systems under limited communication.
Their study, rooted in distributed model predictive control (MPC) for linear systems, extends the existing perturbation analysis for MPC~\cite{lin2021perturbation,lin2022bounded} by involving communication constraints and shows a similar exponential decay as in~\cite{zhang2023optimal}. In contrast, our framework introduces a general approach to functional closed-loop control laws applicable to a wider range of networked linear systems, relaxes assumptions about system structure and communication, and provides a non-monotonic communication-performance trade-off of how predictions and $\kappa$-hop communication influence performance. 

Our work also contrasts with that of Shin et al. \cite{shin2023near}, who developed a near-optimal truncated distributed linear-quadratic regulator (LQR) for disturbance-free networked systems, without making assumptions on open-loop stability and specific $(A,B,Q,R)$ structures. While their approach achieves near-optimality in distributed control for linear systems, our framework further generalizes these results. We use the optimal non-causal policy as a benchmark to evaluate our distributed strategies, instead of the optimal causal gain in~\cite{shin2023near}. Additionally, our approach accommodates a broader class of networked linear systems by relaxing structural assumptions even further with some of the key differences summarized in Table~\ref{tab:comparison}, and  guarantee in the form of non-monotonic communication-performance trade-off is shown, which enhances the practical utility of our results beyond the scope considered in~\cite{shin2023near}.

\textbf{System level synthesis.} The System Level Synthesis (SLS) framework \cite{anderson2019system, wang2018separable, wang2019system, alonso2022distributed,yu2021localized}, provides an effective approach for designing distributed controllers that satisfy locality constraints in networked systems. Recently, SLS has been applied to broader contexts, including the design of predictive safety filters using linear causal controllers \cite{leeman2023predictive}, as well as regret optimal control \cite{didier2022system}. However, the integration of future disturbance predictions into the SLS framework remains largely unaddressed. In contrast, our proposed Predictive SLS (\psls) framework extends SLS by incorporating predictions of future disturbances, offering a more general approach to optimal control in networked linear systems, while still preserving the ability to handle general communication constraints such as locality and delay.

\paragraph{Notational conventions.} Throughout this paper, the set of real numbers and the set of non-negative integers are denoted by $\mathbb{R}$ and $\mathbb{N}$, respectively. $\|\cdot\|$ denotes the $\ell_2$-norm for vectors and the $\ell_2$-induced norm for matrices, whereas $\|\cdot\|_{\infty}$ and $\|\cdot\|_{\mathcal{F}}$ denotes the infinity and Frobenius norm respectively. 
Scalars, vectors, and matrices are denoted by lowercase italic font like ``$x$'', lowercase boldface font like ``$\mathrm{x}$'', and uppercase font like ``$M$'', respectively. Let $[T] \coloneqq \{0,1, \ldots, T-1\}$ denote the discrete control time indices and $[N] \coloneqq \{1, \ldots, N\}$ denote a set of $N$ subsystems. We write a sequence of vectors as $\mathrm{w}_{0:T}\coloneqq (\mathrm{w}_0,\ldots,\mathrm{w}_{T-1})$. We denote the $(i, j)$th component of a matrix $M$ as $M(i,j)$, and use $M(i)$ and $ M(:,j)$ for the $i$th row and $j$th column, respectively. We use the syntax: $M(\mathcal{I},\mathcal{J})\coloneqq [\left[M(i,j)\right]_{j\in\mathcal{J}}^\top]_{i\in\mathcal{I}}$ for a sub-matrix where $\mathcal{I}\coloneqq\{i_1<i_2<\dots<i_q\}$ and $\mathcal{J}\coloneqq\{j_1<j_2<\dots<j_p\}$ are strictly ordered index sets. We use the superscript notation $M^{ij}$ to denote the $(i,j)$-th block matrix. We also use single and double subscripts to represent sub-matrices or sub-vectors arranged in order, such as $M_t$ and $M_{t,k}$ and $k\in[k_1,k_2]$ where 
$$M=\begin{bmatrix}
    M_{t_1}\\
    M_{t_1+1}\\
    \vdots\\
    M_{t_2}
\end{bmatrix} \text{or} \begin{bmatrix}
    M_{t_1,k_1} & M_{t_1,k_1+1} & \cdots & M_{t_1,k_2}\\
    M_{t_1+1,k_1}& M_{t_1+1,k_1+1} & \cdots & M_{t_1+1,k_2}\\
    \vdots& \vdots& \ddots& \vdots\\
    M_{t_2,k_1} & M_{t_2,k_1+1} & \cdots & M_{t_2,k_2}
\end{bmatrix} \text{~for any $t_1\leq t\leq t_2$~and~$k_1\leq k\leq k_2$}.$$ 
Let bold font $\mathbf{x}$ denote signals $\mathbf{x}=\{\mathrm{x}_t\}_{t=0}^\infty$ with $\mathrm{x}_t\in\mathbb{R}^{n}$, $\mathbf{M}$ and $\widehat{\mathbf{M}}$ denote causal and non-causal transfer matrices $\mathbf{M}(z) \coloneqq \sum^{\infty}_{\tau=0} z^{-\tau}M_\tau$ and $\widehat{\mathbf{M}}(z) \coloneqq \sum^{\infty}_{\tau=-\infty} z^{-\tau}M_{\tau}$ respectively with kernel matrices $M_t\in\mathbb{R}^{m\times n}$.

\section{Preliminaries and Problem Setup} \label{sec:problem setting}

We study a networked system with $N$ subsystems over a graph $\mathcal{G} = (\mathcal{V}, \mathcal{E})$ where the nodes $\mathcal{V}\coloneqq [N]$ are the subsystems and $\mathcal{E} \subseteq [N] \times [N]$. Let $\mathrm{diam}(\mathcal{G}) \in [N]$ denote the diameter of $\mathcal{G}$.  We denote $\mathcal{N}(i)\coloneqq\{i\}\cup\{j\in[N]|(i,j)\in\mathcal{E}\}$ as the \textit{neighboring set} of subsystem $i$ and  $d_\mathcal{G}(i,j): [N]\times[N]\to \mathbb{N}_+$ as the shortest distance from subsystem $i$ to $j$. 
The dynamics for each subsystem $i\in [N]$ is governed by
\begin{equation}
    \begin{aligned}
        \mathrm{x}_{t+1}^i = \sum_{j\in\mathcal{N}(i)}\left(A^{ij}\mathrm{x}_t^j +B^{ij}\mathrm{u}_t^j\right) + \mathrm{w}_t^i,
    \end{aligned}
    \label{eq:ddynamics}
\end{equation}
where 
$\mathrm{x}^{i}_t \in \mathbb{R}^{n_i}, \mathrm{w}^{i}_t\in\mathbb{R}^{n_{i}}$, and $\mathrm{u}_t^{i}\in\mathbb{R}^{m_i}$ are the local state, disturbance, and action for the $i$th subsystem, respectively.
Let $n = \sum_{i\in[N]} n_i$, $m = \sum_{i\in[N]} m_i$. We define $\mathrm{x}_t \in \mathbb{R}^{n}$, $\mathrm{u}_t \in \mathbb{R}^{m}$, and $\mathrm{w}_t \in \mathbb{R}^{n}$ as the global vectors of the $N$ agents with $$\mathrm{x}_t = \begin{bmatrix} \left(\mathrm{x}_t^1\right)^\top; \left(\mathrm{x}_t^2\right)^\top; \dots; \left(\mathrm{x}_t^N\right)^\top \end{bmatrix}^\top .$$ 

Vectors $\mathrm{u}_t$ and $\mathrm{w}_t$ are similarly defined. Let 
$A\coloneqq [A^{ij}]_{i,j\in[N]}\in\mathbb{R}^{n\times n}$ and $B\coloneqq [B^{ij}]_{i,j\in[N]}\in\mathbb{R}^{n\times m}$ be the concatenated global system dynamic matrices with $A^{ij} \equiv 0$, $B^{ij} \equiv 0$ for all $j\notin\mathcal{N}(i)$. Dynamics \eqref{eq:ddynamics} can be equivalently represented as
\begin{equation}
    \begin{aligned}
        \mathrm{x}_{t+1} = A\mathrm{x}_{t} + B\mathrm{u}_t + \mathrm{w}_t.
    \end{aligned}
    \label{dynamics}
\end{equation}

\subsection{Assumptions} Without loss of generality, we assume~\eqref{dynamics} is initialized with $\mathrm{x}_0 = \mathbf{0}$. Furthermore, we assume $(A,B)$ is stabilizable and there exists $K\in\mathbb{R}^{m\times n}$ such that $A-BK$ has a spectral radius $\rho <1$~\cite{dullerud2013course}. Thus, the Gelfand’s formula implies that there exist $L>0$, $\gamma\in (0,1)$ such that $\|A-BK\|^{t}\leq L\gamma^t$ for all $t\geq 0$. The system is also said to be $(L,\gamma)$-\textit{stabilizable}. We further assume the system disturbances are uniformly bounded, i.e., 
{there exists a constant $W>0$ such that $\mathrm{w}_{0:T-1}\in \mathcal{W}\coloneqq \{\mathrm{w}_{0:T-1}:\|\mathrm{w}_t^i\|_{\infty}\leq W, i\in [N],t\in [T]\}$}. Uniformly bounded disturbances are standard in the robust control literature~\cite{li2022robustness, yu2020power}, which accommodate a wide range of disturbance models, including those that are stochastic, time-coupling, and adversarial.

\paragraph{$\distance$-localized control.} 
\label{subsec:communication_constraints}
Based on the communication topology $\mathcal{G}$, a controller is \textit{$\kappa$-localized},\footnote{Unlike the $\distance $-distributed controller in~\cite{shin2023near}, which is a linear state feedback controller in $\mathrm{x}_t$, the affine controller in this work depend not only on the state $\mathrm{x}_t$, but also predictions of future disturbances $(\mathrm{w}_{\tau}:\tau\geq t)$.} if for any subsystem $i$, the computation of $\mathrm{u}_t^i$ only depends on the information available at subsystems 
$j\in\mathcal{N}^{\distance}_{\mathcal{G}}(i)\coloneqq\left\{j\in[N]\left|d_{\mathcal{G}}(j, i)\leq\distance\right.\right\}$  
 with some \textit{communication range} $\distance \in \mathbb{N}$. 
The parameter $\distance$ is a design variable.
Our work explores the co-design of $\kappa$ and the distributed controller when there are prediction errors. The following assumption regulates the expansion of the communication graph $\mathcal{G}$.

\begin{assumption}[Sub-exponential expansion]\label{assumption:subexponential}
    \textit{Given a topology $\mathcal{G} = (\mathcal{E},\mathcal{V})$ and a distance metric function $d_{\mathcal{G}}(i,j)$ for nodes $i, j \in \mathcal{V}$, there exists a sub-exponential function $g(\cdot)$ such that}
$
        |\{j\in\mathcal{V}: d_{\mathcal{G}}(i,j)=d\}|\leq g(d), \ \text{ for all } i\in \mathcal{V}.
$
\end{assumption}

{Common topologies, including arbitrary polynomial-growth and regular graphs, satisfy the requirement that the growth rate of each node degree is sub-exponential. This assumption has been adopted in several previous works \cite{shin2023near, xu2024stability, shin2022exponential}.}

Next, we formally present the problem considered in this work.

\subsection{Problem Statement}
In this paper, we consider $\kappa$-localized predictive controllers  for the centralized finite horizon LQR problem, formulated as:
\begin{equation}
\tag{CLQR}
    \begin{aligned}
J^{\star} \coloneqq&
\min_{\mathrm{x}_{1:T},\mathrm{u}_{0:T-1}}\sum^{T-1}_{t=0}\left(\mathrm{x}_t^{\top} Q \mathrm{x}_t^{\phantom{\top}} + \mathrm{u}_t^\top R \mathrm{u}_t^{\phantom{\top}}\right) + \mathrm{x}_T^\top Q_T \mathrm{x}_T^{\phantom{\top}} \\&\text{\,\,\,\,\,\,\,subject to\,\,\,\,\,\,\,~\eqref{dynamics} \ for all  $t\in [T]$.} 
    \end{aligned}
    \label{eq:problem}
\end{equation}

Here, $ Q, Q_T \in \mathbb{R}^{n \times n}$ and $R \in \mathbb{R}^{m \times m}$ are positive definite matrices, with $Q_T$ representing the terminal state cost. When $Q_T = P$ is the Riccati solution, a closed-form solution to \eqref{eq:problem} exists and can be expressed in terms of the optimal linear state-feedback gain and disturbances $\mathrm{w}_{0:T-1}$~\cite{yu2020power,goel2022power}, as detailed in \Cref{sec:benchmark}. 

In this work, we consider a $\kappa$-localized predictive controller that operates with unknown disturbances $\mathrm{w}_{0:T-1}$. In many applications, each subsystem $i$ has access to local predictions of future disturbances, denoted as $\widehat{\mathrm{w}}^i_{0:T-1}$, and share them to its $\distance$-hop neighbors subject to communication constraints. These neighbors are defined to be the set $\mathcal{N}^{\distance}_{\mathcal{G}}(i)$. 
Formally, subsystem $i$ has access to the collection of local predictions $\{ \mathrm{w}_{0:T-1}^j: j \in \mathcal{N}_{\mathcal{G}}^{\distance}(i) \}$, where each $\smash{\mathrm{w}_{0:T-1}^j}$ represents a predicted disturbance trajectory from neighbor $j$. Notably, these predictions are neither global nor guaranteed to be accurate. Furthermore, we assume the predictions are uniformly bounded such that $\widehat{\mathrm{w}}_{0:T-1}\in\mathcal{W}$.

\paragraph{Local prediction error.} For notational simplicity, we write $\mathrm{w}\coloneqq \mathrm{w}_{0:T-1}$ and $\widehat{\mathrm{w}}\coloneqq\widehat{\mathrm{w}}_{0:T-1}$. We quantify the prediction error as the worst-case cumulative error ${\varepsilon}\left(\mathrm{w},\widehat{\mathrm{w}}\right)$ over $T$ as follows
\begin{equation}
    \begin{aligned}
    \label{eq:local_prediction_error}
    {\varepsilon}\left(\mathrm{w},\widehat{\mathrm{w}}\right) \coloneqq \sup_{i\in[N]}\sum^{T-1}_{t=0}\left(e_t^i\right)^2 \text{, with }    e_t^i \coloneqq\left\|\mathrm{w}_t^i - \widehat{\mathrm{w}}_t^i\right\|.
    \end{aligned}
\end{equation}

We denote the set of all predictions that satisfy a local prediction error bound 
$\Bar{\varepsilon}$ by $\overline{\mathcal{W}}(\mathrm{w};\Bar{\varepsilon})\coloneqq \{\mathrm{w}:{\varepsilon}\left(\mathrm{w},\widehat{\mathrm{w}}\right)\leq \Bar{\varepsilon}\}$.
This error bound intuitively describes the worst-case \textit{local prediction error} in the overall networked system. If the local error differences between subsystems are negligible, this indicator reduces to the commonly adopted error definition (see \cite{lin2022bounded}) by $N\Bar{\bm{\varepsilon}}$.

\paragraph{Performance benchmark.}
\label{sec:dynamic_regret}
We use
the \textit{dynamic regret} as the  worst-case performance metric to evaluate controllers. It measures the worst-case gap between the quadratic cost $$J(\pi)=\sum^{T-1}_{t=0}\left(\mathrm{x}_t(\pi)^{\top} Q \mathrm{x}_t(\pi)^{\phantom{\top}} + \mathrm{u}_t(\pi)^\top R \mathrm{u}_t(\pi)^{\phantom{\top}}\right) + \mathrm{x}_T(\pi)^\top Q_T \mathrm{x}_T(\pi)^{\phantom{\top}} $$ 
induced by a controller $\pi$ and the non-causal optimal cost $J^\star$ to \eqref{eq:problem}, defined as
\begin{align}
    \label{eq:dynamic_regret}
    \mathrm{DR}(\pi)\coloneqq \sup_{\mathrm{w}\in \mathcal{W}}\,\,\,\sup_{\widehat{\mathrm{w}}\in\overline{\mathcal{W}}(\mathrm{w};\Bar{\varepsilon})} \left(J(\pi) - J^{\star}\right),
\end{align}
with respect to system disturbances and predictions. The notion of dynamic regret in control has been explored in recent works, such as~\cite{lidisentangling, baby2022optimal}. Unlike the benchmarks in regret-optimal control~\cite{abbasi2011regret, goel2023regret, al2023distributionally} and the cost gap in distributed linear systems~\cite{shin2023near, zhang2023optimal}, which compare performance against a static optimal causal controller $K^{\star}$, dynamic regret above defines a stronger benchmark that can be time-varying.

\section{Predictive System-Level Synthesis \label{section:centralized}}

In this section, we introduce the framework of predictive system-level synthesis ($\psls$). It parameterizes all achievable closed-loop system responses under affine feedback predictive controllers of the form
\begin{align}
\label{eq:noncausal_form}
   \mathrm{u}_t =  \sum_{\tau \leq t} K_{t,\tau}\mathrm{x}_{t-\tau} + \sum^{\infty}_{\tau = 0} L_{t,\tau} \widehat{\mathrm{w}}_{t+\tau},
\end{align}
where $K_{t,\tau}\in \mathbb{R}^{m\times n}$ and $L_{t,\tau}\in \mathbb{R}^{m\times n}$ for $t,\,\tau \in \mathbb{N}$ parameterizes the affine predictive controller \eqref{eq:noncausal_form}. Furthermore, $\psls$ provides a dynamic predictive controller that realizes the prescribed closed-loop system responses (\Cref{theorem constraints}).

We show that \psls naturally lends itself to distributed and localized synthesis and implementation thanks to the spatio-temporally exponential decaying properties of the parameterization, as detailed in Theorem \ref{theorem:temporal}.

\subsection{System-Level Parameterization of Predictive Controllers}
\label{subsec:predsls_intro}

Consider the closed loop of \eqref{dynamics} under the (dynamic) state-feedback non-causal predictive controller \eqref{eq:noncausal_form}, which can be compactly written as $\mathbf{u} = \mathbf{Kx} + \widehat{\mathbf{L}}\widehat{\mathbf{w}}$ where $\mathbf{K}$ and $\widehat{\mathbf{L}}$ are proper and improper transfer matrices.\footnote{Note that the optimal non-causal (offline) controller that minimizes \eqref{eq:problem}
follows this form, as discussed in~\cite{yu2020power,goel2022power}.} We denote the closed-loop mappings from exogenous inputs $\mathbf{w}$ and $\mathbf{ \widehat w}$ to the state and control actions as follows
\begin{equation}
    \begin{aligned} 
    \begin{bmatrix}
         \mathbf{x}\\
          \mathbf{u}
    \end{bmatrix} = \begin{bmatrix}
              \mathbf{\Phi}^{\mathrm{x}} \ & \ \widehat{\mathbf{\Phi}}^{\mathrm{x}}\\
              \mathbf{\Phi}^{\mathrm{u}} \ & \ \widehat{\mathbf{\Phi}}^{\mathrm{u}}
          \end{bmatrix}\begin{bmatrix}
              \mathbf{w}\\
              \widehat{\mathbf{w}}
          \end{bmatrix},
  \end{aligned}\label{closedloop responses}
\end{equation}
where it can be verified via simple algebra that the mappings $\bm{\Phi}^{\mathrm{x}}: \mathbf{w}\to\mathbf{x},  \widehat{\bm{\Phi}}^{\mathrm{x}}: \widehat{\mathbf{w}}\to\mathbf{x},
              \bm{\Phi}^{\mathrm{u}}: \mathbf{w}\to\mathbf{u}$, and 
$\widehat{\bm{\Phi}}^{\mathrm{u}}: \widehat{\mathbf{w}}\to\mathbf{u}$ satisfy
\begin{align}  \label{eq:mappings}
    \begin{bmatrix}
              \mathbf{\Phi}^{\mathrm{x}} \ & \  \widehat{\mathbf{\Phi}}^{\mathrm{x}}\\
              \mathbf{\Phi}^{\mathrm{u}} \ & \ \widehat{\mathbf{\Phi}}^{\mathrm{u}}
          \end{bmatrix} &= \begin{bmatrix}
        U \ \ & \ \ UB\widehat{\mathbf{L}}\\
        \mathbf{K}U \ \ & \ \ (\mathbf{K}UB + I)\widehat{\mathbf{L}}
    \end{bmatrix}
\end{align}
with
$
    U\coloneqq(zI-(A+B\mathbf{K}))^{-1}.
$
We refer to $\mathbf{\Phi}$ as a group of these four \textit{closed-loop mappings} (transfer matrices), and define it as $\mathbf{\Phi} \coloneqq [\mathbf{\Phi}^{\mathrm{x}}, \widehat{\mathbf{\Phi}}^{\mathrm{x}}; \mathbf{\Phi}^{\mathrm{u}}, \widehat{\mathbf{\Phi}}^{\mathrm{u}}]$. The following theorem provides an affine parameterization for all such closed-loop mappings induced by controllers of the form  \eqref{eq:noncausal_form}.

\begin{theorem}[Predictive system-level parameterization]
\label{theorem constraints}
The affine subspace defined by 
\begin{equation}
    \begin{aligned} 
    &\begin{bmatrix}
       zI-A \ \  &  \ \ - B   
    \end{bmatrix}
         \begin{bmatrix}
             \mathbf{\Phi}^\mathrm{x} \ & \ \widehat{\mathbf{\Phi}}^{\mathrm{x}}\\
             \mathbf{\Phi}^\mathrm{u} \ & \ \widehat{\mathbf{\Phi}}^{\mathrm{u}}
         \end{bmatrix} = \begin{bmatrix}
             I \ & \ 0
         \end{bmatrix}
  \end{aligned} \label{eq:cl-constraints}
\end{equation} 
characterizes all achievable closed-loop responses \eqref{closedloop responses} for \eqref{dynamics} under  \eqref{eq:noncausal_form}. Moreover, given any $\mathbf{\Phi}$ that satisfies \eqref{eq:cl-constraints}, the predictive controller \eqref{eq:noncausal_form} instantiated with
\begin{equation}
\label{eq:sls-controller}
    \begin{aligned} 
    \mathbf{K} = \mathbf{\Phi}^\mathrm{u}(\mathbf{\Phi}^{\mathrm{x}})^{-1}, \quad\widehat{\mathbf{L}}=\widehat{\mathbf{\Phi}}^{\mathrm{u}} - \mathbf{\Phi}^\mathrm{u}(\mathbf{\Phi}^{\mathrm{x}})^{-1}\widehat{\mathbf{\Phi}}^{\mathrm{x}},
  \end{aligned}
\end{equation}
achieves the closed-loop response prescribed by $\mathbf{\Phi}$ for \eqref{dynamics} in the sense of \eqref{closedloop responses}.
\end{theorem}

The proof of Theorem~\ref{theorem constraints} can be found in Appendix \ref{appendix:theorem 3.1}. 
A direct consequence of Theorem~\ref{theorem constraints} is the equivalence of~\eqref{eq:noncausal_form} and the synthesized $\mathbf{K}$ and $\widehat{\mathbf{L}}$ parameterized by closed-loop mappings in the affine space described by~\eqref{eq:cl-constraints}. If we impose the constraints $\bm{\Phi}^\mathrm{x}, \bm{\Phi}^\mathrm{u}\in \frac{1}{z}\mathcal{RH}_\infty$ and $\widehat{\bm{\Phi}}^\mathrm{x}, \widehat{\bm{\Phi}}^\mathrm{u}\in \mathcal{RH}_\infty^i$ in addition to~\eqref{eq:cl-constraints}, {where $\mathcal{RH}_\infty$ denotes the space of strictly proper real rational stable transfer matrices while $\mathcal{RH}_\infty^i$} denotes the space of all improper and real rational stable transfer matrices, then the corresponding controller \eqref{eq:sls-controller} is internally stabilizing.

The parameterization in Theorem~\ref{theorem constraints} enables the formulation of a system-level predictive controller optimization with quadratic cost, which we term~\eqref{eq:centralized problem}:  
    \begin{align}
\min_{\left(\mathbf{\Phi}^{\mathrm{x}},\widehat{\mathbf{\Phi}}^{\mathrm{x}},\mathbf{\Phi}^{\mathrm{u}},\widehat{\mathbf{\Phi}}^{\mathrm{u}}\right)}\left\|\begin{bmatrix}
            Q^{\frac{1}{2}} & \mathbf{0} \\ \mathbf{0} & R^{\frac{1}{2}}
        \end{bmatrix}\begin{bmatrix}
            \mathbf{\Phi}^{\mathrm{x}} + \widehat{\mathbf{\Phi}}^{\mathrm{x}}\\\mathbf{\Phi}^{\mathrm{u}} + \widehat{\mathbf{\Phi}}^{\mathrm{u}}
        \end{bmatrix}\right\|_{\mathcal{H}_2}^2\quad\text{subject~to} ~\eqref{eq:cl-constraints}, \tag{\psls}  
    \label{eq:centralized problem}
    \end{align} 
    where for $\bm{G}=\sum_{t=-\infty}^{\infty}z^{-t}G_{t}$, 
$\|\bm{G}\|_{\mathcal{H}_2}\coloneqq \left(\sum_{t=-\infty}^{\infty}\|G_t\|^2_{F}\right)^{1/2}$ is the $\mathcal{H}_2$-norm of a transfer function $\bm{G}$.

It turns out that the optimal closed-loop mappings obtained by solving~\eqref{eq:centralized problem} are also optimal for the problem~\eqref{eq:problem} in Section~\ref{sec:problem setting}, as established by the theorem below, whose proof can be found in Appendix \ref{proof:proposition:1}.

\begin{proposition}[Optimal predictive controller]
\label{proposition:1}
Consider an optimal solution of \eqref{eq:centralized problem}, denoted by $\mathbf{\Phi^\star} = (\PHI^{\mathrm{x},\star},\widehat{\PHI}^{\mathrm{x},\star},\PHI^{\mathrm{u},\star},\widehat{\PHI}^{\mathrm{u},\star})$. For any disturbance sequence $\mathrm{w}_{0:T-1}$, the predictive controller \eqref{eq:noncausal_form} instantiated with
\vspace{-5pt}
\begin{align*}
    \mathbf{K}= \mathbf{\Phi}^\mathrm{u,\star}(\mathbf{\Phi}^{\mathrm{x},\star})^{-1}\text{ and }\,\widehat{\mathbf{L}}=\widehat{\mathbf{\Phi}}^{\mathrm{u},\star} - \mathbf{\Phi}^{\mathrm{u},\star}(\mathbf{\Phi}^{\mathrm{x},\star})^{-1}\widehat{\mathbf{\Phi}}^{\mathrm{x},\star}
\end{align*}
and implemented using exact predictions such that $\widehat{\mathrm{w}}_t = \mathrm{w}_t$ for all $t\in[T]$ achieves the optimal cost of \eqref{eq:problem} with a terminal cost $Q_T=P$ where $P$ is the solution to the algebraic Riccati equation $P = A^\top P A - (A^\top P B)(R + B^\top P B)^{-1}(B^\top P A) + Q$.
\end{proposition}

While our \psls framework is applied to linear time-invariant systems, its principles can be readily extended to linear time-varying systems. 
The synthesized solution in this case will be the optimal non-causal state-feedback LTV controller.

\subsection{Finite Horizon Approximation and Implementation}
\label{subsec:finite_time_approx}

The framework introduced in the previous subsection is formulated for closed-loop mappings that are transfer matrices. Consequently, \eqref{eq:centralized problem} contains infinitely many optimization variables, making it impractical for direct computation using standard off-the-shelf convex optimization tools. To address this limitation, we introduce a finite horizon approximation of $\psls$ by restricting the closed-loop mappings to have finite impulse response (FIR) of horizon $H$, e.g., \begin{equation}
    \label{eq:index}
    \bm{\Phi} \in  \mathcal{F}_H \coloneqq \left\{\bm{\Phi}:\bm{\Phi} = \sum^{H}_{\tau=-H}z^{-\tau}\mathrm{\Phi}_\tau\right\}.
\end{equation}

In particular, the FIR closed-loop mappings are of the form:
\begin{align}
\label{eq:FIR}
\mathbf{\Phi}^{\mathrm{x}}_{\mathcal{F}_H}&\coloneqq\begin{bmatrix}
        \Phi^{\mathrm{x}}_{0,0}  & \mathbf{0} & \cdots & \cdots & \cdots & \mathbf{0}\\
        \vdots & \ddots& \ddots & & &\vdots\\
        \Phi^{\mathrm{x}}_{H,0} & &\ddots&\ddots& &\vdots  \\
        \mathbf{0} & \ddots& &\ddots & \ddots &\vdots \\
        \vdots &\ddots & \ddots &&\ddots &\mathbf{0}\\
        \mathbf{0}&\cdots  &\mathbf{0}  & {\Phi}^{\mathrm{x}}_{T,T-H}&\cdots & {\Phi}^{\mathrm{x}}_{T,T}
    \end{bmatrix}, \widehat{\mathbf{\Phi}}^{\mathrm{x}}_{\mathcal{F}_H} \coloneqq\begin{bmatrix}
        \widehat{\Phi}^{\mathrm{x}}_{0,0}  & \cdots& \widehat{\Phi}^{\mathrm{x}}_{0,H} &\mathbf{0}&\cdots&\mathbf{0}\\
        \vdots & \ddots& &\ddots &\ddots&\vdots \\
        \widehat{\Phi}^{\mathrm{x}}_{H,0} & & \ddots & &\ddots&\mathbf{0}\\
        \mathbf{0}&\ddots &  &\ddots& &\widehat{\Phi}^{\mathrm{x}}_{T-H,T}\\
        \vdots&\ddots & \ddots &&\ddots&\vdots \\
        \mathbf{0}&\cdots &\mathbf{0}  & \widehat{\Phi}^{\mathrm{x}}_{T,T-H}&\cdots & \widehat{\Phi}^{\mathrm{x}}_{T,T}
    \end{bmatrix}, \end{align}
with component matrices $\Phi^{\mathrm{x}}_{t,k}, \widehat{\Phi}^{\mathrm{x}}_{t,k}\in\mathbb{R}^{n\times n}$ where we map the indices $\tau$ from the kernel matrices in \eqref{eq:index} to a double index $t,\,k$ with $\tau = t-k$ for the clarity of the presentation in the later sections. The component matrices of $\bm{\Phi}^{\mathrm{u}}_{\mathcal{F}_H}$ and $\widehat{\bm{\Phi}}^{\mathrm{u}}_{\mathcal{F}_H}$ are similarly defined. Restricting the closed-loop mappings to have FIR, the synthesis problem \eqref{eq:centralized problem} can be equivalently expressed in terms of the component matrices as
    \begin{align} \label{finite problem}
\min_{\mathbf{\Phi}_{\mathcal{F}_T}}&\sum^{T}_{t=0}\sum^{\overline{t}}_{k=\underline{t}}\left\|\begin{bmatrix}
            Q^{\frac{1}{2}} & \mathbf{0} \\ \mathbf{0} & R^{\frac{1}{2}}
        \end{bmatrix}\begin{bmatrix}
            \Phi^{\mathrm{x}}_{t,k} + \widehat{\Phi}^{\mathrm{x}}_{t,k} \\\Phi^{\mathrm{u}}_{t,k} + \widehat{\Phi}^{\mathrm{u}}_{t,k}  
        \end{bmatrix}\right\|_{\mathcal{F}}^2\quad\text{subject to \eqref{eq:constraint1}-\eqref{eq:constraint2}}, \ \text{for all } k\in[T],
    \end{align} 
where $\underline{t}\coloneqq \max(0,t-H)$, and $\overline{t}\coloneqq\min(T,t+H)$ for all $t\in[T]$.
The constraint \eqref{eq:cl-constraints} is also transformed into equivalent time-domain constraints on the component matrices: 
\begin{subequations}
\label{eq:constraints}
    \begin{align}
    \Phi^{\mathrm{x}}_{t,k}&=I,\text{for $t=k$,}\label{eq:constraint1}
    \\
    \Phi^{\mathrm{x}}_{t+1,k}&=\begin{cases}
        A\Phi^{\mathrm{x}}_{t,k}+B\Phi^{\mathrm{u}}_{t,k}, &\text{if $t\in\{k,\ldots,\overline{k}\}$,}\\
        \mathbf{0},&\text{otherwise},
    \end{cases} \\
    \widehat{\Phi}^{\mathrm{x}}_{t,k} &= \mathbf{0},\text{for $t=0$,}\\
        \widehat{\Phi}^{\mathrm{x}}_{t+1,k} &= \begin{cases}          A\widehat{\Phi}^{\mathrm{x}}_{t,k}+B\widehat{\Phi}^{\mathrm{u}}_{t,k},&\text{ if $t\in\{\underline{k},\ldots,\overline{k}\}$},\\
            \mathbf{0}, &\text{ otherwise},
        \end{cases}\label{eq:constraint2}
    \end{align}
\end{subequations}
where $\overline{k}\coloneqq\min(T-1,k+H-1)$, $\underline{k}\coloneqq\max(0,k-H+1)$ for all $k\in[T]$. 
Correspondingly, a time-domain realization for the \psls controller \eqref{eq:sls-controller} is as follows
\begin{subequations}
\label{eq:policy}
    \begin{align}
    \label{eq:policy1}
    \widetilde{\mathrm{w}}_t &\coloneqq \mathrm{x}_t - \sum^{t-1}_{k=\underline{t}}\Phi^{\mathrm{x}}_{t,k}\widetilde{\mathrm{w}}_{k}, \\
    \label{eq:policy2}
    \Bar{\mathrm{w}}_t &\coloneqq \widehat{\mathrm{x}}_t - \sum^{t-1}_{k=\underline{t}}\Phi^{\mathrm{x}}_{t,k}\Bar{\mathrm{w}}_{k}, \quad
    \widehat{\mathrm{x}}_t\coloneqq \sum^{\overline{t}}_{k=\underline{t}}\widehat{\Phi}^{\mathrm{x}}_{t,k}\widehat{\mathrm{w}}_{k-1},\\
        \mathrm{u}_t &= \sum^t_{k=\underline{t}}\Phi^{\mathrm{u}}_{t,k}\left(\widetilde{\mathrm{w}}_{k} - \Bar{\mathrm{w}}_{k}\right) + \sum^{\overline{t}}_{k=\underline{t}}\widehat{\Phi}^{\mathrm{u}}_{t,k}\widehat{\mathrm{w}}_{k-1}, \label{eq:policy3}
    \end{align}
\end{subequations}
where $\widehat{\mathrm{w}}$ is the disturbance prediction; $\widetilde{\mathrm{w}}$ and  $\Bar{\mathrm{w}}$ are controller internal states that keep track of past disturbances up to time~$t$ and future predictions, with $\widetilde{\mathrm{w}}_0 \coloneqq \mathrm{x}_0, \Bar{\mathrm{w}}_0 \coloneqq 0$. 
With simple algebra, it is straightforward to verify that 
$\widetilde{\mathrm{w}}_t-\Bar{\mathrm{w}}_t = \mathrm{w}_{t-1}$.

\begin{remark}[Approximation error]
    Due to the FIR constraint on the closed-loop mappings, the solution to \eqref{finite problem} is sub-optimal. As a first step towards quantifying the sub-optimality, in \Cref{theorem:temporal} we show that the closed-loop mappings present an exponentially decaying structure over the FIR horizon, e.g., $\|\widehat{\Phi}^{\mathrm{x}}_{t,k}\|$ decays exponentially with respect to $|t-k|$.
\end{remark}

\section{Distributed and Localized Synthesis \label{sec:distributed setting}}
For large-scale networks, solving \eqref{eq:centralized problem} and implementing the corresponding centralized controller impose significant communication and computation burdens. In this section, we describe a decomposition of the centralized problem~\eqref{eq:centralized problem} via agent-wise parallel synthesis, thus enabling scalable computation. In what follows, we let the FIR horizon $H=T$ where $T$ is the problem horizon defined in ~\eqref{eq:problem} for simplicity.\footnote{Note that \eqref{finite problem} is equivalent to the $H$-horizon problem \eqref{eq:problem} with the termination condition with $Q_H = P$.}

Specifically, the synthesis problem \eqref{eq:centralized problem} described above can be viewed as an instantiation of the general optimal control problem:
\begin{align}
    &\min_{\mathbf{\Phi}} C\left(\mathbf{\Phi}^{\mathrm{x}}, \widehat{\mathbf{\Phi}}^{\mathrm{x}}, \mathbf{\Phi}^{\mathrm{u}},\widehat{\mathbf{\Phi}}^{\mathrm{u}}\right),\label{eq:general}\\
    \text{subject to }\quad&  \eqref{eq:cl-constraints},\,\,\,\mathbf{\Phi}\in \mathcal{L}\cap\mathcal{F}_T\cap\mathcal{X}.\nonumber
\end{align}

Function $C(\cdot)$ denotes a convex system-level objective, subspace $\mathcal{L}$ represents any convex spatial-locality constraints, and $\mathcal{X}$ is a convex sparsity subspace that can be used to model communication delay. All these definitions follow the classic SLS literature (see \cite{anderson2019system}). In this work, we focus on the following localized communication constraint set $\mathcal{L}_{\distance}$:
\begin{definition} [$\distance$-locality constraint set] 
\label{definition:locality}
The $\distance$-locality set $\mathcal{L}_{\distance}$ is a set of matrices of particular sparsity patterns defined according to the communication topology, where $\mathcal{L}_{\distance}:= \{(\mathrm{\Phi}^{\mathrm{x}}, \mathrm{\Phi}^{\mathrm{u}})\in\mathbb{R}^{m\times m}\times\mathbb{R}^{m\times n}: \mathrm{sp}(\mathrm{\Phi}^{\mathrm{x}})= \mathcal{C}^{\distance}, \,\mathrm{sp}(\mathrm{\Phi}^{\mathrm{u}})= \mathcal{C}^{\distance+1} \}$, with $\mathcal{C}^\kappa\in\{0,1\}^{N\times N}$, $\mathcal{C}^\kappa(i,j)=1$ for $\forall i\in[N]$ and $j\in\mathcal{N}_{\mathcal{G}}^\kappa(i)$, and $\mathcal{C}^\kappa(i,j)=0$ otherwise.
\end{definition}

\textbf{Localized synthesis.}
Let $\mathbf{\Phi}(:,i)\coloneqq[\mathbf{\Phi}^{\mathrm{x}}(:,i),\widehat{\mathbf{\Phi}}^{\mathrm{x}}(:,i);\mathbf{\Phi}^{\mathrm{u}}(:,i),\widehat{\mathbf{\Phi}}^{\mathrm{u}}(:,i)]$ for distributed control. If \eqref{eq:general} is \textit{column-wise separable}~\cite{wang2018separable}, then it can be partitioned into $N$ parallel sub-problems in the form: 
\begin{align}
    & \min_{\mathbf{\Phi}(:,i)} C_i\left(\mathbf{\Phi}^{\mathrm{x}}(:,i), \widehat{\mathbf{\Phi}}^{\mathrm{x}}(:,i), \mathbf{\Phi}^{\mathrm{u}}(:,i),\widehat{\mathbf{\Phi}}^{\mathrm{u}}(:,i)\right),\label{eq:dgeneral}\\
    \text{subject to }\quad&  \begin{bmatrix}
       zI-A \ \  &  \ \ - B   
    \end{bmatrix}
         \mathbf{\Phi}(i) = \begin{bmatrix}
             I(:,i) \ && \ 0
         \end{bmatrix},\nonumber\\
         &\mathbf{\Phi}(:,i)\in \mathcal{L}(:,i)\cap\mathcal{F}_T\cap\mathcal{X}_i, \nonumber
\end{align}
where $C_i(\cdot)$, $\mathcal{L}_d(:,i)$, and $\mathcal{X}_i$ are the column-wise objective function and subspace constraints, respectively. 

We now turn to illustrate a specific case of \eqref{eq:dgeneral}, called \eqref{localized problem}, which naturally leads to an optimal $\kappa$-localized controller defined in \Cref{sec:problem setting}. We first introduce an equivalent representation of the communication constraints for $\kappa$-localized control in terms of the closed-loop mappings $\bm{\Phi}$.
For simplicity,\footnote{For SLS, the communication graph in general does not have to match the system dynamics~\cite{wang2014localized}.} consider a communication graph that shares the same topology as the dynamics $\mathcal{G}=([N],\mathcal{E})$.

Imposing localization on the closed-loop mappings enforces that in the closed loop, disturbances are prevented from propagating outside of the localized region. 
Thanks to \Cref{theorem constraints}, a key feature of the \psls is that any localization constraint on the closed-loop mappings directly translates to the controller realization, e.g., \eqref{eq:policy}, enabling localized controller implementation.

\subsection{Spatial and Temporal Decompositions} 
We now show how \eqref{finite problem} can be split into independent column-wise sub-problems via two forms of decomposition: \textit{spatial} and \textit{temporal}.

\textbf{Spatial decomposition.} 
We decompose the centralized problem \eqref{eq:centralized problem} into $N$ distinct sub-problems. Each sub-problem \eqref{localized problem} is an optimization that corresponds to the $i$th column of the FIR closed-loop mapping matrices, $(\Phix_{t,k},\hPhix_{t,k},\Phiu_{t,k},\hPhiu_{t,k})_{t,k=0}^T$, as defined in \eqref{eq:index}: 
\begin{subequations}
    \begin{align}
&\min_{\bm{\varphi}}\sum^{T}_{t=0}\sum^{T}_{k=0}\left\|\begin{bmatrix}
            Q^{\frac{1}{2}} & \mathbf{0} \\ \mathbf{0} & R^{\frac{1}{2}}
        \end{bmatrix}\begin{bmatrix} \pxl_{t,k} + \hpxl_{t,k} \\ \pul_{t,k} + \hpul_{t,k} 
        \end{bmatrix}\right\|^2 \tag{\text{\dpsls}} \label{localized problem} \\
        \text{subject~to} &\quad  \pxl_{t,k}=\mathrm{e}_i, \text{ for $t=k$, $k\in[T]$},\nonumber\\&\quad
        \pxl_{t+1,k}=\begin{cases}
            A\varphi^{\mathrm{x},i}_{t,k}+B\varphi^{\mathrm{u},i}_{t,k}, &\text{ if $t\in\{k,\ldots,T-1\}$, $k\in[T]$},\\
            \mathbf{0}, &\text{otherwise}. 
        \end{cases}
	  \tag{causal dynamics}\label{eq:causal dynamics}\\&\quad  \hpxl_{t,k}=\mathbf{0}, \text{ for $t=0$, $k\in[T]$},\nonumber\\
         &\quad \hpxl_{t+1,k} = \begin{cases}
             A\widehat{\varphi}^{\mathrm{x},i}_{t,k}+B\widehat{\varphi}^{\mathrm{u},i}_{t,k}, & \text{if $t,k\in[T]$,}\\
             \mathbf{0}, & \text{otherwise},
         \end{cases} \tag{non-causal dynamics}\label{eq:non-causal dynamics}\\
              &\quad\left(\pxl_{t,k},\pul_{t,k}\right), \left(\hpxl_{t,k},\hpul_{t,k}\right) \in \mathcal{L}_{\distance}(:,i), \,\,\text{for $t,k\in[T]$,} \tag{locality constraints} \label{eq:locality constraint}
    \end{align} 
\end{subequations}
where we denote {$(\pxl_{t,k},\hpxl_{t,k},\pul_{t,k},\hpul_{t,k})_{t,k=0}^T$} the corresponding column variables associated with the $i$th sub-problem, with each variable satisfying $\pxl_{t,k},\hpxl_{t,k}\in\mathbb{R}^{n}, \pul_{t,k},\hpul_{t,k}\in\mathbb{R}^{m}$. 
Note that the constraints in \eqref{localized problem} are the column-wise components of the constraints in \eqref{eq:centralized problem}.

\paragraph{Temporal decomposition.} 
To reduce the size of the column sub-problems, we further decompose the \eqref{localized problem} problem temporally along the FIR horizon for each $k \in [T]$ as follows:
\begin{align}&\min\sum^{T}_{t=0}\left\|\begin{bmatrix}
            Q^{\frac{1}{2}} & \mathbf{0} \\ \mathbf{0} & R^{\frac{1}{2}}
        \end{bmatrix}\begin{bmatrix} \pxl_{t,k} + \hpxl_{t,k} \\ \pul_{t,k} + \hpul_{t,k} 
        \end{bmatrix}\right\|^2\label{eq:k-decomp}\tag{\decpsls}\\
        \text{subject to} & \quad 
\eqref{eq:causal dynamics}, \eqref{eq:non-causal dynamics}, \eqref{eq:locality constraint}.\nonumber
\end{align} 

Concatenating the solutions from \eqref{eq:k-decomp} along the temporal index $k\in[T]$ recovers the optimal solution to \eqref{localized problem}, as neither the objective function nor the constraints contain coupling terms with respected to $k$. 

A key advantage of the \psls framework is its ability to directly synthesize the optimal controller under communication constraints, rather than relying on separate design and truncation steps. This is a significant departure from common methods that first design a centralized controller and then truncate it to fit a communication topology~\cite{zhang2023optimal,shin2023near,xu2024stability}. By formulating the controller synthesis as a single optimization problem that incorporates these constraints from the outset, \psls achieves a truly optimal solution for a given communication range, $\kappa$. This unified approach is formally stated in the following remark, which is a direct consequence of~\Cref{theorem constraints}.

\begin{remark}[\ouralg vs. post hoc controllers]
By~\Cref{theorem constraints}, the set of CLMs characterized by the \psls framework consists of the CLMs induced by any post-hoc $\kappa$-truncated LQR controllers (e.g., \cite[Eq.~($\kappa$-distributed)]{shin2023near} and \cite[Eq.~(8)]{zhang2023optimal}). Therefore, the \ouralg controller is guaranteed to outperform existing $\kappa$-localized controllers.
\end{remark}

\section{Performance Analysis of~\ouralg} 

In this section, we present the main results this paper: {(1) optimal solution derivation for \eqref{eq:k-decomp}, (2) decaying properties of the solution to \eqref{eq:centralized problem}, (3) regret analysis of the proposed $\kappa$-localized controller. First, we reformulate \eqref{eq:k-decomp} into two surrogate problems, which facilitates the closed-form derivation of the global optima of the original problem (\Cref{proposition:2}). In the second part, we establish the temporally and spatially decaying properties of the closed-loop mappings (CLMs) provided by solving \eqref{eq:centralized problem} (\Cref{theorem:temporal} and \ref{theorem:spatial} respectively). Subsequently, we demonstrate that the upper bound of the performance difference, induced by the offline centralized optimal policy to \eqref{eq:problem} and the \psls controller computed with \eqref{eq:centralized problem}, can be decomposed into a sum of terms governed by three distinct contributing factors, establishing a nontrivial communication and performance trade-off.

\subsection{Optimum Characterization}
We continue with the distributed setting of \psls from \Cref{sec:distributed setting}. In this section, we establish the equivalence between the optimal solution of the original problem \eqref{eq:k-decomp} and its surrogate problems. To this end, we define 
    \begin{align}
    \label{eq:k-ca}
         & \min_{\psi}\sum^{T-k}_{t=0}\left\|\begin{bmatrix}
            Q^{\frac{1}{2}} & \mathbf{0} \\ \mathbf{0} & R^{\frac{1}{2}}
        \end{bmatrix}\begin{bmatrix}
            \psi^{\mathrm{x}}_{t}\\ \psi^{\mathrm{u}}_t
        \end{bmatrix}\right\|^2\tag{\causalk}\\
        \text{subject~to}\quad& \psi^{\mathrm{x}}_{0} =\mathrm{e}_i, \nonumber\\&\psi^{\mathrm{x}}_{t+1} = A\psi^{\mathrm{x}}_{t} + B\psi^{\mathrm{u}}_{t}, \quad\text{ $\forall t\in[T-k]$},\nonumber\\&\left(\psi_t^{\mathrm{x}}, {\psi}_t^{\mathrm{u}}\right)\in\mathcal{L}_{\distance}(:, i),\quad\text{ $\forall t\in[T-k]$,}\nonumber
    \end{align}
and 
    \begin{align}
    \label{eq:k-no}
        & \min_{\widehat{\psi}}\sum^{T}_{t=0}\left\|\begin{bmatrix}
            Q^{\frac{1}{2}} & \mathbf{0} \\ \mathbf{0} & R^{\frac{1}{2}}
        \end{bmatrix}\begin{bmatrix}
            \widehat{\psi}^{\mathrm{x}}_{t}\\ \widehat{\psi}^{\mathrm{u}}_t
        \end{bmatrix}\right\|^2 \tag{\mixedk}\\
        \text{subject~to}\quad& \widehat{\psi}^{\mathrm{x}}_{0} = \delta_0^{(k)},\nonumber\\&\widehat{\psi}^{\mathrm{x}}_{t+1} = A\widehat{\psi}^{\mathrm{x}}_{t} + B\widehat{\psi}^{\mathrm{u}}_{t} + \delta_{t+1}^{(k)}, \quad\text{ $\forall t\in[T]$},\nonumber\\& \left({\widehat{\psi}}_t^{\mathrm{x}}, {\widehat{\psi}}_t^{\mathrm{u}}\right)\in\mathcal{L}_{\distance}(:, i),\quad\text{ $\forall t\in[T]$}, \nonumber
    \end{align} 
where $\psi$ and $\widehat{\psi}$ is the decision variables of \eqref{eq:k-ca} and \eqref{eq:k-no} respectively;  $\mathrm{e}_i\in \mathbb{R}^n$ represents the standard basis vector and its $i$th component is equal to 1; {$\delta_{t}^{(k)}$ is defined as a $k$-induced vector such that $\delta_{k}^{(k)} =\mathrm{e}_i$ and $\delta_t^{(k)}=\mathbf{0}$ for $k \not = t$}. 

\SetKwComment{tcc}{//}{} 

\begin{algorithm}[t]
\DontPrintSemicolon
\SetAlgoLined
\BlankLine
\tcc{\small{Phase 1: Synthesis of CLMs}}
\For{each subsystem $i \in [N]$ \textbf{in parallel}}{

    \For{each temporal index $k \in [T]$}{
        Solve the optimization problems in~\eqref{eq:k-ca} and~\eqref{eq:k-no} to obtain the system response maps $(\psi^{\mathrm{x}}_{t},\psi^{\mathrm{u}}_{0:T-k})_{t=0}^{T-k}$ and $(\widehat{\psi}^{\mathrm{x}}_{t},\widehat{\psi}^{\mathrm{u}}_{0:T})_{t=0}^T$\;

        \BlankLine
        Define the CLM components for subsystem $i$ via Proposition~\ref{proposition:2}:\;
        $\bm{\varphi}^i_k \coloneqq (\pxl_{t,k},\hpxl_{t,k},\pul_{t,k}, \hpul_{t,k})_{t=0}^T$\;
    }
    Assemble the complete set of CLMs for $i$: $(\bm{\varphi}^i_k)_{k=0}^{T-1}$\;
}
\vspace{0.2cm}

\tcc{\small{Phase 2: Implement distributed controller}}
\For{each subsystem $i \in \{1, \dots, N\}$}{
    Initialize local state $\mathrm{x}_0^i \gets \mathbf{0}$\;
    
    \BlankLine
    Receive local predictions $\bigcup_{j\in\mathcal{N}^\kappa_{\mathcal{G}}(i)}\widehat{\mathrm{w}}^j_{0:T-1}$\;
    
    \For{time step $t = 0, \dots, T-1$}{
        Observe local state $\mathrm{x}^i_{t}$ and disturbances $\bigcup_{j\in\mathcal{N}^\kappa_{\mathcal{G}}(i)}\mathrm{w}^j_t$\;
        
    \BlankLine
        Compute local control action $\mathrm{u}^i_t$ using the synthesized CLMs and received information, according to~\eqref{eq:policy}\;
    }
}
\caption{Distributed Synthesis and Implementation for Problem~\eqref{localized problem}}
\label{alg:1}
\end{algorithm}

Based on the equivalence, the following result characterizes the optimal solution to \eqref{eq:k-decomp}:
\begin{proposition}[Optimal solution to \eqref{eq:k-decomp}] \label{proposition:2}Given a time index $k\in[T]$ and a subsystem $i\in[N]$, let $({\psi}^{\mathrm{x}\star}_{t}, {\psi}^{\mathrm{u}\star}_{t})_{t\in[T-k]}$ and $(\widehat{\psi}^{\mathrm{x}\star}_{t}, \widehat{\psi}^{\mathrm{u}\star}_{t})_{t\in[T]}$ be the solutions to \eqref{eq:k-ca} and \eqref{eq:k-no}. Then 
\begin{align}
\begin{bmatrix}
    \pxl_{t,k}  \ \ & \ \ \hpxl_{t,k} \\  \pul_{t,k} \ \ & \ \ \hpul_{t,k}
\end{bmatrix} = \begin{cases}
    \begin{bmatrix}
        \mathbf{0} \  & \  \widehat{\psi}^{\mathrm{x}\star}_{t}\\ \mathbf{0} \  &  \ \widehat{\psi}^{\mathrm{u}\star}_{t}
    \end{bmatrix} &\mathrm{if}~t<k,\\
    \begin{bmatrix}
        {\psi}^{\mathrm{x}\star}_{t} \ \  &  \ \ \widehat{\psi}^{\mathrm{x}\star}_{t} - {\psi}^{\mathrm{x}\star}_{t}\\ {\psi}^{\mathrm{u}\star}_{t} \ \ & \ \ \widehat{\psi}^{\mathrm{u}\star}_{t} - {\psi}^{\mathrm{u}\star}_{t}
    \end{bmatrix} &\mathrm{otherwise},
\end{cases}\nonumber
\end{align}
is a minimizer of the problem \eqref{eq:k-decomp}.
\end{proposition}

We defer the proof of~\Cref{proposition:2} to Appendix~\ref{proof: proposition 2}. Building on these equivalent formulations, \Cref{alg:1} outlines a procedure to synthesize the localized LQR controller by solving \eqref{eq:k-ca} or \eqref{eq:k-no}. For completeness, \Cref{proposition:optimal solution} in the appendix further presents the derivation of an closed-form solution to \eqref{eq:k-decomp} by leveraging the structure of \eqref{eq:k-ca} and \eqref{eq:k-no}.

\subsection{Decaying Properties}
\label{sec:decaying_properties} 
Fix a communication graph $\mathcal{G}$ and a communication range $\distance \in \mathbb{N}_+$. For a matrix $M\in\mathbb{R}^{m\times n}$, we define its $(i,\distance)$-truncation as $M^{[i,\distance]}\in\mathbb{R}^{m\times n}$ and $(i,\distance)$-boundary $M^{[i,\distance]}_{\perp}\in\mathbb{R}^{m\times n}$ by 
\begin{equation}
    \begin{aligned}
        M^{[i,\distance]}(j) &= \begin{cases}
            M(j), \quad &j\in\mathcal{N}^{\distance}_{\mathcal{G}}(i),\\
            \mathbf{0}, &\text{otherwise},
        \end{cases}\\M^{[i,\distance]}_\perp(j) &=
             \nonumber
    \begin{cases}
            M(j), \quad &j\in\mathcal{N}^{\distance+1}(i)\backslash\mathcal{N}^{\distance}_{\mathcal{G}}(i),\\
            \mathbf{0}, &\text{otherwise}.
        \end{cases}
    \end{aligned}
\end{equation}

Using the preceding definition and for illustrative purposes, we assume the following assumptions:
\begin{assumption}[Model regularity]\label{assumption:networked_system} 
    \textit{ The matrices $(A,B,Q,R)$ in \eqref{eq:ddynamics} satisfy:
    \begin{enumerate}
    \item $\|A\|,\|B\|,\|Q\|,\|R\|\leq L$, and $Q\succeq \mu_Q, R\succeq\mu_R$ for  some $L,\mu_Q,\mu_R>0$.
        \item There exists $\kappa \in \mathbb{N}$ with $1\leq\kappa\leq\mathrm{diam}(\mathcal{G})$ such that for all $i\in[N]$, the system $(A^{[i,\distance]},B^{[i,\distance]})$ is $(L,\gamma)$-stabilizable for some $\gamma\in (0,1)$.
        \item For all $i\in[N]$, matrix $B^{[i,\distance]}_\perp\left(\mathcal{J}^\kappa_{\mathcal{G}}(i), \mathcal{N}^\kappa_{\mathcal{G}}(i)\right)$ is full-rank where $\mathcal{J}^\kappa_{\mathcal{G}}(i)\coloneqq \mathcal{N}^{\distance+1}(i)\backslash\mathcal{N}^{\distance}_{\mathcal{G}}(i)$.
    \end{enumerate}}
\end{assumption}

The first condition is standard in regret-optimal control analysis~\cite{shin2023near,xu2024stability,goel2023regret}. The second and third conditions of Assumption~\ref{assumption:networked_system} are also common assumptions drawn from \cite{xu2024stability,yu2021localized}. In particular, the second condition requires that the $\kappa$-hop system $(A^{[i,\distance]},B^{[i,\distance]})$ of each subsystem $i \in [N]$ is independently stabilizable. The third condition guarantees the \textit{localizability} of the networked system, a property that is critical for underactuated systems.

We now highlight two decaying properties for the closed-loop mapping $\bm{\Phi}$ synthesized from \eqref{eq:centralized problem}: 
\begin{theorem}[Temporally decaying property]
\label{theorem:temporal}
Denote by 
$(\pxl_{t,k},\hpxl_{t,k},\pul_{t,k},\hpul_{t,k})$
an optimal solution to the decomposed problem \eqref{eq:k-decomp}.
Under Assumption~\ref{assumption:networked_system}, there exist constants $C>0$ and $\rho\in (0,1)$ dependent on the system matrices $(A, B, Q, R)$ in~\eqref{eq:problem} such that
\begin{equation}
    \begin{aligned}
        \left\|\begin{bmatrix}
            \pxl_{t,k} \ \ & \ \ \hpxl_{t,k}\\ \pul_{t,k} \ \ & \ \ \hpul_{t,k}
        \end{bmatrix}\right\|^2\leq C\rho^{|t-k|}.
    \end{aligned}
\end{equation}
\end{theorem}
We defer the proof of \Cref{theorem:temporal} to Appendix~\ref{proof:temporal}. 
Intuitively, \Cref{theorem:temporal} implies that when the temporal index  $k\in[T]$ and the time step $t\in[T]$ are far from each other, the $k$th temporal component of the closed-loop mapping is negligible. Therefore, using the FIR constraint to truncate the closed-loop responses as illustrated in \eqref{eq:FIR} results in approximations that are exponentially close to their infinite-horizon counterparts. Moreover, \Cref{theorem:temporal} provides a characterization of the finite-time stability of the closed loop of \eqref{dynamics} under the predictive SLS controller \eqref{eq:policy} with an explicit decaying rate and bounding constants, which have been shown to be essential for the downstream performance analysis \cite{dean2020sample,NEURIPS2018_0ae3f79a,dean2019safely,dean2020robust}.

Next, we confirm that the spatial gap between the localized and centralized subsystem closed-loop mappings also satisfies a similar exponentially decaying property with respect to the communication range $\distance$.   

\begin{theorem}[Spatially decaying property]
\label{theorem:spatial}
{Let localized CLMs $(\pxl_{t,k},\hpxl_{t,k},\pul_{t,k},\hpul_{t,k})$  and~$(\px_{t,k},\hpx_{t,k},\pu_{t,k},\hpu_{t,k})$ denote optimal solutions to the problem \eqref{eq:k-decomp} under the $\kappa$-localized (with \eqref{eq:locality constraint}) and centralized (without \eqref{eq:locality constraint}) settings, respectively.} 
Under Assumption~\ref{assumption:subexponential} and \ref{assumption:networked_system}, there exist constants $D>0$ and  $\vartheta\in (0,1)$ such that
    \begin{equation}
        \begin{aligned}
           \sum_{t=0}^{T}\left\|\begin{bmatrix}
               \px_{t,k}(j) -\pxl_{t,k}(j) \ \ & \ \ \hpx_{t,k}(j)-\hpxl_{t,k}(j) \\ \pu_{t,k}(j) - \pul_{t,k}(j) \ \ & \ \ \hpu_{t,k}(j)-\hpul_{t,k}(j)
           \end{bmatrix} \right\|^2\leq D\vartheta^{\distance}, \forall j\in[N]\nonumber
        \end{aligned}
    \end{equation}
\end{theorem}
We defer the proof of \Cref{theorem:spatial} to Appendix~\ref{appendix:spatial}.
Note that here $D, \vartheta$ depend on the system parameters $(A,B,Q,R)$ and the communication structure $\mathcal{G}$.

\Cref{theorem:spatial} implies that the $\kappa$-localized closed-loop mapping $\bm{\Phi}$ approximates the centralized predictive controller at an exponential rate as the communication range $\distance$ increases. Therefore, the performance of the localized and distributed \psls controller can be near-optimal by setting an appropriately chosen  $\distance\ll N$ in large-scale networked systems. Compared to a similar property in~\cite{zhang2023optimal}, Theorem~\ref{theorem:spatial} does not require the system to be open-loop stable. Our result generalizes that of~\cite{shin2023near}, which considers only a disturbance-free model. Beyond $\kappa$-localization, thanks to the system-level parameterization, any communication constraints that can be expressed as convex constraints in $\bm{\Phi}$ can be directly incorporated into the \psls controller synthesis.

\subsection{Regret Bound}
The two decaying properties with decay rates $\rho,\vartheta\in (0,1)$ from~\Cref{sec:decaying_properties} enable {us to characterize the controller performance of 
\eqref{localized problem} (or \eqref{eq:k-decomp}) }using the \textit{dynamic regret} (defined in~\eqref{eq:dynamic_regret}), which is expressed in terms of the communication range $\kappa$ and prediction error $\bar \varepsilon$:
\begin{theorem}[Performance analysis]\label{theorem:regret}
Under Assumption~\ref{assumption:subexponential} and \ref{assumption:networked_system}, the dynamic regret of the controller derived from the \eqref{localized problem} (or \eqref{eq:k-decomp}) is bounded as
\begin{align}
    &\nonumber\mathrm{DR}(\PredSLS)\leq \mathcal{O}\left(\left(p(\distance)\right)^2\left(C_{1}\overline{\bm{{\varepsilon}}}+C_{2} W^2\rho_0^{\distance}\right)+C_{3}\left(p(\mathrm{diam}({\mathcal{G}})) - p(\distance)\right)^2W^2\right),\nonumber
\end{align}
where $W$ is the uniform upper bound of the disturbances and $\Bar{\varepsilon}$ is the local prediction error bound, both of which are defined in \Cref{sec:problem setting}. Parameter $\rho_0\coloneqq\max(\rho^{1/2}, \vartheta^{1/2})$, and $p(\distance)$ denotes the upper bound of the number of accessible nodes with respected to the distance $\distance$ such that $ p(\distance) = (\distance+1)\sup_{d\in\{1,2,\ldots,\kappa\}} g(d)$. Constants $C_1$, $C_2$ and $C_3$ are defined as
\begin{align}
    C_1&\coloneqq\left(\frac{C(1+\rho)}{1-\rho}\right)^2, \quad C_2\coloneqq\frac{2(CD)^{\frac{1}{2}}(1+(1+\max(\rho^{\frac{1}{4}},\vartheta^{\frac{1}{4}}))^2)}{(1-\max(\rho^{\frac{1}{4}},\vartheta^{\frac{1}{4}}))^2},\quad
    C_3\coloneqq\left(\frac{C(2+\rho)}{(1-\rho)}\right)^2.\nonumber
\end{align}
where $C, D, \rho, \vartheta$ are consistent with their definitions in \Cref{theorem:temporal} and \Cref{theorem:spatial}.
\end{theorem}

We defer the proof of \Cref{theorem:regret} to Appendix~\ref{appendix:regret}.

\begin{remark}
In this bound, we identify three critical terms that delineate the performance landscape of control systems:
\vspace{-4pt}
\begin{align}\small
&\underbrace{C_1 p^2(\distance)\Bar{\varepsilon}}_{\text{error propagation}}  +\underbrace{C_2p^2(\distance) W^2\rho_0^{\distance}}_{\text{sub-optimality}}+\underbrace{C_3\left(p(\mathrm{diam}({\mathcal{G}})) - p(\distance)\right)^2W^2}_{\text{localized communication}}\nonumber
\end{align}
\begin{enumerate}
    \item \textbf{Error propagation}: this term measures the deviation of the prediction $( \widehat{\mathrm{w}}_t:t\in [T])$ from the actual disturbances $( \mathrm{w}_t:t\in [T])$. It also reveals the impact of propagated prediction errors among neighboring nodes so that global communication with larger $\kappa$ may lead to larger total loss.
    \item \textbf{Sub-optimality}: this term represents the performance discrepancy between the centralized optimal solution $\bm{\Phi}^*$ to the predictive SLS problem \eqref{finite problem} and its distributed counterpart $\bm{\Phi}^\kappa$ with $\kappa$-locality constraints.
    \item \textbf{{Localized} communication}: this term reflects the performance loss stemming from the localized information exchange.
\end{enumerate}
Together, these terms highlight a trade-off between performance and communication range in networked systems, indicating the necessity of identifying an optimal $\kappa$ as a nontrivial task. In the next section, we will use experiments to show how this upper bound in~\Cref{theorem:regret} can be utilized empirically for communication range optimization.
\end{remark}

\section{Numerical Experiments}
\label{sec:case_studies}

\begin{figure}[t]
    \centering
    \includegraphics[width=1\linewidth]{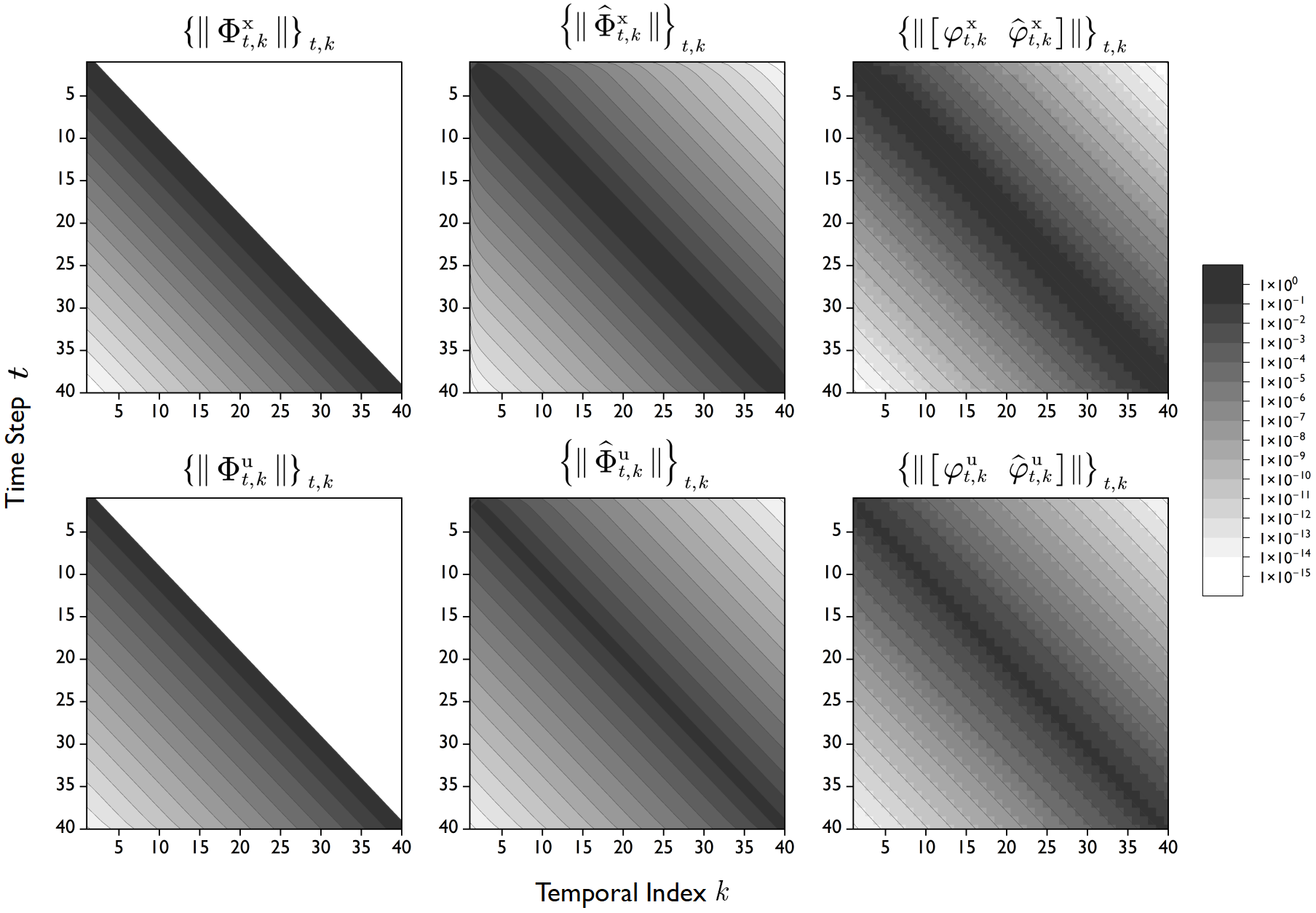}
    \caption{Heatmaps of the CLMs ($\ell^2$-norm) with respect to temporal indexes $t,k\in[40]$ (\textsc{\textbf{Left}} and \textsc{\textbf{Mid}}: the complete CLMs matrices solved by \psls; \textsc{\textbf{Right}}: the columns of CLMs solved by \eqref{eq:k-decomp}). The considered CLMs in the top figures are those related to states. The CLMs in the bottom figures are those related to the control actions. }
    \label{fig:temporal_results}
\end{figure}

\begin{figure}[h]
    \centering
    \includegraphics[width=1\linewidth]{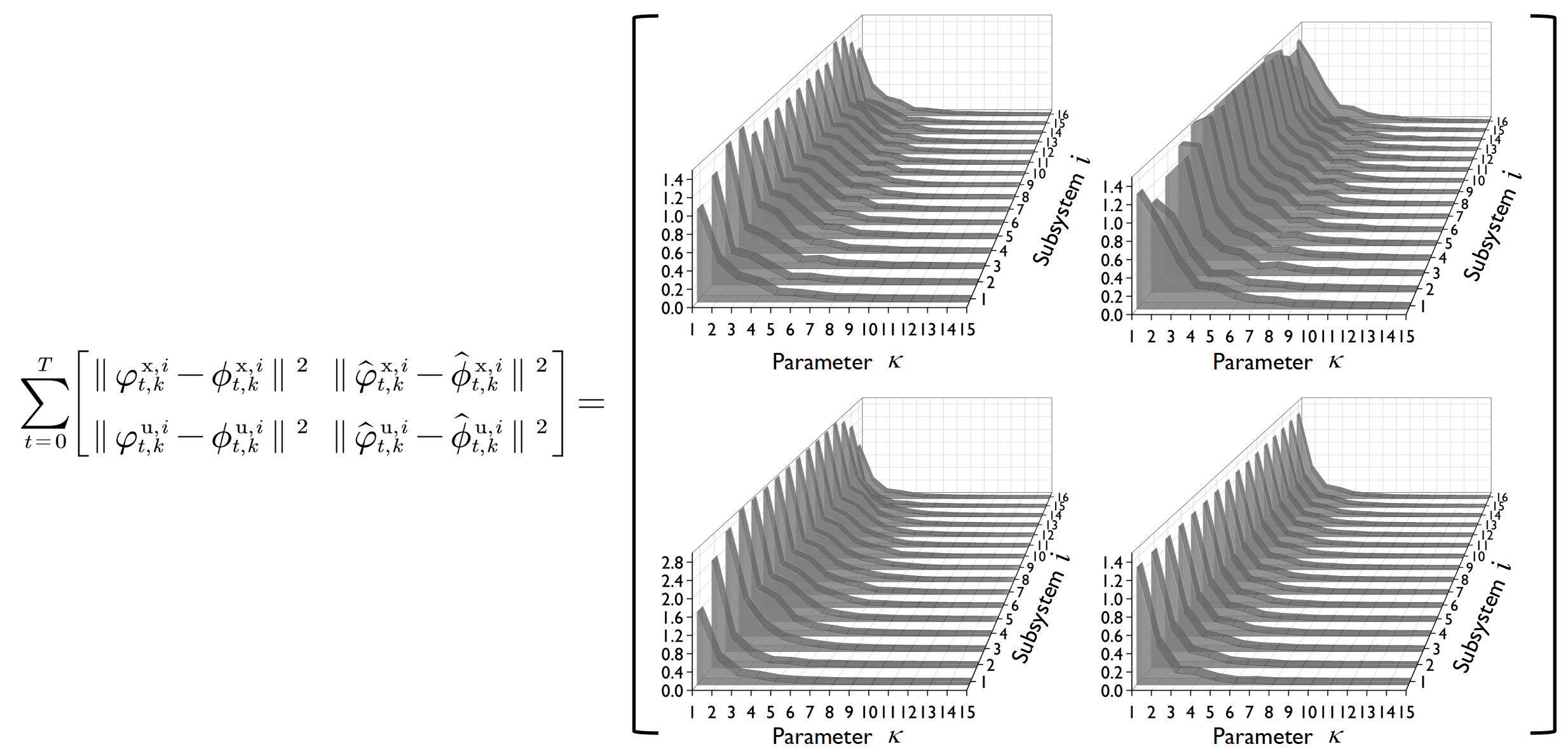}
    \caption{Resulting plots to demonstrate the spatially decaying property in Theorem~\ref{theorem:spatial} with respect to $\distance$. Each $\phi_{t,k}^{i}$ denotes an optimal solution to \eqref{eq:k-decomp} without \eqref{eq:locality constraint} in~\eqref{localized problem} (equivalently, with the maximal $\distance$).}
    \label{fig:spatial_results}
\end{figure}

\begin{figure}[h]
    \centering
    \includegraphics[width=1\linewidth]{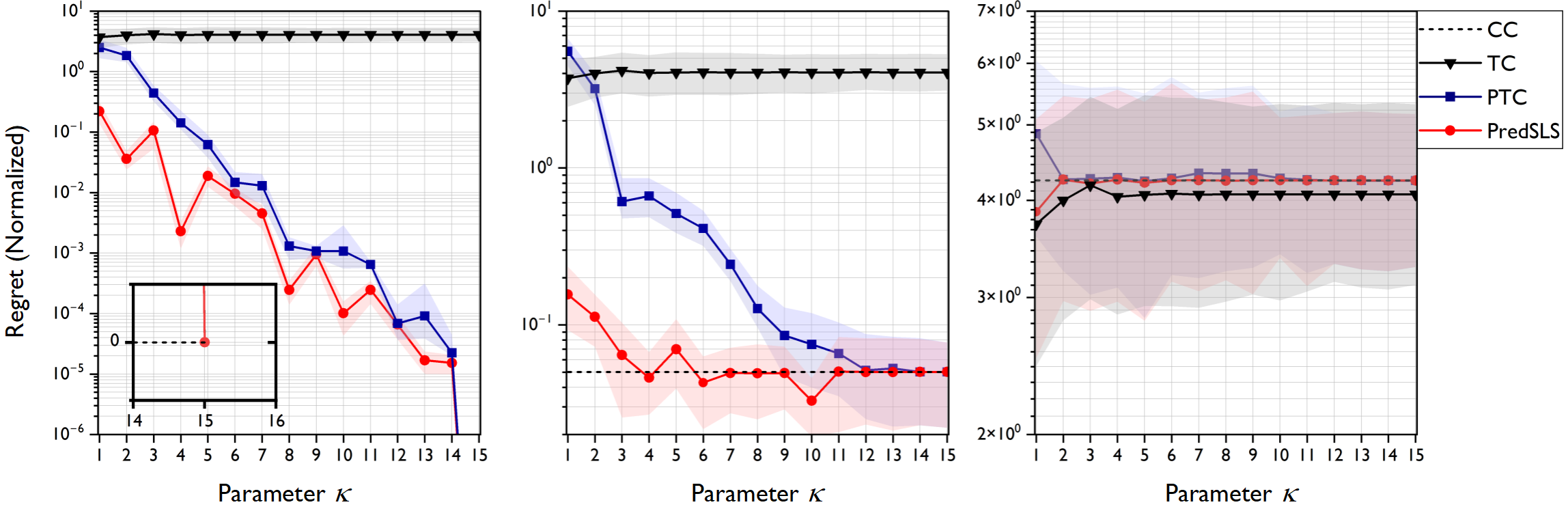}
    \caption{A comparison of different control methods (the centralized controller ($\mathtt{CC}$), $\kappa$-truncated controller ($\mathtt{TC}$), predictive $\kappa$-truncated controller ($\mathtt{PTC}$), and the synthesized controller induced by~\eqref{eq:k-decomp} ($\mathtt{PredSLS}$)) on 100 experiments using normalized regret $\mathrm{DR}(\pi) / J^*$ in three cases. \textsc{Left}: the introduced example with $\mathbb{E}[\|\mathrm{w}_t - \widehat{\mathrm{w}}_t\|]=0$, \textsc{Mid}: $\mathbb{E}[\|\mathrm{w}_t - \widehat{\mathrm{w}}_t\|]=0.1$. \textsc{Right}: $\mathbb{E}[\|\mathrm{w}_t - \widehat{\mathrm{w}}_t\|]=1$.}
    \label{fig:regret}
\end{figure}

In this section, we first visualize the theoretical results presented in Theorem \ref{theorem:temporal}, Theorem \ref{theorem:spatial} and \ref{theorem:regret}. Then, we compare the performance of \psls against existing controllers. To demonstrate the practical implications of our theoretical analysis, we consider four different topologies: chain graph, cyclic graph, tree graph, and mesh graph. For brevity, we focus on the chain graph in this section and leave the experimental details of the other topologies to the extended version of this work in \Cref{appendix:experiments}.

\paragraph{Chain Graph. }  We consider a networked linear system induced by a chain graph with scalar subsystems of the form: 
\begin{align}
    &\begin{bmatrix}
        x^1_{t+1}\\\vdots\\x^{16}_{t+1}
    \end{bmatrix} = {A}\begin{bmatrix}
        x^1_{t}\\\vdots\\x^{16}_{t} 
    \end{bmatrix} + {B} \begin{bmatrix}
        u^1_{t}\\\vdots\\u^{16}_{t} 
    \end{bmatrix} + \begin{bmatrix}
        w^1_{t}\\\vdots\\w^{16}_{t} 
    \end{bmatrix},~\text{where} \nonumber\\  &{A} = {B} = \begin{bmatrix}
        1 & 0.5 &\\
        0.5 & 1 & 0.5 & \\
         & \ddots & \ddots & \ddots &\\
         & & 0.5 & 1 & 0.5  \\
         & & & 0.5 & 1
    \end{bmatrix}, \,\,{Q} = {R} = I \nonumber\\  &w^i_t \sim\mathcal{N}(0, 0.5) + \begin{cases}
    0.08, &t=2, \\
    0.18, &t=4, \\
    0, &\text{otherwise},
    \end{cases}~\text{$\forall i\in\{1,\dots,16\}$}.\nonumber
\end{align}

Here we have 16 scalar subsystems in the network with $\mathrm{x}_t = \begin{bmatrix}
        x^1_{t}&\dots&x^{16}_{t}
    \end{bmatrix}^\top\in\mathbb{R}^{16}$ and $\mathrm{u}_t = \begin{bmatrix}
        u^1_{t}&\dots&u^{16}_{t}
    \end{bmatrix}^\top\in\mathbb{R}^{16}$. The matrix $A$ is induced by a chain graph and $\mathcal{N}(0, 0.5)$ denotes a Gaussian distribution with zero mean and variance of $0.5$. This is an open-loop unstable system where adjacent nodes are weakly connected. Dynamical systems with similar Laplacians appear naturally in related fields like diffusion processes, consensus dynamics, and the practical networked system \cite{NEURIPS2018_0ae3f79a}. In each node $i$, we assume that the predictions $(\widehat{w}^i_t:t\in [T])$ are available to each subsystem over the control horizon $T = 40$. The first node (Node 1) is the only node affected by the prediction error such that $|e_t^i| = |\widehat{w}_t^i - w_t^i|>0$, {while all other nodes receive perfect predictions of the disturbances. }

\paragraph{Visualization of \Cref{theorem:temporal} and \Cref{theorem:spatial}.}
The temporal and spatial properties of the \psls and \eqref{eq:k-decomp} are illustrated in \Cref{fig:temporal_results} and \Cref{fig:spatial_results}.We observe that both the global configuration \psls and distributed subsystem problem \eqref{eq:k-decomp} exhibit an exponentially decaying behavior as shown in \Cref{theorem:temporal} when $|t-k|$ increases. Also, when compared with the centralized algorithm, the solution of \eqref{eq:k-decomp} converges to the optimal solution exponentially fast (\Cref{theorem:spatial}) with an increasing communication range $\kappa$. 

\paragraph{Performance evaluation.}
\label{sec:exp_performance}
Next, we compare the controller synthesized via \eqref{eq:k-decomp} against the noncausal centralized controller \eqref{eq:optimal policy} \cite{goel2022power}, and the predictive $\kappa$-truncated method~\eqref{eq:PTC}. As a simple baseline, we also compare against the $\kappa$-truncated control ($\mathtt{TC}$) in \cite{shin2023near}, which synthesizes the optimal infinite LQR controller by treating future disturbances as zeros. Specifically, for our particular example, the control laws for ($\mathtt{TC}$), \eqref{eq:PTC}, and \eqref{eq:optimal policy} are
\begin{align*}
    (\mathtt{TC}):~{u}^i_t&=\sum_{j\in\mathcal{N}_{\mathcal{G}}^{\kappa}(i)}K^{ij,\star}{x}^j_t,\nonumber\\
    (\mathtt{PTC}):~{u}^i_t&=\sum_{j\in\mathcal{N}_{\mathcal{G}}^{\kappa}(i)}\left(K^{ij,\star}{x}^j_t+\sum_{\tau=0}^{T-t-1}L^{ij,\star}\widehat{{w}}_{\tau+t}^j\right),\nonumber\\
    (\mathtt{CC}):~{u}^i_t&=\sum_{j\in[N]}\left(K^{ij,\star}{x}^j_t+\sum_{\tau=0}^{T-t-1}L^{ij,\star}\widehat{{w}}_{\tau+t}^j\right),\nonumber
\end{align*}
where ${u}^i_t$, ${x}^j_t$, and $\widehat{{w}}_{t}^j$ are local action, state, and disturbance predictions at time $t$ for subsystems $i,j\in[N]$; $\mathcal{N}_{\mathcal{G}}^{\kappa}(i)$ is the $\kappa$-localized communication neighborhood set defined in Section~\ref{sec:problem setting}; matrices $K^\star$ and $L^\star$ are computed control gains using centralized methods (see Appendix~B.1 in \cite{wu2025} for more details).
The trajectory length of $T=40$. We run 100 experiments with different random seeds and plot the mean performance (solid line) and 1 standard deviation (shaded region) in \Cref{fig:regret}. Overall, \eqref{eq:k-decomp} consistently the best performance. In the third subplot of \Cref{fig:regret}, while still outperforming other prediction-based controllers, PredSLS is slightly worse off than ($\mathtt{TC}$), which does not use any predictions. This is because under extremely large prediction errors, rejecting predictive information is more optimal than acting on flawed forecasts from neighbors.

\paragraph{Non-monotonic performance-communication trade-off.\label{sec:non-monotonic_perform}}
We highlight that the performance of the distributed controllers do not simply obey the monotonic exponential performance decay trend as a function of the communication range, which has not been reflected in the theoretical analysis of existing works\cite{lin2022decentralized,shin2023near,xu2024stability,zhang2023optimal}. Rather, we observe nontrivial trade-off between the prediction error and the communication range $\kappa$ as shown in Figure~\ref{fig:bound}. This corroborates with \Cref{theorem:regret}.

\paragraph{Empirical communication and control co-design.\label{sec:exp_codesign}}
Given a fixed prediction error,  \Cref{theorem:regret} provides insights on communication structure design. 
In particular, we compare the theoretical optimal communication range $\kappa$ given by \Cref{theorem:regret} and the actual optimal $\kappa$ that achieves the smallest regret in the experiments in the third panel of \Cref{fig:bound}.
Across various graphs and levels of prediction errors, \Cref{theorem:regret} provides optimal $\kappa$'s that are highly consistent with the actual optimal ranges. 
This suggests that optimizing the upper bound in \Cref{fig:bound} over $\kappa$ as a surrogate is an effective empirical solution to the communication and control co-design challenge for the networked LQR problem with predictions.

\begin{figure}[t]
    \centering
    \includegraphics[width=1\linewidth]{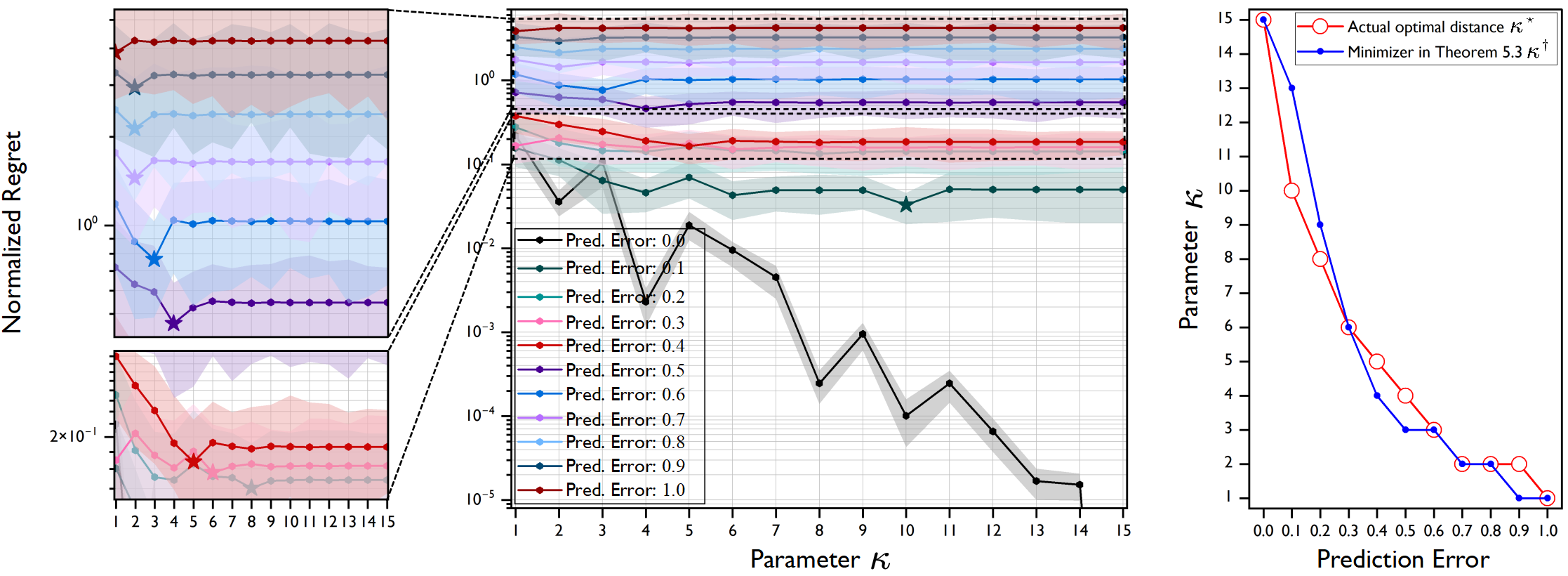}
    \caption{Performance of \psls for the chain-graph-induced LQR. \textbf{\textsc{Left}}: Algorithm performance (normalized regret) versus communication distance $\kappa$ with varying prediction errors $\mathbb{E}\left[\|\mathrm{w}_t-\widehat{\mathrm{w}}_t\|\right]$; \textbf{\textsc{Right}}: Actual optimal communication distance  $\kappa^{\star}$ that minimizes the normalized regret, compared with the minimizer $\kappa^\star$ of the regret bound in \Cref{theorem:regret}.
    }
    \label{fig:bound}
\end{figure}

The source code to reproduce our results is available online.\footnote{\url{https://github.com/Wu-yi-fei/predictive-sls-master}} To demonstrate the generalizability of \psls, additional numerical results on a variety of topological graphs are available. Further details about the system setups and results are in Appendix~\ref{appendix:experiments}.

\section{Conclusion and Future Directions}

The \psls framework presented in this work is a novel system-level parameterization that enables scalable, localized control of distributed systems by jointly integrating communication constraints and perturbation predictions into a tractable convex optimization. Our results extend the theoretical guarantees established in prior work, which typically rely on the assumptions of zero disturbance or zero prediction error.
Several exciting directions for future research emerge from this work. First, inspired by the non-monotonic performance-communication trade-off, we aim to formulate a comprehensive optimization framework for joint controller-topology co-design.  Second,  online learning mechanisms might be helpful to adaptively tune the controller's reliance on predictions in response to real-time errors. This involves open research questions, including verifying the convexity of the \psls controller with respect to some confidence parameter and ensuring controller stability for any adaptive tuning of such a parameter. Finally, it would be interesting to apply the \psls framework to critical large-scale applications, such as distributed voltage control in power grids.

\bibliographystyle{ieeetr}
\bibliography{references}  
\addtocontents{toc}{\protect\setcounter{tocdepth}{2}}
\begin{appendices}

\newpage
\renewcommand*\contentsname{Appendix Outline}
\tableofcontents
\newpage

\section{Preliminaries on Predictive Controllers} \label{appendix:A}
\addcontentsline{toc}{section}{Preliminaries on Predictive Controllers}
This section reviews the clairvoyant offline optimal controller for the 
stochastic Linear Quadratic Regulator (LQR) problem, as introduced by 
Goel et al.~\cite{goel2022power}, and subsequently presents a predictive 
truncated controller.

\subsection{Centralized Offline Optimal Controller}
\label{sec:benchmark}
Within the optimal control paradigm, a key type of control policy considered 
is the non-causal (also termed \textit{predictive} in this paper) policy. 
Such policies are distinguished by their ability to select actions with 
full knowledge of future disturbance and often operate in a Model 
Predictive Control (MPC) manner.

Consider a linear quadratic control problem where
$\widehat{\mathbf{w}}\coloneqq (\widehat{\mathrm{w}}_0,\dots,\widehat{\mathrm{w}}_{T-1})$ denotes a sequence of predictions of the future unknown disturbance 
trajectory $\mathbf{w}\coloneqq (\mathrm{w}_0,\dots,\mathrm{w}_{T-1})$ at the beginning of the control horizon. We define a predictive control algorithm, $\ALG$, as a function mapping the current state $\mathrm{x}_t$ and the available predictions $\widehat{\mathbf{w}}_{t:T-1}$ to the control action $\mathrm{u}_t$. This action is given by $\mathrm{u}_t=\ALG(\mathrm{x_t},\widehat{\mathbf{w}}_{t:T-1};K,
\left\{L_\tau\right\}_{\tau=0}^{T-t-1})$, where the controller parameters 
$K$ and $\left\{L_\tau\right\}_{\tau=0}^{T-t-1}$ are pre-determined such that:
    \begin{align}
        \ALG\left(\mathrm{x_t},\widehat{\mathbf{w}}_{t:T-1};K,\left\{L_\tau\right\}_{\tau=0}^{T-t-1}\right) = K\mathrm{x}_t + \sum^{T-t-1}_{\tau = 0} L_\tau\widehat{\mathrm{w}}_{t+\tau}. \tag{$\mathtt{CC}$} \label{eq:optimal policy}
    \end{align}
Note that  $\OPT\left(\mathrm{x_t}\right)\coloneqq\ALG(\mathrm{x_t},{\mathbf{w}}_{t:T-1};K^\star,\left\{L^{\star}_\tau\right\}_{\tau=0}^{T-t-1})$ is an offline optimal controller that minimizes the total cost in problem \eqref{eq:problem} with $K^\star\coloneqq-(R + B^\top PB)^{-1}B^\top PA$ and ${L^{\star}_{\tau}} \coloneqq -(R + B^\top PB)^{-1}B^\top\left[(A + BK)^\top\right]^\tau P$ denoting the optimal gain obtained by the discrete algebraic Riccati equation (DARE) solution $P$. This type of non-causal policy is also widely considered in previous literature \cite{li2022robustness,yu2020power,goel2022power}.  

Let $F\coloneqq A+BK$ be the closed-loop system matrix. The system can be reparameterized directly to the exogenous disturbances $\mathbf{w}$ and predictions $\widehat{\mathbf{w}}$ by system responses as
\begin{equation}
    \begin{aligned}
        \mathrm{x}_t &= \sum^t_{k=0}F^k\mathrm{w}_{t-k-1} + \sum^{t-1}_{k_1 = 0}F^{k_1}B\sum^{T-t+k_1}_{k_2 = 0}L_{k_2}{}\widehat{\mathrm{w}}_{t-k_1+k_2-1} \\
        &\coloneqq  \sum^t_{k=0}\Phi^1_{t,k}\mathrm{w}_{t-k-1} + \sum^{T}_{k = 0}\Phi^2_{t,k}\widehat{\mathrm{w}}_{k-1},\nonumber
    \end{aligned} \label{DAP-state}
\end{equation}
and
\begin{equation}
    \begin{aligned}
        \mathrm{u}_t &= \sum^t_{k=0}KF^k\mathrm{w}_{t-k-1} + \sum^{t-1}_{k_1 = 0}KF^{k_1}B\sum^{T-t+k_1}_{k_2 = 0}L_{k_2}\widehat{\mathrm{w}}_{t-k_1+k_2-1} + \sum^{T-t-1}_{k_1 = 0}L_{k_1}\widehat{\mathrm{w}}_{t+k_1} \\
        &\coloneqq  \sum^t_{k=0}\Phi^3_{t,k}\mathrm{w}_{t-k-1} + \sum^{T}_{k = 0}\Phi^4_{t,k}\widehat{\mathrm{w}}_{k-1},\nonumber
    \end{aligned} 
\end{equation}
where $\Phi^1_{t,k}, \Phi^2_{t,k}, \Phi^3_{t,k}$ and $\Phi^4_{t,k}$ denote the reparameterized mappings with respect to the temporal indices $t,k\in[T]$, and we absorb the initial state $\mathrm{x}_0$ into $\mathrm{w}_{-1}$ and $0$ into $\widehat{\mathrm{w}}_{-1}$.

It is important to note that the state $\mathrm{x}_t$ exhibits a 
polynomial dependence on the controller gains 
$(K,L_0,\dots,L_{T-1})$. 
This characteristic makes the direct optimization of these gains for LQR 
problems a highly non-convex task. 
However, if we consider the system response mappings $\Phi^1_{t,k}, 
\Phi^2_{t,k}, \Phi^3_{t,k}$, and $\Phi^4_{t,k}$---which relate states and 
inputs to disturbances and predictions---these mappings possess a convex 
characterization. This convexity is particularly helpful for enabling 
scalable distributed control. 
This insight motivates our novel \psls framework: to reparameterize the control 
problem convexly by leveraging these system responses to exogenous 
disturbances and their predictions, inspired by 
SLS.

\subsection{Predictive $\distance$-Truncated Controller}
\label{appendix:A2}
We now introduce another structural controller of interest: the predictive 
$\distance$-truncated controller (PTC). This controller will serve as a benchmark for comparison in our experiments.

Recall the networked system defined in dynamics \eqref{eq:ddynamics}, where each sub-controller $i\in[N]$ obtains the 
neighboring perturbation predictions with a fixed communication distance $\kappa$, i.e., $(\mathrm{w}^j_{0:T-1}:j\in\mathcal{N}^\kappa_{\mathcal{G}}(i))$ is provided. Given the state $\mathrm{x}_t$, the nodal feedback law is expressed as follows:
\begin{align}
u^{i}_{t} &=\sum_{j\in\mathcal{N}^{\kappa}_{\mathcal{G}}(i)}\left(K^{ij,\star}\mathrm{x}^{j}_{t}+\sum^{T-t-1}_{\tau=0}L^{ij,\star}_\tau\widehat{\mathrm{w}}^{j}_{\tau+t}\right), 
\tag{$\mathtt{PTC}$} \label{eq:PTC}
\end{align}
where the matrices $K^{\star}\coloneqq(R + B^\top PB)^{-1}B^\top PA$ and $L^{\star}_\tau\coloneqq(R + B^\top PB)^{-1}[(A+BK)^{\top}]^\tau P$. Note that with exact predictions, \eqref{eq:PTC} is a truncated form of the offline optimal controller $\OPT(\mathrm{x}_t)$, which can be viewed as the \textit{non-causal extension} of the distributed controller analyzed by \cite{shin2023near}. 

\section{Auxiliary Results and Proofs for~\ouralg}
\label{app:proofs_predsls}
\subsection{Proof of Theorem \ref{theorem constraints} \label{appendix:theorem 3.1}}
\begin{proof} Consider an arbitrary predictive controller $\mathbf{u}\coloneqq \mathbf{Kx} + {\widehat{\mathbf{L}}}\widehat{\mathbf{w}}$. Then the closed-loop response of \eqref{dynamics} is the right-hand side of  \eqref{eq:mappings}. It can be verified that all closed-loop mappings (CLMs) generated by this controller satisfy~\eqref{eq:cl-constraints} as follows
\begin{equation}
\nonumber
    \begin{aligned} 
    &\begin{bmatrix}
              zI-A  & -B
          \end{bmatrix}\begin{bmatrix}
             \mathbf{\Phi}^\mathrm{x} & \widehat{\mathbf{\Phi}}^{\mathrm{x}}\\
             \mathbf{\Phi}^\mathrm{u} & \widehat{\mathbf{\Phi}}^{\mathrm{u}}
         \end{bmatrix} =\begin{bmatrix}
              zI-A & -B
          \end{bmatrix}\begin{bmatrix}
        U \ \ & \ \ UB{\widehat{\mathbf{L}}}\\
        \mathbf{K}U \ \  & \ \ (\mathbf{K}UB + I){\widehat{\mathbf{L}}}
    \end{bmatrix} = \begin{bmatrix}
        I \,\,&\,\, 0
    \end{bmatrix},
  \end{aligned}
\end{equation}
where $U\coloneqq(zI-A-B\mathbf{K})^{-1}$.
For the other direction, consider $\mathbf{\Phi}$ satisfying constraint~\eqref{eq:cl-constraints}. First, $(\mathbf{\Phi}^{\mathrm{x}})^{-1}$ exists since \eqref{eq:cl-constraints} implies that $\mathbf{\Phi}$ is a block lower triangular matrix with identity matrices on its diagonal blocks. Now, let  \scalebox{1}{$\smash{\widehat{\mathbf{L}}\coloneqq\widehat{\mathbf{\Phi}} ^{\mathrm{u}}-\mathbf{\Phi}^{\mathrm{u}}(\mathbf{\Phi}^{\mathrm{x}})^{-1}\widehat{\mathbf{\Phi}} ^{\mathrm{x}}}$} and $\mathbf{K}\coloneqq\mathbf{\Phi}^{\mathrm{u}}(\mathbf{\Phi}^{\mathrm{x}})^{-1}$. It follows that
\begin{equation*}
    \begin{aligned} 
    \mathbf{x} &= U \mathbf{w} + UB\widehat{\mathbf{L}}\widehat{\mathbf{w}} \\&= (zI-(A+B \mathbf{\Phi}^\mathrm{u}(\mathbf{\Phi}^{\mathrm{x}})^{-1}))^{-1} \mathbf{w} + (zI-(A+B \mathbf{\Phi}^\mathrm{u}(\mathbf{\Phi}^{\mathrm{x}})^{-1}))^{-1}B(\widehat{\mathbf{\Phi}}^{{\mathrm{u}}} - \mathbf{\Phi}^\mathrm{u}(\mathbf{\Phi}^{\mathrm{x}})^{-1}\widehat{\mathbf{\Phi}}^{\mathrm{x}})\widehat{\mathbf{w}} \\
    & = ((zI-A)\mathbf{\Phi}^{\mathrm{x}}+B \mathbf{\Phi}^\mathrm{u})^{-1} \mathbf{\Phi}^{\mathrm{x}}\mathbf{w} + (zI-(A+B \mathbf{\Phi}^\mathrm{u}(\mathbf{\Phi}^{\mathrm{x}})^{-1}))^{-1}(zI-(A + B\mathbf{\Phi}^\mathrm{u}(\mathbf{\Phi}^{\mathrm{x}})^{-1}))\widehat{\mathbf{\Phi}}^{\mathrm{x}}\widehat{\mathbf{w}} \\
    & =  \mathbf{\Phi}^{\mathrm{x}}\mathbf{w} + \widehat{\mathbf{\Phi}}^{\mathrm{x}}\widehat{\mathbf{w}},\end{aligned}
\end{equation*}
where the third equality is due to \eqref{eq:cl-constraints}. Furthermore, we consider the control actions generated as
\begin{equation*}
    \begin{aligned}
        \mathbf{u} &= \mathbf{K}\mathbf{x}+\widehat{\mathbf{L}}\widehat{\mathbf{w}}
        \\ &= \mathbf{\Phi}^{\mathrm{u}}(\mathbf{\Phi}^{\mathrm{x}})^{-1}(\mathbf{\Phi}^{\mathrm{x}}\mathbf{w} + \widehat{\mathbf{\Phi}}^{\mathrm{x}}\widehat{\mathbf{w}}) + (\widehat{\mathbf{\Phi}} ^{\mathrm{u}}-\mathbf{\Phi}^{\mathrm{u}}(\mathbf{\Phi}^{\mathrm{x}})^{-1}\widehat{\mathbf{\Phi}} ^{\mathrm{x}})\widehat{\mathbf{w}}     
        \\&= \mathbf{\Phi}^{\mathrm{u}}\mathbf{w} + \widehat{\mathbf{\Phi}}^{\mathrm{u}}\widehat{\mathbf{w}},
    \end{aligned}
\end{equation*}
where the third equality is derived from the state trajectory in the last step. Therefore, there exists a predictive controller that realizes the prescribed CLMs $\mathbf{\Phi}$. This concludes the proof.
\end{proof}
\subsection{Proof of Proposition \ref{proposition:1}}\label{proof:proposition:1}
\begin{proof}

  Firstly, we consider a special case where the disturbances $\mathrm{w}_t$'s are i.i.d. Gaussian with zero mean and identity variance and {predictions are random variables satisfying $\widehat{\mathrm{w}}_t = \mathrm{w}_t$ almost surely, e.g., the mean and variance of $\widehat{\mathrm{w}}_t - \mathrm{w}_t$ are zero.}
Then, the following equivalence holds:

\begin{subequations}
    \small
    \begin{align*}\eqref{eq:problem}\xlongequal[]{Q_T =P}&\min_{\mathbf{x},\mathbf{u}}\mathbb{E}\left[\sum^{\infty}_{t=0}\left(\mathrm{x}^\top_t Q\mathrm{x}_t + \mathrm{u}_t^
        \top R\mathrm{u}_t\right)\right], \text{subject~to}~ \eqref{dynamics}\nonumber\\\xlongequal[]{\substack{\text{Theorem \ref{theorem constraints}}}}
        &\min_{\mathbf{\Phi}}\sum^\infty_{\tau=-\infty}\left\|\begin{bmatrix}
            Q^{\frac{1}{2}} & \mathbf{0} \\ \mathbf{0} & R^{\frac{1}{2}}
        \end{bmatrix}\begin{bmatrix}
            \mathrm{\Phi}^{\mathrm{x}}_\tau & \widehat{\mathrm{\Phi}}^{\mathrm{x}}_\tau\\\mathrm{\Phi}^{\mathrm{u}}_\tau & \widehat{\mathrm{\Phi}}^{\mathrm{u}}_\tau 
        \end{bmatrix}\begin{bmatrix}
            I & \Sigma_{\mathbf{{w}}\widehat{\mathbf{{w}}}} \\ \Sigma_{\widehat{\mathbf{{w}}}\mathbf{{w}}} & I  
        \end{bmatrix}^{\frac{1}{2}}\right\|_{\mathcal{F}}^2, \text{subject~to}~ \eqref{eq:constraints} \nonumber
        \\
        \xlongequal[]{\widehat{\mathbf{w}}=\mathbf{w} \ \text{a.s.}}&\min_{\mathbf{\Phi}}\sum^\infty_{\tau=-\infty}\left\|\begin{bmatrix}
            Q^{\frac{1}{2}} & \mathbf{0} \\ \mathbf{0} & R^{\frac{1}{2}}
        \end{bmatrix}\begin{bmatrix}
            \mathrm{\Phi}^{\mathrm{x}}_\tau + \widehat{\mathrm{\Phi}}^{\mathrm{x}}_\tau\\\mathrm{\Phi}^{\mathrm{u}}_\tau + \widehat{\mathrm{\Phi}}^{\mathrm{u}}_\tau  
        \end{bmatrix}\right\|_{\mathcal{F}}^2, \text{subject~to}~\eqref{eq:constraints}\\
        =&\min_{\mathbf{\Phi}}\left\|\begin{bmatrix}
            Q^{\frac{1}{2}} & \mathbf{0} \\ \mathbf{0} & R^{\frac{1}{2}}
        \end{bmatrix}\begin{bmatrix}
            \mathbf{\Phi}^{\mathrm{x}} + \widehat{\mathbf{\Phi}}^{\mathrm{x}}\\\mathbf{\Phi}^{\mathrm{u}} + \widehat{\mathbf{\Phi}}^{\mathrm{u}}  
        \end{bmatrix}\right\|_{\mathcal{H}_2}^2 , \text{subject~to}~ \eqref{eq:constraints}, \nonumber
    \end{align*}
\end{subequations}
where $\Sigma_{\mathrm{\mathbf{w}}\widehat{\mathrm{\mathbf{w}}}}$ denotes the covariance matrix of $\mathrm{w}_t, \widehat{\mathrm{w}}_t$. {This is identity since $\widehat{\mathrm{w}}_t = \mathrm{w}_t$ almost surely.} The third to last equality is derived based on the definition of $\mathcal{H}_2$-norm.
Defining $(\bm{\Phi}^{\mathrm{x}\star}, \widehat{\bm{\Phi}}^{\mathrm{x}\star},\bm{\Phi}^{\mathrm{u}\star},\widehat{\bm{\Phi}}^{\mathrm{u}\star})$ as an optimal solution to \eqref{eq:centralized problem}, we know that by \Cref{theorem constraints}, 
$\mathbf{K} = \bm{\Phi}^{\mathrm{u}\star}(\bm{\Phi}^{\mathrm{x}\star})^{-1}$ and $\widehat{\mathbf{L}} = \widehat{\bm{\Phi}} ^{\mathrm{u}\star}-\bm{\Phi}^{\mathrm{u}\star}(\bm{\Phi}^{\mathrm{x}\star})^{-1}\widehat{\bm{\Phi}} ^{\mathrm{x}\star}$ form an optimal predictive controller to \eqref{eq:problem} when the disturbances are i.i.d. Gaussian with zero mean and identity variance. 

{
According to Theorem 4.1 in \cite{yu2020power}, the controller solved by \eqref{eq:problem} for this specially-designed disturbances and the optimal controller for all general disturbances maintain the identity structure when $\widehat{\mathrm{w}}_t = \mathrm{w}_t$. Correspondingly, a time-domain realization to \eqref{eq:centralized problem} is indeed an optimal controller for all general disturbances when their predictions are exact.  }

Consequently, we can extend the optimality of \psls to all other disturbance sequences. This completes the proof.
\end{proof}

\subsection{Auxiliary Implementation Details}
In this section, we introduce implementation details to facilitate the analysis of certain internal properties of the proposed \psls framework.
First, we present an auxiliary frequency-domain implementation that is equivalent to the time domain realization in~\eqref{eq:policy}.

\begin{figure}[h]
    \centering
    \includegraphics[width=0.6\linewidth]{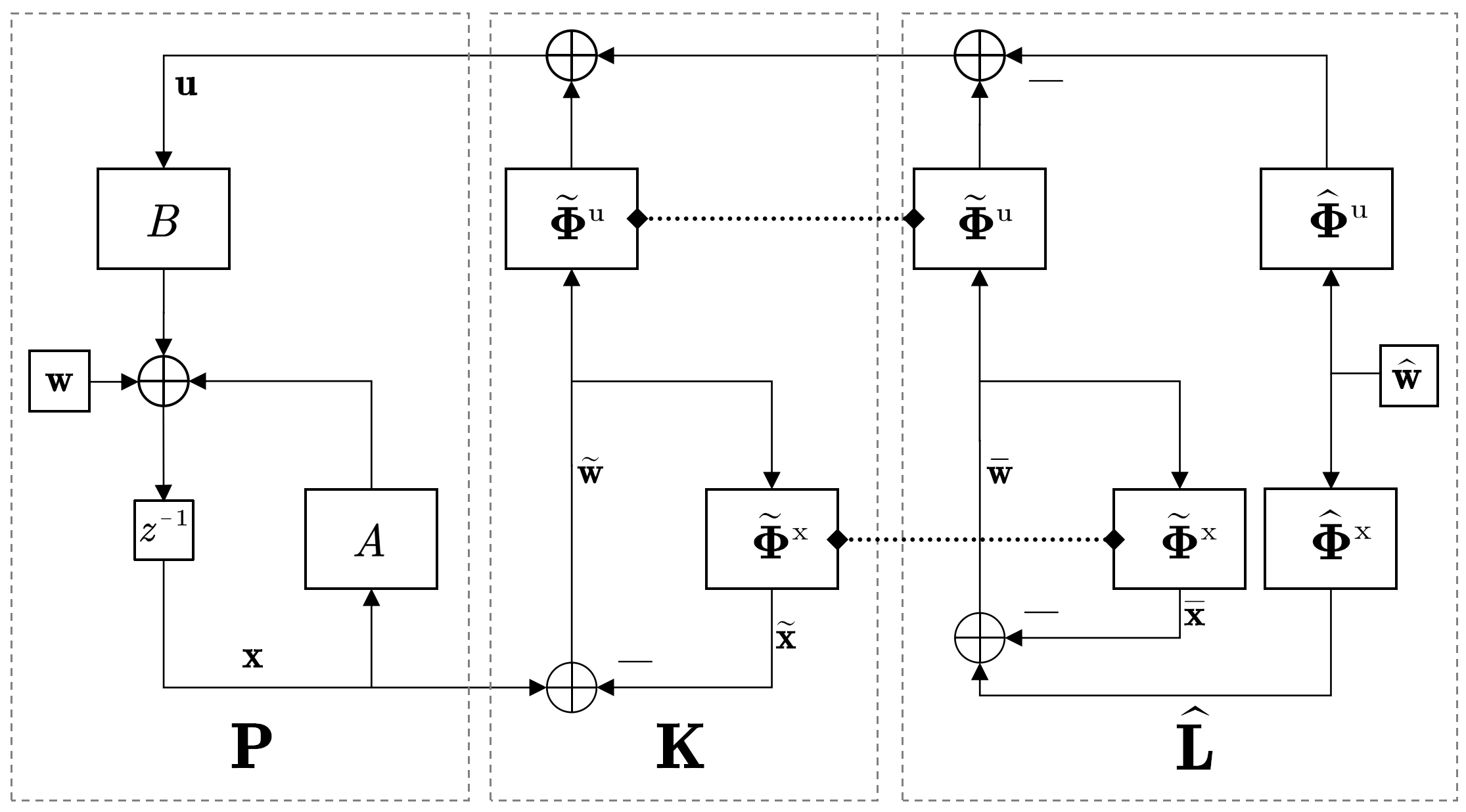}
    \caption{Frequency domain block diagram of the proposed \psls framework structure, with $\widetilde{\mathbf{\Phi}}^{\mathrm{x}} \coloneqq I - z\mathbf{\Phi}^{\mathrm{x}} $, $\widetilde{\mathbf{\Phi}}^{\mathrm{u}} \coloneqq z\mathbf{\Phi}^{\mathrm{u}} $, $\mathbf{P}$ denoting plant, and 
the short dashed line representing the same mapping operator in two connected blocks.}
    \label{fig:block diagram}
\end{figure}

\begin{lemma}[Frequency domain implementation]
Given transfer matrices $\mathbf{\Phi}= [\mathbf{\Phi}^{\mathrm{x}}, \widehat{\mathbf{\Phi}}^{\mathrm{x}}; \mathbf{\Phi}^{\mathrm{u}}, \widehat{\mathbf{\Phi}}^{\mathrm{u}}]$ satisfying the affine constraint \eqref{eq:cl-constraints}, the \psls framework is implemented as shown in block diagram \Cref{fig:block diagram} via the following equations
\begin{subequations}
\label{implementation frequency} 
    \begin{align}
    \widetilde{\mathbf{w}} &= \mathbf{x} -  (z\mathbf{\Phi}^\mathrm{x}\widetilde{\mathbf{w}} - \widetilde{\mathbf{w}}), \label{a}\\
        \Bar{\mathbf{w}} &= \widehat{\mathbf{\Phi}}^{\mathrm{x}}\widehat{\mathbf{w}}-(z\mathbf{\Phi}^\mathrm{x}\Bar{\mathbf{w}} - \Bar{\mathbf{w}}), \label{b}\\
        \mathbf{u} &= z{\mathbf{\Phi}}^\mathrm{u}(\widetilde{\mathbf{w}} - \Bar{\mathbf{w}}) + \widehat{\mathbf{\Phi}}^{{\mathrm{u}}}\widehat{\mathbf{w}}, \label{c}\end{align}
\end{subequations}
where $\mathbf{x}$, $\Bar{\mathbf{x}}$, and $\widetilde{\mathbf{w}}$ are internal estimated disturbance signals, and $\widetilde{\mathbf{w}}$ and $\Bar{\mathbf{w}}$ are internal estimated disturbance signals.
\end{lemma}
\begin{proof}
Given \eqref{a} and \eqref{b}, it suffices to derive the following form 
\begin{equation}
    \begin{aligned}
        \widetilde{\mathbf{w}} = z^{-1}(\mathbf{\Phi}^{\mathrm{x}})^{-1}\mathbf{x},\quad
        \Bar{\mathbf{w}} = z^{-1}(\mathbf{\Phi}^{\mathrm{x}})^{-1}\widehat{\mathbf{\Phi}}^{\mathrm{x}}\widehat{\mathbf{w}}.\nonumber
    \end{aligned}
\end{equation}
Substituting the preceding equations into \eqref{c}, we obtain the following:
\begin{equation*}
    \begin{aligned}
         \mathbf{u} &= {\mathbf{\Phi}}^\mathrm{u}(z^{-1}(\mathbf{\Phi}^{\mathrm{x}})^{-1}\mathbf{x} - z^{-1}(\mathbf{\Phi}^{\mathrm{x}})^{-1}\widehat{\mathbf{\Phi}}^{\mathrm{x}}\widehat{\mathbf{w}}) + \widehat{\mathbf{\Phi}}^{{\mathrm{u}}}\widehat{\mathbf{w}}\\
         &={\mathbf{\Phi}}^\mathrm{u}(\mathbf{\Phi}^{\mathrm{x}})^{-1}\mathbf{x} + (\widehat{\mathbf{\Phi}}^{\mathrm{u}}-{\mathbf{\Phi}}^\mathrm{u}(\mathbf{\Phi}^{\mathrm{x}})^{-1}\widehat{\mathbf{\Phi}}^{\mathrm{x}})\widehat{\mathbf{w}},\\
         &= \mathbf{Kx} + \widehat{\mathbf{L}}\widehat{\mathbf{w}}.
    \end{aligned}
\end{equation*}
By Theorem \eqref{theorem constraints}, this implementation  achieves all the desired controllers.
\end{proof}
Next, we provide the following lemma to clarify the relationships of internal disturbance signals:
\begin{lemma}[Disturbance Relationship]\label{lemma:w-w=w}
    For any internal disturbances $\widetilde{\mathrm{w}}_t$ and $\Bar{\mathrm{w}}_t$ satisfying closed-loop characterization \eqref{eq:policy} with the external disturbance $\mathrm{w}_{t-1}$, we have
\begin{equation}
    \begin{aligned}
        \widetilde{\mathrm{w}}_t - \Bar{\mathrm{w}}_t = \mathrm{w}_{t-1}.
    \end{aligned}
\end{equation}
\end{lemma}

\begin{proof}
Recall the implementation and subspace constraints. For $t\in[T]$, we have
\begin{align}
\nonumber
    \Bar{\mathrm{w}}_t &= \sum^{T}_{k=0}\widehat{\mathrm{\Phi}}^{\mathrm{x}}_{t,k}\widehat{\mathrm{w}}_{k-1} - \sum^{t-1}_{k=0}\mathrm{\Phi}^{\mathrm{x}}_{t,k}\Bar{\mathrm{w}}_{k} \\
    \nonumber
        &= \sum^{T}_{k=0}\left(A\widehat{\mathrm{\Phi}}^{\mathrm{x}}_{t-1,k} + B\widehat{\mathrm{\Phi}}^{\mathrm{u}}_{t-1,k}\right)\widehat{\mathrm{w}}_{k-1} - \sum^{t-1}_{k=0}\left(A\mathrm{\Phi}^{\mathrm{x}}_{t-1,k} \Bar{\mathrm{w}}_{k} + B\mathrm{\Phi}^{\mathrm{u}}_{t-1,k} \Bar{\mathrm{w}}_{k}\right) \\
        \label{barw}
        &= B\sum^{T}_{k=0}\widehat{\mathrm{\Phi}}^{\mathrm{u}}_{t-1,k}\widehat{\mathrm{w}}_{k-1} - B\sum^{t-1}_{k=0}\mathrm{\Phi}^{\mathrm{u}}_{t-1,k} \Bar{\mathrm{w}}_{k},
\end{align}
where in the second equality, we utilize the equation of 
\begin{equation*}
    \Bar{\mathrm{w}}_{t-1}=\sum^{T}_{k=0}\widehat{\mathrm{\Phi}}^{\mathrm{x}}_{t-1,k}\widehat{\mathrm{w}}_{k-1} - \sum^{t-2}_{k=0}\mathrm{\Phi}^{\mathrm{x}}_{t-1,k}\Bar{\mathrm{w}}_{k}. 
\end{equation*}
Then we substitute \eqref{barw} into the remaining equations~\eqref{eq:policy2} and~\eqref{eq:policy3} in \eqref{eq:policy} and obtain
\begin{equation*}
    \begin{aligned}
\widetilde{\mathrm{w}}_t 
        =&  A\mathrm{x}_{t-1} +B\mathrm{u}_{t-1}+\mathrm{w}_{t-1}-\sum^{t-1}_{k=0}(A{\mathrm{\Phi}}^{\mathrm{x}}_{t-1,k}+B{\mathrm{\Phi}}^{\mathrm{u}}_{t-1,k})\widetilde{\mathrm{w}}_k\\
        =&A\sum^{t-1}_{k=0}{\mathrm{\Phi}}^{\mathrm{x}}_{t-1,k}\widetilde{\mathrm{w}}_k+B\left(\sum^{t-1}_{k=0}\Phi^{\mathrm{u}}_{t-1,k}(\widetilde{\mathrm{w}}_{k}-\Bar{\mathrm{w}}_{k})+\sum^T_{k=0}\widehat{\mathrm{\Phi}}^{\mathrm{x}}_{t-1,k}\widehat{\mathrm{w}}_{k-1}\right)+\mathrm{w}_{t-1} \\ &- \sum^{t-1}_{k=0}(A{\mathrm{\Phi}}^{\mathrm{x}}_{t-1,k}+B{\mathrm{\Phi}}^{\mathrm{u}}_{t-1,k})\widetilde{\mathrm{w}}_k\\
        =&B\sum^{T}_{k=0}\widehat{\mathrm{\Phi}}^{\mathrm{u}}_{t-1,k}\widehat{\mathrm{w}}_{k-1} - B\sum^{t-1}_{k=0}\mathrm{\Phi}^{\mathrm{u}}_{t-1,k} \Bar{\mathrm{w}}_{k} + \mathrm{w}_{t-1}
        \\=& \Bar{\mathrm{w}}_t +\mathrm{w}_{t-1}
    \end{aligned}
\end{equation*}
as desired. This completes the proof of Lemma \ref{lemma:w-w=w}.
\end{proof}

\section{Auxiliary Results for~\Cref{sec:distributed setting}}
In this section, we continue with~\Cref{sec:distributed setting} and provide supplementary contents on distributed \psls implementation and dimension reduction.
\subsection{Distributed Implementation}\label{sec:distributed implementation}
Recall that in \dpsls, once the local optimal CLMs are obtained, the local controller 
for subsystem $i \in [N]$ must receive information from other subsystems 
within the communication distance $\kappa$. It then executes the distributed 
action $\mathrm{u}^i_t$ at each time step $t \in [T]$ using a similar internal 
structure as in~\eqref{eq:policy}:
\begin{subequations}\nonumber
    \begin{align}
    \widehat{\mathrm{x}}^i_t&\coloneqq  \sum_{j\in\mathcal{N}_{\mathcal{G}}^{\distance}(i)}\sum_{k=0}^{T} \widehat{\varphi}^{\mathrm{x},j}_{t,k}(i)\widehat{\mathrm{w}}^j_{k-1},\\
    \widetilde{\mathrm{w}}_t^i &\coloneqq \mathrm{x}_t^i - \sum_{j\in\mathcal{N}_{\mathcal{G}}^{\distance}(i)}\sum^{t-1}_{k=0}\varphi^{\mathrm{x},j}_{t,k}(i)\widetilde{\mathrm{w}}^j_{k}, \\
    \Bar{\mathrm{w}}_t^i &\coloneqq \widehat{\mathrm{x}}_t^i - \sum_{j\in\mathcal{N}_{\mathcal{G}}^{\distance}(i)}\sum^{t-1}_{k=0}\varphi^{\mathrm{x},j}_{t,k}(i)\Bar{\mathrm{w}}^j_{k},\\
        \mathrm{u}_t^i &\coloneqq \sum_{j\in\mathcal{N}_{\mathcal{G}}^{\distance}(i)}\left(\sum^t_{k=0}\varphi^{\mathrm{u},j}_{t,k}(i)\left(\widetilde{\mathrm{w}}_{k}^j - \Bar{\mathrm{w}}_{k}^j\right) + \sum^{T}_{k=0}\widehat{\varphi}^{\mathrm{u},j}_{t,k}(i)\widehat{\mathrm{w}}_{k-1}^j \right),
    \end{align}
\end{subequations} 
where $\mathrm{x}_t^i$, $\mathrm{u}_t^i$, $\widetilde{\mathrm{w}}_t^i$, and $\Bar{\mathrm{w}}_t^i$ are the local state, control action, and internal disturbances, respectively. 

\subsection{Dimension Reduction}
Subject to the column-wise locality condition $\mathcal{L}_{\kappa}(:,i)$, we need to enforce the sparsity of CLMs. 
Such sparsity shows redundant to make computation on the full-size system, allowing for the dimension reduction of \eqref{eq:k-decomp}. Specifically, we take the causal part as an example and rearrange \eqref{eq:causal dynamics} with the nonzero components grouped together as follows
\begin{equation}
    \begin{aligned}
        \underbrace
        {\begin{bmatrix}
            \phi^{\mathrm{x},i}_{t+1,k}\\
            \mathbf{0}\\
            \mathbf{0}
        \end{bmatrix}}_{=\mathcal{P}_{\mathrm{x}}\pxl_{(t+1,k)}}=\underbrace{\begin{bmatrix}
            A^{(i)}_{\text{nn}} & A^{(i)}_{\text{nb}} & A^{(i)}_{\text{nz}} \\
            A^{(i)}_{\text{bn}} & A^{(i)}_{\text{bb}} & A^{(i)}_{\text{bz}} \\
            A^{(i)}_{\text{zn}} & A^{(i)}_{\text{zb}} & A^{(i)}_{\text{zz}}
        \end{bmatrix}}_{= \mathcal{P}_{\mathrm{x}}A\mathcal{P}_{\mathrm{x}}^\top}\underbrace{\begin{bmatrix}
            \phi^{\mathrm{x},i}_{t,k}\\
            \mathbf{0}\\
            \mathbf{0}
        \end{bmatrix}}_{=\mathcal{P}_{\mathrm{x}}\pxl_{t,k}} + \underbrace{\begin{bmatrix}
            B^{(i)}_{\text{nn}} & B^{(i)}_{\text{nz}} \\
            B^{(i)}_{\text{bn}} & B^{(i)}_{\text{bz}} \\
            B^{(i)}_{\text{zn}} & B^{(i)}_{\text{zz}}
        \end{bmatrix}}_{=\mathcal{P}_{\mathrm{x}}B\mathcal{P}_{\mathrm{u}}^\top}\underbrace{\begin{bmatrix}
            \phi^{\mathrm{u},i}_{t,k}\\
            \mathbf{0}
        \end{bmatrix}}_{=\mathcal{P}_{\mathrm{u}}\pul_{t,k}}, \text{for $t\in\{k,\ldots,T\}$}, 
    \end{aligned} \label{eq:reduction equation}
\end{equation}
where 
\begin{align*}\small
&(\cdot)_{\text{nn}}^{(i)}\coloneqq(\cdot)^{[i,\distance]}_\perp\left(\mathcal{N}^\kappa_{\mathcal{G}}(i), \mathcal{N}^\kappa_{\mathcal{G}}(i)\right), \,\,(\cdot)_{\text{nb}}^{(i)}\coloneqq(\cdot)^{[i,\distance]}_\perp\left(\mathcal{N}^\kappa_{\mathcal{G}}(i), \mathcal{J}^\kappa_{\mathcal{G}}(i)\right),\,\,(\cdot)_{\text{nz}}^{(i)}\coloneqq(\cdot)^{[i,\distance]}_\perp\left(\mathcal{N}^\kappa_{\mathcal{G}}(i), \mathcal{Z}^\kappa_{\mathcal{G}}(i)\right),\\
&(\cdot)_{\text{bn}}^{(i)}\coloneqq(\cdot)^{[i,\distance]}_\perp\left(\mathcal{J}^\kappa_{\mathcal{G}}(i), \mathcal{N}^\kappa_{\mathcal{G}}(i)\right),\,\,
(\cdot)_{\text{bb}}^{(i)}\coloneqq(\cdot)^{[i,\distance]}_\perp\left(\mathcal{J}^\kappa_{\mathcal{G}}(i), \mathcal{J}^\kappa_{\mathcal{G}}(i)\right),\,\,
(\cdot)_{\text{bz}}^{(i)}\coloneqq(\cdot)^{[i,\distance]}_\perp\left(\mathcal{J}^\kappa_{\mathcal{G}}(i), \mathcal{Z}^\kappa_{\mathcal{G}}(i)\right),
\\
&(\cdot)_{\text{zn}}^{(i)}\coloneqq(\cdot)^{[i,\distance]}_\perp\left(\mathcal{Z}^\kappa_{\mathcal{G}}(i), \mathcal{N}^\kappa_{\mathcal{G}}(i)\right),\,\,
(\cdot)_{\text{zb}}^{(i)}\coloneqq(\cdot)^{[i,\distance]}_\perp\left(\mathcal{Z}^\kappa_{\mathcal{G}}(i), \mathcal{J}^\kappa_{\mathcal{G}}(i)\right),\,\,
(\cdot)_{\text{zz}}^{(i)}\coloneqq(\cdot)^{[i,\distance]}_\perp\left(\mathcal{Z}^\kappa_{\mathcal{G}}(i), \mathcal{Z}^\kappa_{\mathcal{G}}(i)\right),
\end{align*}
here $\mathcal{J}^\kappa_{\mathcal{G}}(i), \mathcal{N}^\kappa_{\mathcal{G}}(i)$ follow the definition in \Cref{assumption:networked_system}, and  $Z^\kappa_{\mathcal{G}}(i)\coloneqq  [N]\backslash\left(\mathcal{J}^\kappa_{\mathcal{G}}(i)\cup\mathcal{N}^\kappa_{\mathcal{G}}(i)\right)$;
$\mathcal{P}_{\mathrm{x}}, \mathcal{P}_{\mathrm{u}}$ denote the permutation operators that permute $\pxl_{t,k}$ and $\pul_{t,k}$ to $\begin{bmatrix}
        (\phi^{\mathrm{x},i}_{t,k})^\top &
            \mathbf{0} & 
            \mathbf{0}
        \end{bmatrix}^\top$ by $\mathcal{P}_{\mathrm{x}}\pxl_{t,k}$ and $\begin{bmatrix}
        (\phi^{\mathrm{u},i}_{t,k})^\top &
            \mathbf{0}
        \end{bmatrix}^\top$ by $\mathcal{P}_{\mathrm{u}}\pul_{t,k}$.  $\phi^{\mathrm{x},i}_{t,k}$ denotes a vector with nonzero entries in $\pxl_{t,k}$, and  $\phi^{\mathrm{u},i}_{t,k}$ denotes a vector with nonzero entries in $\pul_{t,k}$, respectively. Since the predictive part of CLMs shares the same locality constraints as the causal part, the aforementioned treatment also applies to $\hpxl,\hpul$.
        
Note that here exists a ``boundary" component in $\phi^{\mathrm{u},i}_{t,k}$ since a wider distance is considered to enforce the sparsity of state maintained on $\kappa$-locality. We present the following lemma to illustrate the reduction problem with a necessary constraint \eqref{eq:boundary} that enables the state to transition from nonzero to zero as follows
\begin{lemma}[Reduced problem] \label{lemma:reduced problem}
    \begin{subequations}
    \label{reduced problem}
    \eqref{eq:k-decomp} can be rewritten to an equivalent problem with the reduction system focused on the non-zero structure as  \begin{align}&\min_{\phi}\sum^{T}_{t=0}\left\|\begin{bmatrix}
            Q^{\frac{1}{2}}_{\text{n}} & \mathbf{0} \\ \mathbf{0} & R^{\frac{1}{2}}_{\text{n}}
        \end{bmatrix}\begin{bmatrix} \px_{t,k} + \hpx_{t,k} \\ \pu_{t,k} + \hpu_{t,k} 
        \end{bmatrix}\right\|^2\nonumber \\
\mathrm{subject~to} 
         \quad & \px_{k,k}=\widetilde{\mathrm{e}}_i, 
         \quad  \px_{t+1,k}=A^{(i)}_{\text{nn}}\px_{t,k}+B^{(i)}_{\text{nn}}\pu_{t,k}, \text{ for $t\in\{k,\ldots,T\}$,}\nonumber \\
         & \hpx_{k,k} = \mathbf{0},\quad \hpx_{t+1,k}=A^{(i)}_{\text{nn}}\hpx_{t,k}+B^{(i)}_{\text{nn}}\hpu_{t,k}, \text{ for $t\in[T]$},
         \nonumber\\
         &\left\{\begin{array}{ll}
              A^{(i)}_{\text{bn}}\px_{t,k}+B^{(i)}_{\text{bn}}\pu_{t,k}=\mathbf{0},&\text{ for $t\in\{k,\ldots,T\}$},\\
              A^{(i)}_{\text{bn}}\hpx_{t,k}+B^{(i)}_{\text{bn}}\hpu_{t,k}=\mathbf{0},&\text{ for $t\in[T]$},
        \end{array}\right\} \tag{boundary constraints} \label{eq:boundary}
    \end{align} 
\end{subequations}
where $\widetilde{\mathrm{e}}_i$ denotes the rearrangement of $\mathrm{e}_i$ according to $\phi^{\mathrm{x},i}_{t,k}$; $Q_{\text{n}}$ and $R_{\text{n}}$ denote the reduced matrices rearranged from $Q$ and $R$.
\end{lemma}
\begin{proof}
    The proof is straightforward by substituting equation \eqref{eq:reduction equation} into problem \eqref{localized problem}.
\end{proof}
The key role of boundary constraints lies in applying control to prevent disturbances from propagating beyond the localized regions. The connectivity of subsystem~$i$ within the network $\mathcal{G}$ admits two mutually exclusive cases: either \textit{isolated} from or \textit{coupled} with other subsystems. 
Isolated nodes do not involve considerations of localization or communication. Conversely, for coupled nodes, it is necessary to ensure the solvability of their boundary constraints. A sufficient condition for this is $B_{\mathrm{bn}}^{(i)} (B_{\mathrm{bn}}^{(i)})^\dag = I$ \cite{yu2021localized}, which is satisfied under \Cref{assumption:networked_system}.

\section{Optimal Solution to \eqref{eq:k-decomp} \label{appendix D}}

In this section, we analyze optimal closed-loop solutions to \eqref{eq:k-decomp}. 

\subsection{An Auxiliary Lemma} \label{appendix:affine form}

\begin{lemma}[Recursive solution]
    Let $\mathcal{L}_{\distance}(:,i)$ denote {any $\kappa$-localized} locality constraints as defined in \cref{definition:locality}. Consider the following optimal control problem:
     \begin{align*}
         & \min_{\psi}\sum^{T}_{t=0}\left\|\begin{bmatrix}
            Q^{\frac{1}{2}} & \mathbf{0} \\ \mathbf{0} & R^{\frac{1}{2}}
        \end{bmatrix}\begin{bmatrix}
            \psi^{\mathrm{x}}_{t}\\ \psi^{\mathrm{u}}_t
        \end{bmatrix}\right\|^2\\
        \mathrm{subject~to}\quad& \psi^{\mathrm{x}}_{0} =\delta_0, \nonumber\\& \psi^{\mathrm{x}}_{t+1} = A\psi^{\mathrm{x}}_{t} + B\psi^{\mathrm{u}}_{t} + \delta_{t+1}, \quad\text{ $\forall t\in[T]$},\nonumber\\& \left(\psi_t^{\mathrm{x}}, {\psi}_t^{\mathrm{u}}\right)\in\mathcal{L}_{\distance}(:, i),\quad\text{ $\forall t\in[T]$,}\nonumber
    \end{align*}
    where $A \in \mathbb{R}^{n\times n}$, $B\in \mathbb{R}^{n \times m}$ and $\{\delta_t\}_{t=0}^{T}$ is a given sequence. Then the {optimal control policy} at any $t\in[T]$ is given by the following affine forms \label{lemma:affine form}
   \begin{equation}
       {\psi}^{\mathrm{u}}_t = \Bar{K}_t{\psi}^{\mathrm{x}}_t+\sum^{T-t-1}_{\tau=0}\Bar{M}_{t,\tau}\delta_{t+\tau}, \nonumber
   \end{equation}
   where $\Bar{K}_t$ and $\Bar{M}_{t,\tau}$ denote the optimal finite-horizon LQR gain matrices induced by the system $(A,B,Q,R)$ as defined in \eqref{eq:KM} and $\mathcal{L}_{\distance}(:,i)$ is the $i$th column locality constraint defined via any $\kappa$-localized constraint in \Cref{definition:locality}.
\end{lemma}

\begin{proof}

In the following computations, we consider a scalar system derived from \eqref{dynamics} for simplicity, i.e., $n_i=m_i=1$ for all $i \in [N]$. 
The generalization to higher-dimensional systems is straightforward. 
Let $N^{i}_{\mathrm{x}}$ and $N^{i}_{\mathrm{u}}$ be the number of nonzero elements in the $i$-th column of the matrices $\mathcal{C}^{\distance}$ and 
$\mathcal{C}^{\distance+1}$, respectively. Then, there exist a surjective matrix $M_x \in \mathbb{R}^{(N-N^{i}_\mathrm{x})\times N}$ 
and an injective matrix $M_u \in \mathbb{R}^{N\times N^{i}_\mathrm{u}}$ such that 
the locality constraint \eqref{eq:locality constraint} is explicitly equivalent to
\begin{equation}
    \begin{aligned}
        M_x\varphi^{\mathrm{x}}_t = 0, \quad \varphi^{\mathrm{u}}_t = M_uq_{t}, \quad \forall t \in [T].
    \end{aligned} \label{eq:matrix equation}
\end{equation}
where $q_{t}$ is a free variable. Given the sparsity structure, $M_x$ and $M_u$ can be obtained by stacking standard basis vectors \cite{kjellqvist2022infinite}. Substitute \eqref{eq:matrix equation} into the dynamical constraints in \eqref{eq:k-no}, 
    \begin{align}\label{eq:equations}
        M_xA\widehat{\psi}^{\mathrm{x}}_t + M_xBM_uq_{t}& = 0,\quad  \forall t\in[T].
    \end{align}
Let $Y\coloneqq M_xBM_u$, by adding an unconstrained variable $r_{t}\in\mathbb{R}^{N_r}$, the solution to \eqref{eq:equations} becomes
    \begin{align}
         q_{t} &= - Y^\dag M_xA\widehat{\psi}^{\mathrm{x}}_{t} + N_Y r_{t},\quad \forall t\in[T]. \nonumber
    \end{align}
where $N_Y\in\mathbb{R}^{N^{i}_{\mathrm{u}}\times N_r}$ is a bijection matrix onto the nullspace of $Y$. By substituting \eqref{eq:matrix equation} and \eqref{eq:equations} into \eqref{eq:k-no}, we arrive at an equivalent optimization problem as follows:
\begin{equation}
    \begin{aligned}
        &\min_{r_t}\sum^{T}_{t=0} \left(\widehat{\psi}^{\mathrm{x}\top}_t \widetilde{Q}\widehat{\psi}^{\mathrm{x}}_t + 2r_t^\top \widetilde{Z}\widehat{\psi}^{\mathrm{x}}_t + r_t^\top \widetilde{R}r_t\right),\\
        \text{subject~to}\quad&
        \varphi^{\mathrm{x}}_0 = \delta^{(k)}_0, \quad\widehat{\psi}^{\mathrm{x}}_{t+1} = \widetilde{A}\widehat{\psi}^{\mathrm{x}}_t+\widetilde{B}r_t + \delta_{t},
    \end{aligned}\nonumber
\end{equation}
where $(\widetilde{A}, \widetilde{B}, \widetilde{Q}, \widetilde{R})$ are defined as
\begin{equation*}
    \begin{aligned}
        \widetilde{A} &= (I-BM_uY^\dag M_x)A,\quad\widetilde{B} = BM_uN_Y,\quad
        \widetilde{Q} = Q + A^\top M_x^\top (Y^\dag)^\top M_u^\top R M_uY^\dag M_xA\\
        \widetilde{R} &= (M_uN_Y)^\top RM_uN_Y,\quad
        \widetilde{Z} = -(M_uN_Y)^\top R M_uY^\dag M_xA \, .
    \end{aligned}
\end{equation*}
This problem can be addressed by solving DARE to obtain the standard solution $\widetilde{P}_t$ and affine feedback gains $(\widetilde{K}_t, \widetilde{M}_{t,\tau})$ expressed as follows
\begin{equation}
    \begin{aligned}
        \begin{bmatrix}
            \widetilde{P}_t\\
            \widetilde{K}_t \\
            \widetilde{M}_{t,\tau}
        \end{bmatrix} = \begin{bmatrix}
            \widetilde{{Q}} + \widetilde{{A}}^\top \widetilde{P}_{t+1}\widetilde{{A}} -(\widetilde{{A}}^\top \widetilde{P}_{t+1}\widetilde{{B}} + \widetilde{{Z}}^\top)(\widetilde{R} + \widetilde{{B}}^\top \widetilde{P}_{t+1}\widetilde{{B}})^{-1} (\widetilde{{B}}^\top\widetilde{P}_{t+1}\widetilde{{A}} + \widetilde{{Z}})\\
            -(\widetilde{{R}} + \widetilde{{B}}^\top \widetilde{P}_{t+1}\widetilde{{B}})^{-1} (\widetilde{{B}}^\top\widetilde{P}_{t+1}\widetilde{{A}} + \widetilde{{Z}}) \\
             -(\widetilde{{R}} + \widetilde{{B}}^\top \widetilde{P}_{t+1}\widetilde{{B}})^{-1} \prod_{s=0}^\tau(\widetilde{{A}}+\widetilde{{B}}\widetilde{K}_{t+s})^{\top}\widetilde{P}_{t+\tau+1}
        \end{bmatrix}.\nonumber
    \end{aligned}
\end{equation}

Let $\Bar{K}_{t}\coloneqq M_uN_Y\widetilde{K}_{t} - M_uY^\dag M_xA$ and ${\Bar{M}}_{t,\tau} \coloneqq M_uN_Y\widetilde{M}_{t,\tau}$, and we obtain the optimal solutions to \eqref{eq:k-no} with ${\psi}^{\mathrm{u}}_t = \Bar{K}_t{\psi}^{\mathrm{x}}_t+\sum^{T-t-1}_{\tau=0}\Bar{M}_{t,\tau}\delta_{t+\tau} $, which completes the proof. \end{proof}

\subsection{Optimal Solution to \eqref{eq:k-decomp}}
Now it suffices to discuss the following lemma:
\begin{lemma}[Optimal solution to \eqref{eq:k-decomp}] \label{proposition:optimal solution} \textit{The optimal solution that minimizes \eqref{eq:k-decomp}, represented by $\bm{\varphi} \coloneqq (\pxl_{t,k}, \hpxl_{t,k}, \pul_{t,k}, \hpul_{t,k})_{t\in[T]}$, is unique and satisfies:
\begin{equation}
\label{eq:pul}
    \begin{aligned}
    \begin{bmatrix}
        \pul_{t,k} \\
        \hpul_{t,k}
    \end{bmatrix} = \begin{bmatrix}
        \Bar{K}_{t,t-k}\pxl_{t,k}\\ 
        \Bar{K}_t\hpxl_{t,k} + \Bar{M}_{t, k-t-1}e_i 
    \end{bmatrix}, \quad t \in [T], 
    \end{aligned}
\end{equation}
where $\Bar{K}_t, \Bar{M}_{t,\tau}$ are defined in \Cref{lemma:affine form} and $\pxl_{t,k}, \hpxl_{t,k}$ are defined according to \eqref{eq:causal dynamics} and \eqref{eq:non-causal dynamics} based on \eqref{eq:pul}. }
\end{lemma}

\begin{proof}[Proof of 
\Cref{proposition:optimal solution}]
Recall the problem of \eqref{eq:k-decomp} as following: 
    \begin{align}\label{decomposed problem 1}
    &\min_{\bm{\varphi}}\begin{array}{cc}
         J_1(\bm{\varphi})
    \end{array}\\
    \textrm{subject to} \,\,\,\,& \eqref{eq:causal dynamics}, \eqref{eq:non-causal dynamics}, \eqref{eq:locality constraint}, \nonumber
    \end{align} 
where $J_1(\bm{\varphi})\coloneqq\sum^{T}_{t=0}\left\|\begin{bmatrix}
            Q^{\frac{1}{2}} & \mathbf{0} \\ \mathbf{0} & R^{\frac{1}{2}}
        \end{bmatrix}\begin{bmatrix} \pxl_{t,k} + \hpxl_{t,k} \\ \pul_{t,k} + \hpul_{t,k} 
        \end{bmatrix}\right\|^2.$

{To solve the problem, we now construct an identical problem of \eqref{eq:k-decomp} with relaxed dynamic constraints:}
\begin{align}
    &\min_{\bm{\varphi}}\,\, J_1(\bm{\varphi}), \tag{\texttt{PredSLS}$(i,k)$-\texttt{relax}}\label{eq:mixed problem}\\\text{subject to}\,\,\,\,& ~
      \pxl_{t+1,k} + \hpxl_{t+1,k} = A(\pxl_{t,k} + \hpxl_{t,k}) + B(\pul_{t,k} + \hpul_{t,k}) + \delta_{t+1}^{(k)},\nonumber\\
       & \pxl_{0,k} +  \hpxl_{0,k} = \delta_0^{(k)}, \ \text{ and $\forall$ $t\in[T]$,} \nonumber\\
       & \eqref{eq:locality constraint}.\nonumber
\end{align}
    
    Here, we rewrite \eqref{eq:causal dynamics} and \eqref{eq:non-causal dynamics} with one equality constraint, and $\{\delta_t\}$ is the sequence of vectors with  $\delta_{t} \coloneqq \mathrm{e}_i$ if $t=k$ and $\delta_t\coloneqq\mathbf{0}$ otherwise.
    
    Denote $\widehat{\psi}_t^{\mathrm{x}}\coloneqq \pxl_{t,k} + \hpxl_{t,k}$ and $\widehat{\psi}_t^{\mathrm{u}}\coloneqq \pul_{t,k} + \hpul_{t,k}$ as the aggregated variables.  
    Note that after such a change of variable, \eqref{eq:mixed problem} transform to the problem \eqref{eq:k-no}. Here we show  \eqref{eq:k-no} for reference:
    \begin{align}
        \label{eq:noncausal}&\min_{\widehat{\psi}}\sum^{T}_{t=0}\left\|\begin{bmatrix}
            Q^{\frac{1}{2}} & \mathbf{0} \\ \mathbf{0} & R^{\frac{1}{2}}
        \end{bmatrix}\begin{bmatrix}
            \widehat{\psi}^{\mathrm{x}}_{t}\\\widehat{\psi}^{\mathrm{u}}_t
        \end{bmatrix}\right\|^2 \\
         \text{subject to}~& \widehat{\psi}^{\mathrm{x}}_{0} = \delta_0^{(k)},\,\, \widehat{\psi}^{\mathrm{x}}_{t+1} = A\widehat{\psi}^{\mathrm{x}}_{t} + B\widehat{\psi}^{\mathrm{u}}_{t} + \delta_{t+1}^{(k)}, \,\,\text{ $\forall t\in[T]$},\nonumber\\& \left({\widehat{\psi}}_t^{\mathrm{x}}, {\widehat{\psi}}_t^{\mathrm{u}}\right)\in\mathcal{L}_{\distance}(:, i),\,\,\text{ $\forall t\in[T]$}. \nonumber
    \end{align} 

    By \Cref{lemma:affine form}, the solution to the preceding problem is 
   \begin{equation}\small
   \label{eq:mpc_solution}
   \begin{aligned}
       \widehat{\psi}^{\mathrm{u}}_t &= \Bar{K}_t\widehat{\psi}^{\mathrm{x}}_t+\Bar{M}_{t,k-t}\mathrm{e}_i,\,\,\forall t \in [T],\\
       \widehat{\psi}^{\mathrm{x}}_{t+1} &= A\widehat{\psi}^{\mathrm{x}}_{t} + B\widehat{\psi}^{\mathrm{u}}_{t} + \delta_{t+1}^{(k)},\,\,\forall t \in [T],
   \end{aligned}
   \end{equation}
 where $\bar M_{t,\tau}$ and $\bar K_t$ satisfy the following forms:
 \begin{align}
     \label{eq:KM}
     \Bar{K}=M_uN_Y\widetilde{K}_t-M_uY^\dag M_xA,\quad\Bar{M}_{t,\tau}=M_uN_Y\widetilde{M}_{t,\tau},
 \end{align}
and $M_x, M_u, Y, N_Y$ are locality constraint matrices and $(\widetilde{K}_t,\widetilde{M}_{t,\tau})$ denotes the DARE solution of the defined LQR system $(\widetilde{A},\widetilde{B},\widetilde{Q},\widetilde{R},\widetilde{Z})$, and all of them are defined in the proof of \Cref{lemma:affine form}. Substituting $\widehat{\psi}_t^{\mathrm{x}}= \pxl_{t,k} + \hpxl_{t,k}$ and $\widehat{\psi}_t^{\mathrm{u}}= \pul_{t,k} + \hpul_{t,k}$ into \eqref{eq:mpc_solution}, we see that the optimal solution is
\begin{align}\small
        \pul_{t,k} + \hpul_{t,k} &=   \Bar{K}_t\left(\pxl_{t,k} + \hpxl_{t,k}\right)  +\Bar{M}_{t,k-t-1}e_i\nonumber\\ \Rightarrow\hpul_{t,k} &= \Bar{K}_t\hpxl_{t,k} +\Bar{M}_{t,k-t-1}e_i + \left(\Bar{K}_t\pxl_{t,k} -  \xi_{t,k}\right),\label{eq:mixed solution}\\
        \pxl_{t+1,k} + \hpxl_{t+1,k} &=A\left(\pxl_{t,k} + \hpxl_{t,k}\right) + B\left(\pul_{t,k} + \hpul_{t,k}\right) + \delta^{(k)}_{t+1},\nonumber
\end{align}
here we assume $\pul_{t,k}=\xi_{t,k}$ and $\xi_{t,k} \in\mathbb{R}^{n_i}$ denotes a free vector variable. In other words, for all $\xi_{t,k} \in\mathbb{R}^{m_i}$, \eqref{eq:mixed solution} minimizes \eqref{eq:mixed problem}.

Next, we consider the optimal solutions to the following problem that is transformed from \eqref{eq:mixed problem} by setting $\hpxl_{t,k}=\mathbf{0}, \hpul_{t,k}=\mathbf{0}, \forall t\in[T]$:
\begin{align}
    &\min_{\bm{\varphi}}\,\, \min_{\bm{\varphi}}\,\, \sum^{T}_{t=k}\left\|\begin{bmatrix}
            Q^{\frac{1}{2}} & \mathbf{0} \\ \mathbf{0} & R^{\frac{1}{2}}
        \end{bmatrix}\begin{bmatrix} \pxl_{t,k}  \\ \pul_{t,k}
        \end{bmatrix}\right\|^2,\tag{\texttt{PredSLS}$(i,k)$-\texttt{cau}}\label{eq:causal problem}
    \\
       \text{subject~to}\quad  &\pxl_{t+1,k} = A\pxl_{t,k} + B\pul_{t,k}, \text{ for $t\in \{k,\ldots, T-1\}$},\nonumber\\&\varphi^{\mathrm{x}}_{k,k}  = \mathrm{e}_i, \varphi^{\mathrm{x}}_{t,k} = \mathbf{0}, \varphi^{\mathrm{u}}_{t,k} = \mathbf{0}, \text{ for $t\in[k]$},\nonumber\\& \eqref{eq:locality constraint}.\nonumber
\end{align}
Note that \eqref{eq:causal problem} is exactly \eqref{eq:k-ca} by a notation transformation with $\varphi\to\psi$ and a time step drift from $t+k$ to $t$ for $t\in\{0,\ldots,T-k\}$. By \Cref{lemma:affine form}, we obtain the following solution: 
\begin{equation}
\begin{aligned}
    \varphi^{\mathrm{u}}_{t,k} &= \Bar{K}_t \varphi^{\mathrm{x}}_{t,k}, \,\, \text{for all }t\in \{k,\ldots,T-1\}, \\  \pxl_{t+1,k} &= A\pxl_{t,k} + B\pul_{t,k}, \,\, \text{for all }t\in \{k,\ldots,T-1\},
\end{aligned} \label{eq:causal solution}
\end{equation}
where $\Bar{K}_t$ is an identical gain to that of \eqref{eq:mixed solution}. Since $\xi_{t,k}$ in \eqref{eq:mixed solution} is a free variable, we will substitute \eqref{eq:causal solution} into $\xi_{t,k}$, e.g., set the free variable as $\xi_{t,k}=\Bar{K}_t \varphi^{\mathrm{x}}_{t,k}$ in \eqref{eq:mixed solution} and arrive at an optimal solution to \eqref{eq:mixed problem} at $t$ as follows:
\begin{equation}
    \begin{aligned}
    \begin{bmatrix}
    \pxl_{t+1,k} &
        \hpxl_{t+1,k} \\
        \pul_{t,k} &
        \hpul_{t,k}
    \end{bmatrix} = \begin{bmatrix}
    A\pxl_{t,k} + B\pul_{t,k} \ & \
        A\widehat{\psi}^{\mathrm{x}}_{t} + B\widehat{\psi}^{\mathrm{u}}_{t} + \delta_{t+1}^{(k)}\\
        \Bar{K}_{t,k}\pxl_{t,k} \ & \
        \Bar{K}_t\hpxl_{t,k} +\Bar{M}_{t,k-t-1}e_i 
    \end{bmatrix}, 
    \end{aligned}\label{eq:decomp solution}
\end{equation}
where $\Bar{K}_{t,\tau} = \Bar{K}_t$ for any index $\tau\geq0$ and $\Bar{K}_{t,\tau} = \mathbf{0}$ otherwise. The vectors $\pxl_{t,k}, \hpxl_{t,k}$ can be computed based on \eqref{eq:decomp solution} and the update rules specified by the constraints in \eqref{eq:causal problem} and \eqref{eq:mixed problem}. Therefore, \eqref{eq:decomp solution}  minimizes \eqref{eq:mixed problem}.

Note that the feasible region of \eqref{eq:mixed problem} is larger than \eqref{eq:k-decomp} while keeping everything the same. Therefore, \eqref{eq:decomp solution} is the optimal solution to \eqref{eq:k-decomp}.

Finally, it is straightforward to confirm that the objective function in \eqref{eq:k-decomp} is strictly convex (quadratic function with $(Q,R)$) and the feasible region is convex, illustrating the uniqueness of the optimal solution \eqref{eq:decomp solution}. 
\end{proof}

{An immediate consequence of \Cref{proposition:optimal solution} is that we now have an explicit solution to \eqref{eq:k-decomp}.}

\begin{corollary}
    [Explicit solution to \eqref{eq:k-decomp}] \label{lemma:explicit form}
    \textit{Given the optimal gains $\Bar{K}_t\in\mathbb{R}^{m\times n}$, $\Bar{M}_{t,\tau}\in\mathbb{R}^{m\times n}$ defined in \Cref{lemma:affine form} and Assumption~\ref{assumption:networked_system}, there exist $C_1 >0$, $L>0$ and $\rho \in (0,1)$ such that $\|\Bar{K}_t\|\leq L$, $\|\Bar{M}_{t,\tau}\|\leq L_1\gamma^\tau$, and $\|A+B\Bar{K}_t\|^\tau\leq L\gamma^\tau$ for all $t,\tau\in[T]$. Furthermore, the optimal solution to \eqref{eq:k-decomp} can be written as
    \begin{align} 
    \begin{bmatrix}
        \varphi_{t,k}^{\mathrm{x},i} & \quad \widehat{\varphi}_{t,k}^{\mathrm{x},i}\\
        \varphi_{t,k}^{\mathrm{u},i} & \quad \widehat{\varphi}_{t,k}^{\mathrm{u},i}
    \end{bmatrix} \!\!\!=\!\!\! \begin{bmatrix}
        \Bar{F}_{t,k} & \Bar{N}_{t,k}\\\Bar{K}_{t}\Bar{F}_{t,k} \ & \  \Bar{K}_{t}\Bar{N}_{t,k} + \Bar{M}_{t,k-t-1}\mathrm{e}_i
    \end{bmatrix},
    \nonumber
    \end{align}
     where 
     \begin{equation*}
         \begin{aligned}
             \Bar{F}_{t,k}&\coloneqq \left(\prod_{\tau=k}^{t-1}(A+B\Bar{K}_\tau)\right)(\mathrm{e}_i\circ \mathbf{1}(t-k)),\\
             \Bar{N}_{t,k}&\coloneqq \sum^{t-1}_{\tau=0}\left(\prod_{h=t-\tau}^{t-1}(A+B\Bar{K}_h)\right)B\Bar{M}_{\tau,k-t+\tau}\mathrm{e}_i + \Bar{F}_{t,k},
         \end{aligned}
     \end{equation*}
     and indices $t, k\in\mathbb{N}_+$; $\mathbf{1}(\tau), \tau\in\mathbb{N}$ denotes the Heaviside step function; $\Bar{M}_{(\cdot),\tau}=0$ for all $\tau<0$.}
\end{corollary}

\subsection{Proof of Proposition \ref{proposition:2}}\label{proof: proposition 2}
\begin{proof}
Let $(\psi^{\mathrm{x}}_t,\psi^{\mathrm{u}}_t)$ and $(\widehat{\psi}^{\mathrm{x}}_t,\widehat{\psi}^{\mathrm{u}}_t)$ be the optimal solutions to \eqref{eq:k-ca} and \eqref{eq:k-no} respectively. By the proof of \Cref{proposition:optimal solution}, $(\widehat{\psi}^{\mathrm{x}}_t,\widehat{\psi}^{\mathrm{u}}_t)$ are characterized by \eqref{eq:mpc_solution} and $(\psi^{\mathrm{x}}_t,\psi^{\mathrm{u}}_t)$ are solved as 
\begin{align*}
    \psi^{\mathrm{u}}_{t,k} &= \Bar{K}_t \psi^{\mathrm{x}}_{t,k}, \,\, \text{for all }t\in [T-k], \\  \psi_{t+1,k}^{\mathrm{x}} &= A\psi_{t,k}^{\mathrm{u}} + B\psi_{t,k}^{\mathrm{x}}, \,\, \text{for all }t\in [T-k],
\end{align*}
Then given \eqref{eq:decomp solution}, it can be directly obtained that \begin{align}
\begin{bmatrix}
    \pxl_{t,k}  \ \ & \ \ \hpxl_{t,k} \\  \pul_{t,k} \ \ & \ \ \hpul_{t,k}
\end{bmatrix} = \begin{cases}
    \begin{bmatrix}
        \mathbf{0} \  & \  \widehat{\psi}^{\mathrm{x}\star}_{t}\\ \mathbf{0} \  &  \ \widehat{\psi}^{\mathrm{u}\star}_{t}
    \end{bmatrix} &\mathrm{if}~t<k,\vspace{1pt}\\
    \begin{bmatrix}
    {\psi}^{\mathrm{x}\star}_{t} \ \  &  \ \ \widehat{\psi}^{\mathrm{x}\star}_{t} - {\psi}^{\mathrm{x}\star}_{t}\\ {\psi}^{\mathrm{u}\star}_{t} \ \ & \ \ \widehat{\psi}^{\mathrm{u}\star}_{t} - {\psi}^{\mathrm{u}\star}_{t}
    \end{bmatrix} &\mathrm{otherwise},
\end{cases}\nonumber
\end{align}
is an optimal solution to \eqref{eq:k-decomp}.
\end{proof}

\section{Proof for Theorem \ref{theorem:temporal}} \label{proof:temporal}
We have already explicitly constructed an optimal solution to \eqref{eq:k-decomp} in Appendix \ref{appendix D}. In this proof, the solution to \eqref{eq:k-decomp} is denoted as $(\pxl_{t,k}, \hpxl_{t,k}, \pul_{t,k}, \hpul_{t,k})$. Now consider the following inequality:
\begin{align}
    \left\|\begin{bmatrix}
        \pxl_{t,k}  & \quad \hpxl_{t,k}\\ \pul_{t,k} & \quad\hpul_{t,k}
    \end{bmatrix}\right\|^2\leq &\underbrace{\left\|\pxl_{t,k}\right\|^2+\left\|\pul_{t,k}\right\|^2}_{\text{(a)}}+\underbrace{\left\|\hpxl_{t,k}\right\|^2+\left\|\hpul_{t,k}\right\|^2}_{\text{(b)}}\nonumber ,
    \end{align}
here we use $ \left\|\begin{bmatrix}
    M_1 & M_2
\end{bmatrix}\right\|^2\leq\|M_1\|^2 + \|M_2\|^2$. 
We now bound (a) and (b) separately. By \Cref{lemma:explicit form}, it follows that
\begin{align}
    \text{(a)}&\leq \left\|\prod^{t-1}_{\tau=k}(A+B\Bar{K}_\tau)\right\|^2  + 
        \|\Bar{K}_t\|^2\left\|\prod^{t-1}_{\tau=k}(A+B\Bar{K}_\tau)\right\|^2  
        \label{eq:upper_a}
         \leq{(1+L^2)L^2\gamma^{2(|t-k|-1)}}.
\end{align}

For simplification, we denote $\Bar{F}_\tau\coloneqq(A+B\Bar{K}_\tau)$. Then
\begin{align}
\small
\nonumber
    \text{(b)}&\leq\left( \left\|\sum^{t-1}_{\tau=0}\left(\prod_{h=t-\tau}^{t-1}\Bar{F}_h\right)B\Bar{M}_{\tau,k-t+\tau}\right\| + \left\|\prod_{\tau=k}^{t-1}\Bar{F}_\tau\right\| \right)^2\\ & \quad + \left(
        \left\|\Bar{K}_{t}\right\|\left(\left\|\sum^{t-1}_{\tau=0}\left(\prod_{h=t-\tau}^{t-1}\Bar{F}_h\right)B\Bar{M}_{k-t+\tau}\right\| + \left\|\prod_{\tau=k}^{t-1}\Bar{F}_\tau\right\|\right) + \left\|\Bar{M}_{t,k-t-1}\right\|\right)^2\nonumber\\
        \nonumber
        &\leq \left(L^2L_1\sum^{t-1}_{\tau=0}\gamma^{2\tau-1+k-t} + L\gamma^{t-k-1}\right)^2 + \left(L^3L_1\sum^{t-1}_{\tau=0}\gamma^{2\tau-1+k-t} + L^2\gamma^{t-k-1} + L_1\gamma^{k-t-1}\right)^2\\
        \label{eq:upper_b}
        &\leq \frac{((L^2L_1+(1-\gamma^2)L)^2+(L^3L_1+(1-\gamma^2)(L^2+L_1))^2)\gamma^{2(|t-k|-1)}}{(1-\gamma^2)^2}.
\end{align}
Combining the upper bounds~\eqref{eq:upper_a} and~\eqref{eq:upper_b} in (a) and (b), we obtain that 
\begin{align}
    \left\|\begin{bmatrix}
        \pxl_{t,k} & \quad\hpxl_{t,k}\\ \pul_{t,k}& \quad \hpul_{t,k}
    \end{bmatrix}\right\|^2\leq C\gamma^{2|t-k|}.
\end{align}
where $C\coloneqq\frac{1}{\gamma^2}\left((1+L^2)L^2 + \frac{(L^2L_1+(1-\gamma^2)L)^2+(L^3L_1+(1-\gamma^2)(L^2+L_1))^2}{(1-\gamma^2)^2}\right)$. Let $\rho\coloneqq \gamma^2\in(0,1)$. We complete the proof.

\section{Proof of Theorem \ref{theorem:spatial}} 
\label{appendix:spatial}
\subsection{Necessary Pre-Processing}

This subsection proceeds in three steps. As a proof outline, we first \textbf{reformulate problems \eqref{eq:k-ca} and \eqref{eq:k-no}}. Second, we derive the \textbf{KKT conditions and KKT matrix $H$}. Finally, we analyze the \textbf{decaying property of $H^{-1}$}. The proofs of all lemmas presented in this section are deferred to later contexts.

\paragraph{Step 1: Reformulate \eqref{eq:k-ca} and \eqref{eq:k-no}.} We first re-formulate \eqref{eq:k-ca} and \eqref{eq:k-no} to equivalent optimization problems whose constraints are in the standard KKT condition's form. In particular, we will absorb the locality constraints into new optimization variables through a change of variable.  

Similar to the dimension reduction procedure in \eqref{eq:reduction equation}, the constraints of \eqref{eq:k-no} can be written as:
    \begin{align}
        \nonumber&\left\{(\widehat{\psi}^{\mathrm{x}}_t,\widehat{\psi}^{\mathrm{u}}_t)_{t\in[T]}\,|\,\widehat{\psi}^{\mathrm{x}}_{t+1} = A\widehat{\psi}^{\mathrm{x}}_t + B\widehat{\psi}^{\mathrm{u}}_t + \delta_{t+1}^{(k)}, \quad \left(\widehat{\psi}^{\mathrm{x}}_t, \widehat{\psi}^{\mathrm{u}}_t\right)\in\mathcal{L}_{\distance}(:, i)\right\}\\
        \nonumber&=\left\{(\widehat{\psi}^{\mathrm{x}}_t,\widehat{\psi}^{\mathrm{u}}_t)_{t\in[T]}\left|
        \begin{array}{c}
        \begin{bmatrix}
            \Psi^{\mathrm{x}}_{t+1}\\
            \mathbf{0}\\
            \mathbf{0}
        \end{bmatrix}=\begin{bmatrix}
            A^{(i)}_{\text{nn}} & A^{(i)}_{\text{nb}} & A^{(i)}_{\text{nz}} \\
            A^{(i)}_{\text{bn}} & A^{(i)}_{\text{bb}} & A^{(i)}_{\text{bz}} \\
            A^{(i)}_{\text{zn}} & A^{(i)}_{\text{zb}} & A^{(i)}_{\text{zz}}
        \end{bmatrix}\begin{bmatrix}
            \Psi^{\mathrm{x}}_{t}\\
            \mathbf{0}\\
            \mathbf{0}
        \end{bmatrix} + \begin{bmatrix}
            B^{(i)}_{\text{nn}} & B^{(i)}_{\text{nz}} \\
            B^{(i)}_{\text{bn}} & B^{(i)}_{\text{bz}} \\
            B^{(i)}_{\text{zn}} & B^{(i)}_{\text{zz}}
        \end{bmatrix}\begin{bmatrix}
            \Psi^{\mathrm{u}}_{t}\\
            \mathbf{0}
        \end{bmatrix} + \begin{bmatrix}
            \widetilde{\delta}_{t+1}^{(k)} \\ \mathbf{0}\\\mathbf{0}
        \end{bmatrix}\\\widehat{\psi}^{\mathrm{x}}_t=\mathcal{P}_{\mathrm{x}} \begin{bmatrix}
            \Psi^{\mathrm{x}}_{t}\\
           \mathbf{0}\\
            \mathbf{0}
        \end{bmatrix},\quad\widehat{\psi}^{\mathrm{u}}_t=  \mathcal{P}_{\mathrm{u}}\begin{bmatrix}
            \Psi^{\mathrm{u}}_{t}\\
            \mathbf{0}
        \end{bmatrix}
        \end{array}
        \right.\right\}\\
        \nonumber&= \left\{(\widehat{\psi}^{\mathrm{x}}_t,\widehat{\psi}^{\mathrm{u}}_t)_{t\in[T]}\left|\begin{array}{c}
            \widehat{\psi}^{\mathrm{x}}_{t+1}-\mathcal{P}_{\mathrm{x}}\begin{bmatrix}
            A^{(i)}_{\text{nn}} & A^{(i)}_{\text{nb}} & A^{(i)}_{\text{nz}}\\
            \mathbf{0} & \mathbf{0} & \mathbf{0} \\
            \mathbf{0} & \mathbf{0} & \mathbf{0}
        \end{bmatrix}\mathcal{P}_{\mathrm{x}}^\top \widehat{\psi}^{\mathrm{x}}_{t} - \mathcal{P}_{\mathrm{x}} \begin{bmatrix}
            B^{(i)}_{\text{nn}} & B^{(i)}_{\text{nz}}\\
            \mathbf{0} & \mathbf{0}\\
            \mathbf{0} & \mathbf{0}
        \end{bmatrix}\mathcal{P}_{\mathrm{u}}^\top\widehat{\psi}^{\mathrm{u}}_{t} = {\delta}_{t+1}^{(k)},\\
        \mathcal{P}_{\mathrm{x}}\begin{bmatrix}
        \mathbf{0} & \mathbf{0} & \mathbf{0} \\
            A^{(i)}_{\text{bn}} & A^{(i)}_{\text{bb}} & A^{(i)}_{\text{bz}}\\
            \mathbf{0} & \mathbf{0} & \mathbf{0}
        \end{bmatrix}\mathcal{P}_{\mathrm{x}}^\top\widehat{\psi}^{\mathrm{x}}_{t} +\mathcal{P}_{\mathrm{x}} \begin{bmatrix}
        \mathbf{0} & \mathbf{0} \\
            B^{(i)}_{\text{bn}} & B^{(i)}_{\text{bz}}\\
            \mathbf{0} & \mathbf{0}
        \end{bmatrix}\mathcal{P}_{\mathrm{u}}^\top\widehat{\psi}^{\mathrm{u}}_{t}=\mathbf{0},\\\widehat{\psi}^{\mathrm{u}}_t=  \mathcal{P}_{\mathrm{u}}\begin{bmatrix}
            \Psi^{\mathrm{u}}_{t}\\
            \Psi^{\mathrm{u}, \mathrm{z}}_{t}
        \end{bmatrix}, \quad \Psi^{\mathrm{u}, \mathrm{z}}_{t}=\mathbf{0}
        \end{array}\right.\right\}
        \\&=\left\{(\widehat{\psi}^{\mathrm{x}}_t,\widehat{\psi}^{\mathrm{u}}_t)_{t\in[T]}\left|\underbrace{\begin{bmatrix}
            I & \quad - A^{[i,\distance]}& \quad -B^{[i,\distance]}\\ \mathbf{0}& \quad \widetilde{A}_\perp^{[i,\distance]} & \quad\widetilde{B}_\perp^{[i,\distance]}&\\ \mathbf{0}& \quad\mathbf{0}& \quad I_ \lnot^{[i,\distance]}
        \end{bmatrix}}_{Z_{i,\kappa}}\begin{bmatrix}
            \widehat{\psi}^{\mathrm{x}}_{t+1}\\ \widehat{\psi}^{\mathrm{x}}_t\\ \widehat{\psi}^{\mathrm{u}}_t
        \end{bmatrix}=\begin{bmatrix}
            \delta_{t+1}^{(k)}\\\mathbf{0}\\
            \mathbf{0}
        \end{bmatrix}\right.\right\}, \label{eq:rearrangement constraints}
    \end{align} 
where $\mathcal{P}_{\mathrm{x}}$ and $\mathcal{P}_{\mathrm{u}}$ denote the permutation matrices introduced in \eqref{eq:reduction equation}; $\widetilde{A}_\perp^{[i,\kappa]}$ and $\widetilde{B}_\perp^{[i,\kappa]}$ represent the largest possible row-independent sub-matrices extracted from ${A}_\perp^{[i,\kappa]}$ and ${B}_\perp^{[i,\kappa]}$ respectively. Furthermore, the notation $\widetilde{I}_{\lnot}^{[i,\kappa]}\coloneqq \mathrm{extr}({I-I_{\perp}^{[i,\kappa+1]}})$ denotes the 
residual version of identity matrix $I$, and $\mathrm{extr}(\cdot)$ is the operation of extracting non-zero rows. In the first equality, we follow the similar derivation as that of \eqref{eq:reduction equation}. In the third equality, we use a matrix $Z_{i,\kappa}$ to compactly express the second equality. 

Next, we show the following results of the block matrix $Z_{i,\kappa}$, which demonstrates the linear independent constraint qualification (LICQ) satisfaction, defined as:
\begin{definition}
    LICQ for an equality-constrained 
optimization problem, where the constraint set is defined as 
$\{\mathrm{x} \in \mathbb{R}^n \mid h_r(\mathrm{x}) = 0, \text{ for } r=1,\dots,r_{\max}\}$, 
is satisfied at an optimal solution $\mathrm{x}^*$ if and only if the gradients of all the constraints are linearly independent,   that is, the set
    $\{\nabla_{\mathrm{x}} h_r(\mathrm{x}^*) \mid r=1,\dots,r_{\max}\}$
is linearly independent.
\end{definition}

In our context, the constraints $h_r$ of \eqref{eq:k-no} are characterized by \eqref{eq:rearrangement constraints}.

\begin{lemma}[LICQ for \eqref{eq:k-no}]
    \label{lemma: full row rank} Under \Cref{assumption:networked_system}, 
    $Z_{i,\kappa}$ defined in \eqref{eq:rearrangement constraints} has full row rank.
\end{lemma}
\begin{proof}
     The sparsity structure of $Z_{i,\kappa}$ ensures the linear independence 
between its first and second row blocks, as well as between its first and 
third row blocks. Furthermore, the second and third row blocks are 
linearly independent due to the non-zero term 
$B^{(i)}_{\mathrm{bn}}$ in the second row block.

The subsequent analysis addresses the internal linear independence of each 
row block. Specifically, the internal independence of the first and third 
row blocks is evident from their respective inclusion of an identity 
sub-matrix $I$ and non-zero sub-matrix $I^{[i,\kappa]}_{\lnot}$. The internal 
independence of the second row block is a direct consequence of the 
definitions of $\widetilde{A}_\perp^{[i,\kappa]}$ and 
$\widetilde{B}_\perp^{[i,\kappa]}$.

Consequently, the linear independence of all row blocks constituting 
$Z_{i,\kappa}$ is established.
\end{proof}

Note that, the matrix $Z_{i.\kappa}$ formulates a recursive mapping matrix of the complete constraints. Therefore, the preceding lemma directly demonstrates LICQ.

We now restate the problem \eqref{eq:k-no} as
    \begin{align}
        & \min_{(\widehat{\psi}^{\mathrm{x}}_t,\widehat{\psi}^{\mathrm{u}}_t)_{t\in[T]}}\sum^{T}_{t=0}\left\|\begin{bmatrix}
            Q^{\frac{1}{2}} & \mathbf{0} \\ \mathbf{0} & R^{\frac{1}{2}}
        \end{bmatrix}\begin{bmatrix}
            \widehat{\psi}^{\mathrm{x}}_{t}\\ \widehat{\psi}^{\mathrm{u}}_t
        \end{bmatrix}\right\|^2 \label{eq:noc_refor}\\\nonumber
         \text{subject~to}\quad&  \varphi^{\mathrm{x}}_{0} = \delta_0,\\&\begin{bmatrix}
            I & \quad - A^{[i,\distance]}& \quad-B^{[i,\distance]}\\ &\widetilde{A}_\perp^{[i,\distance]} & \widetilde{B}_\perp^{[i,\distance]}&\\&&I_ \lnot^{[i,\distance]}
        \end{bmatrix}\begin{bmatrix}
            \widehat{\psi}^{\mathrm{x}}_{t+1}\\\widehat{\psi}^{\mathrm{x}}_t\\\widehat{\psi}^{\mathrm{u}}_t
        \end{bmatrix}=\begin{bmatrix}
            \delta_{t+1}^{(k)}\\\mathbf{0}\\
            \mathbf{0}
        \end{bmatrix},\,\,t\in[T]. \nonumber
    \end{align}

Using the same reformulation, \eqref{eq:k-ca} can be restated as
    \begin{align}
        & \min_{({\psi}^{\mathrm{x}}_t,{\psi}^{\mathrm{u}}_t)_{t\in[T-k]}}\sum^{T-k}_{t=0}\left\|\begin{bmatrix}
            Q^{\frac{1}{2}} & \mathbf{0} \\ \mathbf{0} & R^{\frac{1}{2}}
        \end{bmatrix}\begin{bmatrix}
            {\psi}^{\mathrm{x}}_{t}\\ {\psi}^{\mathrm{u}}_t
        \end{bmatrix}\right\|^2 \label{eq:cau_refor} \\\nonumber
         \text{subject~to}\quad &   \psi^{\mathrm{x}}_{0} = \mathrm{e}_i,\\&\begin{bmatrix}
            I & \quad - A^{[i,\distance]}& \quad-B^{[i,\distance]}\\ &\quad\widetilde{A}_\perp^{[i,\distance]} & \widetilde{B}_\perp^{[i,\distance]}&\\&&I_ \lnot^{[i,\distance]}
        \end{bmatrix}\begin{bmatrix}
            {\psi}^{\mathrm{x}}_{t+1}\\{\psi}^{\mathrm{x}}_t\\{\psi}^{\mathrm{u}}_t
        \end{bmatrix}=\begin{bmatrix}
            \mathbf{0}\\\mathbf{0}\\
            \mathbf{0}
        \end{bmatrix},\,\,t\in[T-k]. \nonumber
    \end{align}
Note that, when no locality constraints are considered, the recursive matrix $Z_{i,\kappa}$ will reduce to the dynamic matrix as $Z = \begin{bmatrix}
        I & \quad -A & \quad -B
    \end{bmatrix}$.

\paragraph{Step 2: KKT conditions.} In this part, we will define several quantities for the construction of KKT conditions.

First, the Jacobian matrix $J^{[i,\kappa]}_{\mathrm{noc}}$ for \eqref{eq:k-no} can be defined as
\begin{equation}
    J^{[i,\kappa]}_{\mathrm{noc}}=
        \left[\begin{array}{ccccccc}
             I \\
            -{A}^{[i,\distance]} & -{B}^{[i,\distance]} & I \\
            &&-{A}^{[i,\distance]} & -{B}^{[i,\distance]} & I \\
             &&& & \ddots & & \\
            && & &   -{A}^{[i,\distance]} & -{B}^{[i,\distance]} & I\\
            \hline
              {A}^{[i,\distance]}_\perp & {B}^{[i,\distance]}_\perp &  \mathbf{0} \\&&{A}^{[i,\distance]}_\perp & {B}^{[i,\distance]}_\perp & \mathbf{0}  \\& & & & \ddots& \\
             &   &&& {A}^{[i,\distance]}_\perp  & {A}^{[i,\distance]}_\perp & \mathbf{0}\\\hline
             \mathbf{0} &{I}_\lnot^{[i,\distance]} & \mathbf{0} \\&&\mathbf{0} & {I}_\lnot^{[i,\distance]} & \mathbf{0}\\& & & & \ddots& \\ &   &&& \mathbf{0} &{I}_\lnot^{[i,\distance]} & \mathbf{0}
        \end{array}\right]\coloneqq \left[\begin{array}{c}
             J_1^{[i,\kappa]}\\\hline J_2^{[i,\kappa]}\\\hline J_3^{[i,\kappa]}
        \end{array}\right]. \label{eq:jacobian_matrix}
\end{equation}

 Recall Lemma \ref{lemma: full row rank}, it is straightforward to check that $J^{[i,\kappa]}_{\mathrm{noc}}$ always has full row rank. Given any disturbance trajectory $\bm{\delta}\coloneqq(\delta_1^\top,\ldots,\mathrm{\delta}_{T}^\top,\mathbf{0}^\top,\ldots,\mathbf{0}^\top)^\top$, one can always find the solution trajectory $\widehat{\bm{\psi}}\coloneqq(\widehat{\psi}_0^{\mathrm{x}\top},\widehat{\psi}_0^{\mathrm{u}\top},\widehat{\psi}_1^{\mathrm{x}\top},\ldots,\widehat{\psi}_{T-1}^{\mathrm{u}\top},\widehat{\psi}_T^{\mathrm{x}\top})^\top$ satisfying the linear nonhomogeneous system $J^{[i,\kappa]}_{\mathrm{noc}} \widehat{\bm{\psi}} =\bm{\delta} $.
Accordingly, the Jacobian matrix $J^{[i,\kappa]}_{\mathrm{cau}}$ for \eqref{eq:k-ca} shows the same block structure to $J^{[i,\kappa]}_{\mathrm{noc}}$. 

Subsequently, we lower bound the singular values of $J^{[i,\kappa]}_{\mathrm{cau}}$ and $J^{[i,\kappa]}_{\mathrm{noc}}$. By \Cref{assumption:networked_system}, there exists a constant $\mu_{B_\perp}>0$ such that ${B}_\perp ^{[i,\distance]}$ as $\underline{\sigma}({B}_\perp ^{(i)})\geq\mu_{B_\perp}$. Then we have the following lower bound on $J^{[i,\kappa]}_{\mathrm{cau}}$ and $J^{[i,\kappa]}_{\mathrm{noc}}$: 
\begin{lemma}[]\label{lemma:jacobian}
    Under \Cref{assumption:networked_system}, the Jacobian matrices $J^{[i,\kappa]}_{\mathrm{cau}}$ and $J^{[i,\kappa]}_{\mathrm{noc}}$ satisfy
    \begin{equation}
    \underline{\sigma}(J^{[i,\kappa]}_{\mathrm{cau}}), \underline{\sigma}(J^{[i,\kappa]}_{\mathrm{noc}})\geq\frac{\left((1-\gamma)^2+\left(1 + \mu_{B_{\perp}}^2\right)L^2(1+L)^2\right)^\frac{1}{2}}{L(1+L)},\nonumber
\end{equation}
where $\underline{\sigma}(\cdot)$ denotes the smallest singular value.
\end{lemma}
We defer the proof of \Cref{lemma:jacobian} to \Cref{app:proof_singular_lower_bound}.
As a short remark, when there are no locality constraints, the problem 
reverts to its centralized counterpart, where the lower bound was shown to be 
$\underline{\sigma}(J) \geq (1-\gamma)/\left(L(1+L)\right)$ \cite{shin2023near}.

Next consider the objective Hessian matrices for \eqref{eq:k-ca} and \eqref{eq:k-no}, which can be written as:
\begin{equation}
    \begin{aligned}
        G_{\mathrm{cau}}&=\mathrm{bdiag}\left(Q,R,\dots,Q,R,Q_{T-k}\right), \quad G_{\mathrm{noc}}=\mathrm{bdiag}\left(Q,R,\dots,Q,R,Q_T\right)
    \end{aligned}
\end{equation}
where $\mathrm{bdiag}(\cdot)$ denotes the block diagonal operation.

For the KKT conditions of \eqref{eq:cau_refor} and \eqref{eq:noc_refor}, we define the following sets of matrices and vectors.
For the causal problem (\texttt{Causal(i,k)}), we define the primal-dual block matrices, solutions, and bias vector as:
\begin{align*}
    H_{\mathrm{cau}}^{[i,\distance]} &\coloneqq 
    \begin{bmatrix}
        G_{\mathrm{cau}} & (J_{\mathrm{cau}}^{[i,\distance]})^\top \\
        J_{\mathrm{cau}}^{[i,\distance]} & \mathbf{0}
    \end{bmatrix} \\
    \bm{\delta}_{\mathrm{cau}} &\coloneqq 
    \begin{bmatrix}
        \mathrm{e}_i^\top & \mathbf{0} & \cdots & \mathbf{0}
    \end{bmatrix}^\top \\
    \chi_{\mathrm{cau}} &\coloneqq 
    \begin{bmatrix}
        \bm{\psi}_{\mathrm{cau}} \\ \bm{\eta}_{\mathrm{cau}}
    \end{bmatrix} \\
    \bm{\psi}_{\mathrm{cau}} &\coloneqq 
    \begin{bmatrix}
        (\psi_0^{\mathrm{x}})^\top & (\psi_0^{\mathrm{u}})^\top & \cdots & (\psi_{T-k-1}^{\mathrm{u}})^\top & (\psi_{T-k}^{\mathrm{x}})^\top
    \end{bmatrix}^\top \\
    \bm{\eta}_{\mathrm{cau}} &\coloneqq 
    \begin{bmatrix}
        (\eta_0)^\top & \cdots & (\eta_{T-k})^\top & (\eta'_0)^{\top} & \cdots & (\eta''_{T-k-1})^{\top}
    \end{bmatrix}^\top
\end{align*}
Similarly, for the non-causal problem (\texttt{Noncau(i,k)}), the corresponding terms are defined as:
\begin{align*}
    H_{\mathrm{noc}}^{[i,\distance]} &\coloneqq 
    \begin{bmatrix}
        G_{\mathrm{noc}} & (J_{\mathrm{noc}}^{[i,\distance]})^\top \\
        J_{\mathrm{noc}}^{[i,\distance]} & \mathbf{0}
    \end{bmatrix} \\
    \bm{\delta}_{\mathrm{noc}} &\coloneqq 
    \begin{bmatrix}
        (\delta_{1}^{(k)})^{\top} & \cdots & (\delta_{T}^{(k)})^{\top} & \mathbf{0}^\top & \cdots & \mathbf{0}^\top
    \end{bmatrix}^\top \\
    \widehat{\chi} &\coloneqq 
    \begin{bmatrix}
        \widehat{\bm{\psi}} \\ \widehat{\bm{\eta}}
    \end{bmatrix} \\
    \widehat{\bm{\psi}} &\coloneqq 
    \begin{bmatrix}
        (\widehat{\psi}_0^{\mathrm{x}})^\top & (\widehat{\psi}_0^{\mathrm{u}})^\top & \cdots & (\widehat{\psi}_{T-1}^{\mathrm{u}})^\top & (\widehat{\psi}_T^{\mathrm{x}})^\top
    \end{bmatrix}^\top \\
    \widehat{\bm{\eta}} &\coloneqq 
    \begin{bmatrix}
        (\widehat{\eta}_0)^\top & \cdots & (\widehat{\eta}_{T})^\top & (\widehat{\eta}_0')^{\top} & \cdots & (\widehat{\eta}''_{T-1})^{\top}
    \end{bmatrix}^\top
\end{align*}
where $H, \chi, \bm{\delta}$, $\bm{\psi}$ and $\bm{\mu}$ denote the primal-dual block matrices, primal-dual solutions, bias, primal solutions, and dual solutions, respectively; $\eta$, $\eta'$ and $\eta''$ denote the dual solutions determined by the first, second, and third row matrix constraints in $Z_{i,\kappa}$.

In the future contexts we call $H$ KKT matrices. We now show the upper and lower bounds of singular value of KKT matrices.
\begin{lemma}
    \label{lemma:KKT matrix bound}
    Under \Cref{assumption:networked_system}, the KKT matrices $H_{\mathrm{noc}}^{[i,\distance]}$ and $H_{\mathrm{cau}}^{[i,\distance]}$ satisfy
    \begin{equation}
    \left(\left(1+\frac{3L+1}{\mu_G}\right)^2\frac{L^2(1+L)^2(3L+1)}{(1-\gamma)^2+\mu_{B_\perp}^2L^2(1+L)^2}+\frac{1}{\mu_G}\right)^{-1}\leq\underline{\sigma}(H)\leq\overline{\sigma}(H)\leq\max(3L+1,L_P+1),\nonumber
\end{equation}
where $\mu_G\coloneqq\min(\mu_Q,\mu_R)$; $\underline{\sigma}(\cdot)$ and $\overline{\sigma}(\cdot)$ denote the smallest and greatest singular values.
\end{lemma}
We defer the proof of \Cref{lemma:KKT matrix bound} to \Cref{app:proof_KKT_matrix_bound}.
It now suffices to write the complete KKT conditions of \eqref{eq:k-ca} and \eqref{eq:k-no} as:
    \begin{align}
        H_{\mathrm{noc}}^{[i,\distance]}\widehat{\chi}= \begin{bmatrix}
            \mathbf{0} \\
            \bm{\delta}_{\mathrm{noc}}
        \end{bmatrix}, \quad H_{\mathrm{cau}}^{[i,\distance]}\chi= \begin{bmatrix}
            \mathbf{0} \\
            \bm{\delta}_{\mathrm{cau}}
        \end{bmatrix}.\label{eq: KKT condition}
    \end{align}

\paragraph{Step 3: Spatially decaying property of $H^{-1}$.} In the last step we clarify an upper bound for the inverse sub-matrix $H^{-1}(i,j)$ with respect to $i,j\in[N]$ and some spatial structure $\mathcal{G}$ as:
\begin{lemma}
    [] \label{lemma:inverse matrix}Under Assumption \ref{assumption:networked_system}, the following inequalities hold for the norm of the block sub-matrices of $H_{\mathrm{cau}}^{[i,\distance],-1}$ and $H_{\mathrm{noc}}^{[i,\distance],-1}$ such that
    \begin{equation}
        \begin{aligned}
            \left\|\left(H_{\mathrm{cau}}^{[i,\distance]}\right)^{-1}(i,j)\right\|, \left\|\left(H_{\mathrm{noc}}^{[i,\distance]}\right)^{-1}(i,j)\right\|\leq C_{H}\alpha_{H}^{d_\mathcal{G}(i,j)},\quad\forall i,j\in[N],
        \end{aligned}
    \end{equation}
    where $C_H>0$ and $\alpha_H\in(0,1)$ are given by 
    \begin{equation*}
        \begin{aligned}
           C_H&\coloneqq\frac{L_H}{\mu_H^2\alpha_H},\quad\alpha_H\coloneqq\left(\frac{L_H^2-\mu_H^2}{L_H^2+\mu_H^2}\right)^\frac{1}{2},\quad L_H\coloneqq\max(3L+1,L_P+1),\\\mu_H&\coloneqq\left(\left(1+\frac{3L+1}{\mu_G}\right)^2\cdot\frac{L^2(1+L)^2(3L+1)}{(1-\gamma)^2+\mu_{B_\perp}^2L^2(1+L)^2}+\frac{1}{\mu_{B_\perp}}\right)^{-1}.
        \end{aligned}
    \end{equation*}
\end{lemma}
We defer the proof of \Cref{lemma:inverse matrix} to \Cref{app:proof_inverse_matrix}.
Note that, when the communication distance $\distance$ is  large enough such that no locality constraints are active, the terms $L_H$ and $\mu_H$ become
\begin{equation*}
        \begin{aligned}
         &L_H\coloneqq\max(2L+1,L_P+1),\\&\mu_H\coloneqq\left(\left(1+\frac{2L+1}{\mu_G}\right)^2\cdot\frac{L^2(L+1)^2(2L+1)}{(1-\gamma)^2}+\frac{1}{\mu_G}\right)^{-1},
        \end{aligned}
    \end{equation*}
which has been presented in \cite{shin2023near}.
\subsection{Proof of Theorem \ref{theorem:spatial}}
\begin{proof}
In this proof, we use the distributed and centralized KKT conditions to distinguish between problems with and without locality constraints. We first consider the difference between the centralized and distributed KKT conditions \footnote{
To distinguish from the KKT conditions with locality constraints, we use the notation ${H}_{\mathrm{noc}}^{c}, \widehat{\chi}^{c}, {\bm{\delta}}_{\mathrm{noc}}^{c}$ to denote the corresponding matrices under the centralized settings.} of \eqref{eq:k-no} as 
    \begin{align*}
    &\begin{bmatrix}{H}_{\mathrm{noc}}^{c}&\mathbf{0}\\\mathbf{0}&\mathbf{0}\end{bmatrix}\begin{bmatrix}
            \widehat{\chi}^{\text{c}}\\\mathbf{0}
        \end{bmatrix}-{H}_{\mathrm{noc}}^{[i,\kappa]} \widehat{\chi} = \begin{bmatrix}
           \begin{bmatrix}
               \mathbf{0}\\ {\bm{\delta}}_{\mathrm{noc}}^c
           \end{bmatrix}\\\mathbf{0}
        \end{bmatrix}-\begin{bmatrix}
               \mathbf{0}\\ {\bm{\delta}}_{\mathrm{noc}}
           \end{bmatrix},\\
        \Rightarrow & \begin{bmatrix}
            \widehat{\chi}^{\text{c}}\\\mathbf{0}
        \end{bmatrix} -\widehat{\chi}=\left({H}_{\mathrm{noc}}^{[i,\kappa]}\right)^{-1}\left(\left(\begin{bmatrix}
           \begin{bmatrix}
               \mathbf{0}\\ {\bm{\delta}}_{\mathrm{noc}}^c
           \end{bmatrix}\\\mathbf{0}
        \end{bmatrix}-\begin{bmatrix}
               \mathbf{0}\\ {\bm{\delta}}_{\mathrm{noc}}
           \end{bmatrix}\right)+\left({H}_{\mathrm{noc}}^{[i,\kappa]}-\begin{bmatrix}{H}_{\mathrm{noc}}^{c}&\mathbf{0}\\\mathbf{0}&\mathbf{0}\end{bmatrix}\right)\begin{bmatrix}
            \widehat{\chi}^{\text{c}}\\\mathbf{0}
        \end{bmatrix}\right).
    \end{align*}

By applying the KKT conditions of \eqref{eq:k-ca} and \eqref{eq:k-no} and spatially decaying property of $H^{-1}$ in \Cref{lemma:inverse matrix}, we see the following norm inequality holds as:
    \begin{align}
        &\left\|\left(\begin{bmatrix}
            \widehat{\chi}^{\text{c}}\\\mathbf{0}
        \end{bmatrix} -\widehat{\chi}\right)(\ell)\right\|=\left\|\sum_{j\in[N]}\left({H}_{\mathrm{noc}}^{[i,\kappa]}\right)^{-1}(\ell,j)\cdot\sum_{l\in[N]}\underbrace{\left({H}_{\mathrm{noc}}^{[i,\kappa]}(j,l)-\begin{bmatrix}
        {H}_{\mathrm{noc}}^{c}&\mathbf{0}\\\mathbf{0}&\mathbf{0}\end{bmatrix}(j,l)\right)}_{\coloneqq H_\perp(j,l)}\begin{bmatrix}
            \widehat{\chi}^{\text{c}}\\\mathbf{0}
        \end{bmatrix}(l)\right\|\nonumber
        \\&
        \leq C_H\sum_{j\in\left([N]\backslash \mathcal{N}_{\mathcal{G}}^{\distance}(\ell)\right)}\alpha_H^{d_\mathcal{G}(\ell,j)}\sum_{l\in[N]}\left\|H_\perp(j,l)\left(\sum_{o\in[N]}({H}_{\mathrm{noc}}^{c})^{-1}(l,o)\cdot\begin{bmatrix}
            \mathbf{0}\\\bm{\delta}_{\mathrm{noc}}
        \end{bmatrix}(o)\right)\right\|\nonumber\\
        &\leq C_H(2L+1)\sum_{j\in\left([N]\backslash \mathcal{N}_{\mathcal{G}}^{\distance}(\ell)\right)}\alpha_H^{d_\mathcal{G}(\ell,j)}\sum_{l\in[N]}\left\|({H}_{\mathrm{noc}}^{c})^{-1}(l,\ell)\cdot\begin{bmatrix}
            \mathbf{0}\\\bm{\delta}_{\mathrm{noc}}
        \end{bmatrix}(\ell)\right\|\nonumber\\&\leq C_HC_H^{(\text{c})}(2L+1)\sum_{d=\distance+1}^{\infty}g(d)\left(\frac{\alpha_H}{\vartheta_{H}}\right)^{d}\vartheta_{H}^d\sum_{s=0}^{\infty}g(s)\left(\frac{\alpha_H^{(c)}}{\vartheta_{H}}\right)^{s}\vartheta_{H}^s   \nonumber     \\&\leq\underbrace{\frac{ C_HC_H^{(c)}(2L+1)\vartheta_{H}}{(1-\vartheta_{H})^2}\cdot\sup_{d\in\{1,2,\ldots\mathrm{diam}({\mathcal{G}})\}}\left(g(d)\left(\frac{\alpha_H}{\vartheta_{H}}\right)^{d}\right)\cdot\sup_{s\in\{1,2,\ldots\mathrm{diam}({\mathcal{G}})\}}\left(g(s)\left(\frac{\alpha_H^{(c)}}{\vartheta_{H}}\right)^{s}\right)}_{\coloneqq D_H}\cdot\vartheta_{H}^\distance,\label{eq:primal-dual bound}
    \end{align}
where $\alpha_{\text{S}} \coloneqq (\max(\alpha_H^{\text{c}},\alpha_H)+1)/2 \in (0,1)$. 
The first inequality employs (1) Lemma~\ref{lemma:inverse matrix} for $H^{[i,\kappa]}_{\mathrm{noc}}$ 
and, (2) for any $j \in \mathcal{N}^\kappa_\mathcal{G}(\ell)$, the $j$th entry of 
$H_\perp(j,l)\cdot(\widehat{\chi}^{\top}, \mathbf{0}^{\top})^{\top}(l)$ is zero. 
In the second inequality, we use the facts that $\bm{\delta}_{\mathrm{noc}}(\ell)$ is the 
only nonzero component in $\ell$, and $\|H_\perp\| \leq 2L+1$. 
The third and fourth inequalities employ the sub-exponential expansion of
graph requirement described in \Cref{assumption:subexponential}.

Similarly, it is straightforward to obtain the upper bound of $\left\|
    \left(\left((\chi^c)^\top, \mathbf{0}^\top\right)^\top
-\chi\right)(\ell)\right\|, \ell\in[N]$ for \eqref{eq:k-ca} by the same value $D\vartheta_H^{\distance}$.

Given optimal CLMs $(\pxl_{t,k},\hpxl_{t,k},\pul_{t,k},\hpul_{t,k})$ and $(\phi_{t,k}^{\mathrm{x},i},\widehat{\phi}_{t,k}^{\mathrm{x},i},\phi_{t,k}^{\mathrm{u},i},\widehat{\phi}_{t,k}^{\mathrm{u},i})$ solved by \eqref{eq:k-decomp} with and without the locality constraints respectively, we have the following inequalities:
\begin{equation}
\small
    \begin{aligned}
        &\sum_{t=0}^{T}\left\|\begin{bmatrix}
               \px_{t,k}(j) -\pxl_{t,k}(j) \ \ & \ \ \hpx_{t,k}(j)-\hpxl_{t,k}(j) \\ \pu_{t,k}(j) - \pul_{t,k}(j) \ \ & \ \ \hpu_{t,k}(j)-\hpul_{t,k}(j)
           \end{bmatrix} \right\|^2\\&\leq\sum_{t=0}^{T}\left(\left\|\px_{t,k}(j) -\pxl_{t,k}(j)\right\|^2+\left\|\hpx_{t,k}(j)-\hpxl_{t,k}(j)\right\|^2+\left\|\pu_{t,k}(j) - \pul_{t,k}(j)\right\|^2+\left\|\hpu_{t,k}(j)-\hpul_{t,k}(j)\right\|^2\right)\\
        &=\sum_{t=k}^{T}\left(\left\|\psi^{\mathrm{x}}_{t,k}(j) -\psi^{\mathrm{x}(c)}_{t,k}(j)\right\|^2+\left\|\psi^{\mathrm{u}}_{t,k}(j) -\psi^{\mathrm{u}(c)}_{t,k}(j)\right\|^2\right) + \sum_{t=0}^{k-1}\left(\left\|\widehat{\psi}^{\mathrm{x}}_{t,k}(j) -\widehat{\psi}^{\mathrm{x}(c)}_{t,k}(j)\right\|^2+\left\|\widehat{\psi}^{\mathrm{u}}_{t,k}(j) -\widehat{\psi}^{\mathrm{u}(c)}_{t,k}(j)\right\|^2\right)\\
        & \quad + \sum^{T}_{t=k} \left(\left\|\widehat{\psi}^{\mathrm{x}}_{t,k}(j) -\widehat{\psi}^{\mathrm{x}(c)}_{t,k}(j) - \left(\psi^{\mathrm{x}}_{t,k}(j) -\psi^{\mathrm{x}(c)}_{t,k}(j)\right)\right\|^2+\left\|\widehat{\psi}^{\mathrm{u}}_{t,k}(j) -\widehat{\psi}^{\mathrm{u}(c)}_{t,k}(j) - \left(\psi^{\mathrm{u}}_{t,k}(j) -\psi^{\mathrm{u}(c)}_{t,k}(j)\right)\right\|^2\right)
        \\
        &\leq3\left\|\left(\begin{bmatrix}
            {\chi}^{\text{c}}\\\mathbf{0}
        \end{bmatrix} -{\chi}\right)(j)\right\|^2 + 2\left\|\left(\begin{bmatrix}
            \widehat{\chi}^{\text{c}}\\\mathbf{0}
        \end{bmatrix} -\widehat{\chi}\right)(j)\right\|^2\leq 5D^2_H\vartheta^{2\distance}_H, \nonumber
    \end{aligned}
\end{equation}
where $({\psi}^{\mathrm{x}(c)},{\psi}^{\mathrm{u}(c)},\widehat{\psi}^{\mathrm{x}(c)},\widehat{\psi}^{\mathrm{u}(c)})$ denotes an optimal solution to \eqref{eq:k-ca} and \eqref{eq:k-no} without any locality constraints.
In the first equality, we use the equivalent results in \Cref{proposition:2}; the third inequality employs the upper bound of \eqref{eq:primal-dual bound}.  Writing $D\coloneqq5D_H^2$ and $\vartheta\coloneqq \vartheta^2_H$, the proof is completed. 
\end{proof}

\subsection{Proof of Lemma \ref{lemma:jacobian}}
\label{app:proof_singular_lower_bound}
\begin{proof}
    Given that \eqref{eq:k-ca} and \eqref{eq:k-no} possess an identical 
Jacobian block structure, it is sufficient to prove the bound 
for only one of them. Recalling the definition of the Jacobian matrix 
in \eqref{eq:jacobian_matrix}, we first consider establishing a lower 
bound for the singular value of its sub-matrix \(J_2^{[i,\kappa]}\). 
This derivation is straightforward:
    \begin{equation}
    \small
        \begin{aligned}
            J_2^{[i,\kappa]}J_2^{[i,\kappa]\top}&=\left[\begin{array}{cccc}
                 A^{[i,\kappa]}_\perp A^{[i,\kappa]\top}_\perp+B^{[i,\kappa]}_\perp B^{[i,\kappa]\top}_\perp \\
                 & A^{[i,\kappa]}_\perp A^{[i,\kappa]\top}_\perp+B^{[i,\kappa]}_\perp B^{[i,\kappa]\top}_\perp\\
                 &&\ddots\\
                 &&&A^{[i,\kappa]}_\perp A^{[i,\kappa]\top}_\perp+B^{[i,\kappa]}_\perp B^{[i,\kappa]\top}_\perp
            \end{array}\right]\\&\succeq \underline{\lambda}\left(A^{[i,\kappa]}_\perp A^{[i,\kappa]\top}_\perp+B^{[i,\kappa]}_\perp B^{[i,\kappa]\top}_\perp\right)I\succeq \underline{\sigma}_{B_\perp}^2 I\nonumber
        \end{aligned}
    \end{equation}
    where $\underline{\lambda}(\cdot)$ denotes an operator of the smallest eigenvalue. 
    
    Next, it is straightforward to obtain $J^{[i,\kappa]}_3J^{[i,\kappa]\top}_3= I$.
    We then derive the lower bound of $J_1$. Consider the following column operation 
    \begin{equation}\small
        \begin{aligned}
            &\underbrace{\left[\begin{array}{ccccc}
         I \\
            -{A}^{[i,\distance]} &\quad -{B}^{[i,\distance]} & I \\
             & \ddots &   \ddots& \ddots  \\
             & &   -{A}^{[i,\distance]} & \quad -{B}^{[i,\distance]}& \quad I
    \end{array}\right]}_{J_1^{[i,\distance]}}
        \underbrace{\begin{bmatrix}
            I \quad & \quad \mathbf{0} \\
            {K}  & \quad I \quad & \quad \mathbf{0} \\
             & \ddots & \quad \ddots \quad & \quad \ddots   \\
              &&\quad {K} \quad  & \quad  I  \quad  & \quad \mathbf{0}  \\
             && & \quad  \mathbf{0} \quad  & \quad I \\
        \end{bmatrix}}_{M_1^{-1}}= 
        \underbrace{\left[\begin{array}{ccccc}
             I \\
            -\Bar{K}^{[i,\distance]} & \quad -{B}^{[i,\distance]} & I \\
            & \ddots & \ddots & \ddots \\
            & & -\Bar{K}^{[i,\distance]} & \quad -{B}^{[i,\distance]} & \quad I 
        \end{array}\right]}_{M_2},\nonumber
        \end{aligned}
    \end{equation}
    where blank entries represent zero block matrices and $\Bar{K}^{[i,\distance]}\coloneqq{A}^{[i,\distance]}+{B}^{[i,\distance]}K$ denotes the sparse closed-loop gain and $K$ is the feedback gain solved by Riccati equation with matrices $(A,B,Q,R)$. This indicates the following lower bound
    \begin{equation}
        \begin{aligned}
            J_1^{[i,\distance]}J_1^{[i,\distance]\top}=M_2M_1M_1^\top M_2^\top\succeq \underline{\lambda}\left(M_2M_2^\top\right) M_1M_1^\top\succeq\underline{\lambda}\left(M_3M_3^\top\right) M_1M_1^\top\succeq\frac{I}{\|M_1^{-1}\|^2\|M_3^{-1}\|^2},\nonumber
        \end{aligned}
    \end{equation}
    where the last inequality uses the property $M\succeq\|M^{-1}\|^{-1}I$. Moreover, $M_3$ and its inverse are
    \begin{equation}
        \begin{aligned}
            M_3\coloneqq\left[\begin{array}{ccccc}
             I \\
            -\Bar{K}^{[i,\distance]}  & \quad I \\
            &  \ddots & \\
            & & -\Bar{K}^{[i,\distance]}  & I 
        \end{array}\right], \quad M_3^{-1}\coloneqq\left[\begin{array}{ccccc}
             I \\
            \Bar{K}^{[i,\distance]}  & \quad I \\
            \vdots&  \ddots & \ddots \\
             \left(\Bar{K}^{[i,\distance]}\right)^\top& \quad \cdots& \quad-\Bar{K}^{[i,\distance]}  & \quad I 
        \end{array}\right]. \nonumber
        \end{aligned} 
    \end{equation}
    It is straightforward to verify that $\underline{\lambda}(M_2M_2^\top)\geq\underline{\lambda}(M_3M_3^\top)$. Then we apply the following fact in Lemma A.5 of \cite{shin2023near} to complement our proof: for any block matrix $M\coloneqq[M(i,j)]_{i,j\in[N]}$, we have the following inequality holds,
        \begin{equation}
        \|M\|\leq\left(\max_{i\in[n]}\sum_{j\in[n]}\|M(i,j)\|\right)^{\frac{1}{2}}\left(\max_{j\in[n]}\sum_{i\in[n]}\|M(i,j)\|\right)^{\frac{1}{2}}.\nonumber
        \end{equation}
        
    From the model assumptions and preceding fact, we obtain $\|M_1^{-1}\|\leq L/(1-\gamma)$ and $\|M_3^{-1}\|\leq L+1$. It suffices to derive $J_2^{[i,\kappa]}J_2^{[i,\kappa]\top}\succeq(1-\gamma)^2/\left(L^2(1+L)^2\right)$.

    Given the full row rank condition discussed before, we then consider the definition of smallest singular value of $J^{[i,\kappa]}_{\mathrm{noc}}$, we have
    \begin{equation}
        \begin{aligned}
            \underline{\sigma}(J^{[i,\kappa]}_{\mathrm{noc}})=\inf_{\|\mathrm{v}\|=1}\left\|\begin{bmatrix}
                J_1^{[i,\kappa]} \\ J_2^{[i,\kappa]} \\ J_3^{[i,\kappa]}
            \end{bmatrix}\mathrm{v}\right\|&= \inf_{\|\mathrm{v}\|=1} \sqrt{\left\|
                J_1^{[i,\kappa]}\mathrm{v}\right\|
                ^2 +\left\|
                J_2^{[i,\kappa]}\mathrm{v}\right\|^2+\left\|
                J_3^{[i,\kappa]}\mathrm{v}\right\|^2}\nonumber\\&\geq \sqrt{\underline{\lambda}(J_1^{[i,\kappa]}J_1^{[i,\kappa]^\top}) + \underline{\lambda}(J_2^{[i,\kappa]}J_2^{[i,\kappa]\top})+\underline{\lambda}(J_3^{[i,\kappa]}J_3^{[i,\kappa]\top}}) \\&\
                \geq \frac{\left((1-\gamma)^2+\left(1+\mu_{B_\perp}\right)^2L^2(1+L)^2\right)^{\frac{1}{2}}}{L(1+L)}.\end{aligned}\label{eq:J lower bound}
    \end{equation}
    This finishes the proof of Lemma \ref{lemma:jacobian}.
\end{proof}

\subsection{Proof of Lemma \ref{lemma:KKT matrix bound}}
\label{app:proof_KKT_matrix_bound}
\begin{proof}
\eqref{eq:k-ca} and \eqref{eq:k-no} hold the same block structure of the KKT matrices, so we only prove one of them. For convenience, we will use the simplified symbols \( H \), \( G \), and \( J \) to represent the defined KKT matrix, objective Hessian matrix, and Jacobian matrix, respectively.

We first show the upper bound of $H$. Without loss of generality, let terminal horizon $T=2$. By rearranging the rows and columns of matrix in KKT conditions, we derive a linear system 
\begin{equation}
\small
    \begin{aligned}
        \left[\begin{array}{ccccc|ccccc|cc}
            Q &\mathbf{0}&I&A^{(i)\top}_\perp&\mathbf{0}&\mathbf{0}&\mathbf{0}&-A^{[i,\distance]\top}&\mathbf{0}&\mathbf{0}&\mathbf{0}&\mathbf{0}\\\mathbf{0}
            &R&\mathbf{0}&B^{[i,\distance]\top}_\perp&I^{[i,\distance]\top}_\lnot&\mathbf{0}&\mathbf{0}&-B^{[i,\distance]\top}&\mathbf{0}&\mathbf{0}&\vdots&\vdots\\
I&\mathbf{0}&\mathbf{0}&\mathbf{0}&\mathbf{0}&\vdots&\vdots&\mathbf{0}&\vdots&\vdots&&\\A^{[i,\distance]}_\perp&B^{[i,\distance]}_\perp&\mathbf{0}&\mathbf{0}&\mathbf{0}&&&\vdots&&\\\mathbf{0}&I^{[i,\distance]}_\lnot&\mathbf{0}&\mathbf{0}&\mathbf{0}&&&&&&\\
            \hline \mathbf{0}&\mathbf{0}&\cdots&&&Q &\mathbf{0}&I&A^{[i,\distance]\top}_\lnot&\mathbf{0}&\mathbf{0}&-A^{[i,\distance]\top}\\\mathbf{0}&\mathbf{0}&\cdots&&&\mathbf{0}&R&\mathbf{0}&B^{[i,\distance]\top}_\perp&I^{[i,\distance]\top}_\lnot&\mathbf{0}&-B^{[i,\distance]\top}\\-A^{[i,\distance]}& -B^{[i,\distance]}&\mathbf{0}&\cdots&&I&\mathbf{0}&\mathbf{0}&\mathbf{0}&\mathbf{0}&\mathbf{0}&\mathbf{0}\\\mathbf{0}&\mathbf{0}&\cdots&&&A^{[i,\distance]}_\perp&B^{[i,\distance]}_\perp&\mathbf{0}&\mathbf{0}&\mathbf{0}&\vdots&\vdots\\\mathbf{0}&\mathbf{0}&\cdots&&&\mathbf{0}&I^{[i,\distance]}_\lnot&\mathbf{0}&\mathbf{0}&\mathbf{0}&\\\hline\mathbf{0}&\cdots&&&&\mathbf{0}&\mathbf{0}&\mathbf{0}&\cdots&&P &I\\\mathbf{0}
            &\cdots&&&&-A^{[i,\distance]}& -B^{[i,\distance]}& \mathbf{0} &\cdots&&I&\mathbf{0}
            \end{array}\right]\left[\begin{array}{c}
             \widehat{\psi}^{\mathrm{x}}_0\\
             \widehat{\psi}^{\mathrm{u}}_0\\
             \widehat{\eta}_0\\
             \widehat{\eta'}_0\\
             \widehat{\eta''}_0\\
             \hline
             \widehat{\psi}^{\mathrm{x}}_1\\
             \widehat{\psi}^{\mathrm{u}}_1\\
             \widehat{\eta}_1\\
             \widehat{\eta'}_1\\
             \widehat{\eta''}_1\\\hline
             \widehat{\psi}^{\mathrm{x}}_2\\
             \widehat{\eta}_2
        \end{array}\right]=\left[\begin{array}{c}
             \mathbf{0}  \\
             \mathbf{0}  \\
             \delta_{0}  \\
             \mathbf{0}  \\
             \mathbf{0}  \\\hline
             \mathbf{0}  \\
             \mathbf{0}  \\
             \delta_{1}  \\
             \mathbf{0}  \\
             \mathbf{0}  \\\hline
             \mathbf{0}  \\
             \delta_{2}  
        \end{array}\right],\nonumber
    \end{aligned}
\end{equation}
where $P$ denotes the terminal matrix. Based on this linear system, we see that  $\overline{\sigma}(H)\leq\max(3L+1,L_P+1)$.

To prove the lower bound of $H$,  we use the Schur complement to write down the inverse of $H$:
\begin{equation}
    \begin{aligned}
        H^{-1} = \begin{bmatrix}
            G & J^\top\\J 
        \end{bmatrix}^{-1} = \begin{bmatrix}
            G^{-1} - G^{-1}J^\top(G_J)^{-1}J G^{-1} & G^{-1}J^\top(G_J)^{-1} \\ (G_J)^{-1}JG^{-1}  & -(G_J)^{-1}\nonumber
        \end{bmatrix},    \end{aligned}
\end{equation}
where $G_J \coloneqq JG^{-1}J^{\top}$. We then bound the operator norm $\|H^{-1}\|$ from above such that
\begin{equation}
    \begin{aligned}
        \|H^{-1}\| &\leq  \left\|G^{-1} - G^{-1}J^\top(G_J)^{-1}J G^{-1}\right\|  + 2 \|G^{-1}J^\top(G_J)^{-1}\| +\|(G_J)^{-1}\|  \\&\leq\left\|G^{-1}\right\| + \|(G_J)^{-1}\|\left(1+\|JG^{-1}\|\right)^2,          
        \end{aligned}\label{H}
\end{equation}
where $\|G^{-1}\|\leq {1}/{\min(\mu_Q,\mu_R)}$ and $\|JG^{-1}\|\leq {\overline{\sigma}(H)}/ {\min(\mu_Q,\mu_R)}$. In addition,
\begin{equation}
    \begin{aligned}
        \|(G_J)^{-1}\| &\leq\frac{1}{\underline{\lambda}\left(JG^{-1}J^\top\right)}
        \leq\frac{\|G\|}{\underline{\lambda}\left(JJ^\top\right)}\leq \frac{\overline{\sigma}(H)}{\underline{\sigma}^2(J)}
        \end{aligned}
\end{equation}
where the second inequality uses the property such that $M'MM'^\top\succeq\underline{\lambda}(M)M'M'^\top$. By combining all of these, we obtain the desired lower bound on the singular values of $H$. 
\end{proof}

\subsection{Proof of \Cref{lemma:inverse matrix}} \label{app:proof_inverse_matrix} Before presenting the proof, we require the following lemma for an inverse graph-induced matrix:
\begin{lemma}[Bound of inverse graph-induced matrix]

\label{lemma: graph matrix}
    Consider a non-singular matrix ${M}\in \mathbb{R}^{n\times n}$ satisfying $M_{ij}=0$ if the  distance $d_{\mathcal{G}}(i,j)$ is not greater than $d_M\geq 1$ induced by a graph $\mathcal{G}=(\mathcal{V},\mathcal{E})$. With singular value bounds $0<{\mu}_M\leq \underline{\sigma}(M)\leq\overline{\sigma}(M)\leq L_M$, the following holds
\begin{equation*}
    \begin{aligned}
        \left\|M^{-1}(i,j)\right\|\leq\frac{L_M}{\mu_M^2}\left(\frac{L_M^2 - \mu_M^2}{L_M^2 + \mu_M^2}\right)^{\left\lceil \frac{d_{\mathcal{G}}(i,j)-d_M}{2 d_M}\right\rceil_+}, \quad i,j\in\mathcal{V}
    \end{aligned}
\end{equation*}
where $\lceil\cdot\rceil_+$ denotes the smallest non-negative integer that is greater than or equal to the argument.
\end{lemma}
\begin{proof}[Proof of \Cref{lemma: graph matrix}]
    By definition of singular values and the assumptions on $\mu_M, L_M$, we have $\mu_M^2 I\preceq MM^\top \preceq L_M^2 I$. It suffices to derive
\begin{equation*}
    \begin{aligned}
        \frac{\mu_M^2 - L_M^2}{\mu_M^2 + L_M^2} I \preceq I - \frac{2}{\mu_M^2 + L_M^2} MM^\top \preceq \frac{-\mu_M^2 + L_M^2}{\mu_M^2 + L_M^2} I.
    \end{aligned}
\end{equation*}
To show this, we apply the following lemma:
\begin{lemma}
    [Basic Properties of Graph-induced Matrix, Lemma 3.5 in \cite{shin2022exponential}] \label{lemma:graph-induced} Consider $M\in \mathbb{R}^{m\times n}$ with distance not greater than $d_M$ induced by a topology set $(\mathcal{G}, \mathcal{I}, \mathcal{J})$ where $\mathcal{I}$  and $\mathcal{J}$ denote the subset of $\mathcal{V}$; we have that: 
    \begin{enumerate}
        \item $M^\top$ has distance not greater than $d_M$ induced by $(\mathcal{G}, \mathcal{J}, \mathcal{I})$;
        \item if $N\in \mathbb{R}^{m\times n}$ has distance not greater than BY induced by $(\mathcal{G}, \mathcal{I}, \mathcal{J})$, then $M+N$ has distance not greater than $\max(d_M, d_N)$ induced by $(\mathcal{G}, \mathcal{I}, \mathcal{J})$;
        \item if $N\in \mathbb{R}^{n\times k}$ has distance not greater than $d_M$ induced by $(\mathcal{G}, \mathcal{J}, \mathcal{K})$, then $MN$ has distance not greater than $d_M+d_N$ induced by $(\mathcal{G}, \mathcal{I}, \mathcal{K})$.
    \end{enumerate}
\end{lemma} 

Since Lemma \ref{lemma:graph-induced} illustrates $MM^\top$ has distance not greater than $2d_M$ induced by $(\mathcal{G}, \mathcal{I}, \mathcal{I})$, we have
\begin{equation*}
    \begin{aligned}
        M^{-1} &= \frac{2}{\mu_M^2 + L_M^2} M^\top \left(\frac{2}{\mu_M^2 + L_M^2} MM^\top\right)^{-1}\\
         & = \frac{2}{\mu_M^2 + L_M^2} \sum^{\infty}_{q = 0}M^\top\left(I - \frac{2}{\mu_M^2 + L_M^2} MM^\top\right)^q .
    \end{aligned}
\end{equation*}
Again, from Lemma \ref{lemma:graph-induced}, we see $M\left(I - \frac{2}{\mu_M^2 + L_M^2} MM^\top\right)^q$ has distance  not greater than $(2q + 1)d_M$ induced by $(\mathcal{G}, \mathcal{I}, \mathcal{J})$. By the extraction operation $(\cdot)(i,j)$, we obtain
\begin{equation*}
    \begin{aligned}
        M^{-1}(i,j) 
         & = \frac{2}{\mu_M^2 + L_M^2} \sum^{\infty}_{q = q_0(i,j)}\left(M^\top\left(I - \frac{2}{\mu_M^2 + L_M^2} MM^\top\right)^q\right) (i,j),
    \end{aligned}
\end{equation*}
where the range of summation starts from $q_0(i,j)$ instead of $0$ since $q_0(i,j) \coloneqq \lceil \frac{d_{\mathcal{G}}(i,j)-d_M}{2d_M}\rceil_+$, and when $q = 0, \ldots , q_0(i, j) - 1$, the $ij$th entry inside the summation is zero because such $q$ satisfies $(2q + 1)d_M < d_{\mathcal{G}}(i,j)$. Applying the triangle inequality, we obtain the following bound:
\begin{equation*}
    \begin{aligned}
        \|M^{-1}(i,j)\| &\leq \frac{2}{\mu_M^2 + L_M^2} \sum^{\infty}_{q = q_0(i,j)}\left\|M\left(I - \frac{2}{\mu_M^2 + L_M^2} MM^\top\right)^q\right\|
        \\&\leq \frac{2}{\mu_M^2 + L_M^2} \sum^{\infty}_{q = q_0(i,j)} L_M\left(\frac{-\mu_M^2 + L_M^2}{\mu_M^2 + L_M^2}\right)^q\\
        & \leq \frac{L_M}{\mu^2_M} \left(\frac{-\mu_M^2 + L_M^2}{\mu_M^2 + L_M^2}\right)^{\lceil \frac{d_{\mathcal{G}}(i,j)-d_M}{2d_M}\rceil_+}. 
    \end{aligned}
\end{equation*}
\end{proof}

\begin{proof}
    Note that  \Cref{lemma: graph matrix} presents a more general result, of which Lemma \ref{lemma:inverse matrix} is a specific instance. Consequently, using the arguments of \Cref{lemma:KKT matrix bound} and \Cref{lemma: graph matrix},  Lemma \ref{lemma:inverse matrix} can be directly derived.
\end{proof}

\section{Proof of \Cref{theorem:regret}}
\label{appendix:regret}
In this section, we provide a complete proof of \Cref{theorem:regret}.

\subsection{Proof Outline}

 Without loss of generality, we consider scalar subsystems, i.e., $n_i=m_i=1$, and therefore $n=m=N$. We use $\PredSLS, \PredSLSE$, and $\OPT$ to denote the policy generated by \eqref{eq:k-decomp} with imperfect predictions, the policy by \eqref{eq:k-decomp} with exact predictions, and the offline optimal policy \eqref{eq:optimal policy}, respectively. 

Specifically, our analysis involves bounding the discrepancies between sub-states (and similarly sub-actions) induced by $\PredSLS$ and $\OPT$, i.e., $|\mathrm{x}_t^i(\PredSLS)-\mathrm{x}_t^i(\OPT)|$ and $|\mathrm{u}_t^i(\PredSLS)-\mathrm{u}_t^i(\OPT)|$. We note that, according to \Cref{proposition:1}, $\OPT$ is exactly $\PredSLS$ with perfect predictions and without any locality constraints.

\begin{lemma}
    [Subsystem discrepancy bound] \label{lemma: discrepancy bound} Let  Assumption~\ref{assumption:subexponential} and~\ref{assumption:networked_system} hold. 
    The sub-state and sub-action discrepancies for $\PredSLS$ with respect to the offline $\OPT$ are bounded by
    \begin{align}
        &\left|\mathrm{x}_t^i(\PredSLS)-\mathrm{x}_t^i(\OPT)\right|\nonumber\leq Cp(\distance) {\mathrm{err}}(\widehat{\mathbf{w}},\mathbf{w}) +Wp(\distance){\mathrm{loc}}(\bm{\varphi}^{\mathrm{x}})+W\left(p(\mathrm{diam}({\mathcal{G}}))-p(\distance)\right){\mathrm{con}},\nonumber\\
        &\left|\mathrm{u}_t^i(\PredSLS)-\mathrm{u}_t^i(\OPT)\right|\nonumber\leq Cp(\distance)  {\mathrm{err}}(\widehat{\mathbf{w}},\mathbf{w}) +Wp(\distance){\mathrm{loc}}(\bm{\varphi}^{\mathrm{u}})+W\left(p(\mathrm{diam}({\mathcal{G}}))-p(\distance)\right){\mathrm{con}}.\nonumber
    \end{align}
    Here, $p(\distance)$, ${\mathrm{err}}$, ${\mathrm{loc}}$, and ${\mathrm{con}}$ denote functions determined by the communication topology, prediction error, locality difference, respectively. Specifically, they are given by
    \begin{equation*}
        \begin{aligned}
        p(\distance) &\coloneqq (\distance+1)\sup_{d\in\{1,2,\dots,\kappa\}}g(d),\\ 
        {\mathrm{err}}(\widehat{\mathbf{w}},\mathbf{w}) &\coloneqq \sup_{j\in\mathcal{N}^{\distance}_{\mathcal{G}}(i)}\sum^{T}_{k=0}\rho^{|t-k|}\left|\widehat{\mathrm{w}}^j_{k-1}-\mathrm{w}^j_{k-1}\right|,\\
        {\mathrm{loc}}(\bm{\varphi}^{(\cdot)}) &\coloneqq\sup_{j\in\mathcal{N}^{\distance}_{\mathcal{G}}(i)}\left(\sum^t_{k=0}\left|{\varphi}^{(\cdot),j}_{t,k}(i)- {\phi}^{(\cdot),j}_{t,k}(i)\right|+\sum^T_{k=0}\left|\widehat{\varphi}^{(\cdot),j}_{t,k}(i)- \widehat{\phi}^{(\cdot),j}_{t,k}(i)\right|\right),\\    
        {\mathrm{con}}&\coloneqq \frac{C(2+\rho)}{1-\rho},
    \end{aligned}
    \end{equation*}
     where $(\cdot)$ is a shorthand for $\mathrm{x}$ and $\mathrm{u}$;  $\left(\bm{\varphi}^{\mathrm{x},j},\widehat{\bm{\varphi}}^{\mathrm{x},j},\bm{\varphi}^{\mathrm{u},j},\widehat{\bm{\varphi}}^{\mathrm{u},j}\right)_{j\in\mathcal{N}^{\distance}_{\mathcal{G}}(i)}$ are the distributed CLMs from \eqref{eq:k-decomp}, and $\left(\bm{\phi}^{\mathrm{x},j},\widehat{\bm{\phi}}^{\mathrm{x},j},\bm{\phi}^{\mathrm{u},j},\widehat{\bm{\phi}}^{\mathrm{u},j}\right)_{j\in[N]}$ are the distributed CLMs from the offline optimal controller.
\end{lemma}

We defer the proof of \Cref{lemma: discrepancy bound} to Appendix~\ref{appendix:discrepancy bound}. Now, we are ready to prove the performance bound in Theorem \ref{theorem:regret}. 

\begin{proof}
    By \Cref{lemma: discrepancy bound}, we see that
    \begin{align}
        \nonumber&\sum^{T-1}_{t=0}|\mathrm{x}_t^i(\PredSLS)-\mathrm{x}_t^i(\OPT)|^2\\
        &\leq \sum^{T-1}_{t=0}\left(Cp(\distance) \cdot {\mathrm{err}}(\widehat{\mathbf{w}},\mathbf{w}) +Wp(\distance)\cdot{\mathrm{loc}}(\bm{\varphi}^{\mathrm{x}})+\left(p(\mathrm{diam}({\mathcal{G}}))-p(\distance)\right)\cdot{\mathrm{con}}\right)^2\nonumber
        \\
        &\leq 3C^2\left(p(\distance)\right)^2 \cdot \sum^{T-1}_{t=0}{\mathrm{err}}^2(\widehat{\mathbf{w}},\mathbf{w}) + 3W^2\left(p(\distance)\right)^2\cdot \sum^{T-1}_{t=0}\mathrm{loc}^2(\bm{\varphi}^\mathrm{x}) + 3T\left(p(\mathrm{diam}({\mathcal{G}})) - p(\distance)\right)^2\cdot \mathrm{con}^2\label{eq:proof_prop2_bound2}
    \end{align}
   where the first inequality is obtained by applying the subsystem discrepancy bound in~\Cref{lemma: discrepancy bound}, we have used the Cauchy-Schwarz inequality to derive the second inequality in~\eqref{eq:proof_prop2_bound2}. In other words, the RHS of~\eqref{eq:proof_prop2_bound2} is
     \begin{align}
        & 3C^2\left(p(\distance)\right)^2 \underbrace{\sum^{T-1}_{t=0}\left(\sup_{j\in\mathcal{N}^{\distance}_{\mathcal{G}}(i)}\left(\sum^{T}_{k=0}\rho^{|t-k|}\cdot\left|\widehat{\mathrm{w}}^j_{k-1}-\mathrm{w}^j_{k-1}\right|\right)\right)^2}_{\text{(a)}}\nonumber\\&\qquad+3W^2\left(p(\distance)\right)^2 \underbrace{\sum^{T-1}_{t=0}\left(\sup_{j\in\mathcal{N}^{\distance}_{\mathcal{G}}(i)}\left(
        \sum^t_{k=0}\left|{\varphi}^{\mathrm{x},j}_{t,k}(i)- {\phi}^{\mathrm{x},j}_{t,k}(i)\right|+\sum^T_{k=0}\left|\widehat{\varphi}^{\mathrm{x},j}_{t,k}(i)- \widehat{\phi}^{\mathrm{x},j}_{t,k}(i)\right|\right)\right)^2}_{\text{(b)}}\nonumber\\&\qquad+3TW^2\left(p(\mathrm{diam}({\mathcal{G}})) - p(\distance)\right)^2\cdot\frac{C^2(2+\rho)^2}{(1-\rho)^2}, \label{eq:bound 0}
    \end{align}
        
We now consider (a) and (b) separately. For term (a), let $j_t^{\max \text{(a)}}\in\mathcal{N}^{\kappa}_{\mathcal{G}}(i)$ denote the subsystem index that makes the supremum in (a) hold. Applying the Cauchy-Schwarz inequality again, we obtain
    \begin{align}
        \nonumber\text{(a)}&=\sum^{T-1}_{t=0}\left(\sum^{T}_{k=0}\rho^{|t-k|}\left|\widehat{\mathrm{w}}^{j^{\max}_t}_{k-1}-\mathrm{w}^{j^{\max}_t}_{k-1}\right|\right)\left(\sum^{T}_{k=0}\rho^{|t-k|}\left|\widehat{\mathrm{w}}^{j^{\max}_t}_{k-1}-\mathrm{w}^{j^{\max}_t}_{k-1}\right|\right)\nonumber\\
        &\leq\sum^{T-1}_{t=0}\left(\sum^{T}_{k=0}\rho^{|t-k|}\right)\left(\sum^{T}_{k=0}\rho^{|t-k|}\left|\widehat{\mathrm{w}}^{j^{\max}_t}_{k-1}-\mathrm{w}^{j^{\max}_t}_{k-1}\right|^2\right) \nonumber
        \\
        &\leq\sum^{T-1}_{t=0}\left(\sum^{T}_{k=0}\rho^{|t-k|}\right)\left(\sup_{j\in\mathcal{N}^{\distance}_{\mathcal{G}}(i)}\sum^{T}_{k=0}\rho^{|t-k|}\left|\widehat{\mathrm{w}}^j_{k-1}-\mathrm{w}^j_{k-1}\right|^2\right) \nonumber
        \\&\leq \frac{1+\rho}{1-\rho}\cdot\left(\sum^{T-1}_{t=0}\sup_{j\in\mathcal{N}^{\distance}_{\mathcal{G}}(i)}\sum^{T}_{k=0}\rho^{|t-k|}\left|\widehat{\mathrm{w}}^j_{k-1}-\mathrm{w}^j_{k-1}\right|^2\right)\nonumber
        \\&\leq \left(\frac{1+\rho}{1-\rho}\right)^2T\left(\sup_{j\in\mathcal{N}^{\distance}_{\mathcal{G}}(i)}\sum^{T}_{k=0}\left|\widehat{\mathrm{w}}^j_{k-1}-\mathrm{w}^j_{k-1}\right|^2\right)= \left(\frac{1+\rho}{1-\rho}\right)^2T\Bar{\bm{\varepsilon}}.\label{eq:bound 1}
    \end{align} For (b), we have
    \begin{align}\small
        \nonumber\text{(b)}&=\sum^{T-1}_{t=0}\left(
        \sum^t_{k=0}\left|{\varphi}^{\mathrm{x},{j^{\max}_t}}_{t,k}(i)- {\phi}^{\mathrm{x},{j^{\max}_t}}_{t,k}(i)\right|+\sum^T_{k=0}\left|\widehat{\varphi}^{\mathrm{x},{j^{\max}_t}}_{t,k}(i)- \widehat{\phi}^{\mathrm{x},{j^{\max}_t}}_{t,k}(i)\right|\right)^2
        \\&\leq2\sum^{T-1}_{t=0}\left(
        \sum^t_{k=0}\left|{\varphi}^{\mathrm{x},{j^{\max}_t}}_{t,k}(i)- {\phi}^{\mathrm{x},{j^{\max}_t}}_{t,k}(i)\right|\right)^2+2\sum^{T-1}_{t=0}\left(\sum^T_{k=0}\left|\widehat{\varphi}^{\mathrm{x},{j^{\max}_t}}_{t,k}(i)- \widehat{\phi}^{\mathrm{x},{j^{\max}_t}}_{t,k}(i)\right|\right)^2\nonumber\\
        &\leq4\sum^{T-1}_{t=0}\left(\sum^t_{k=0}(CD)^{\frac{1}{4}}\max(\rho^{\frac{1}{4}},\vartheta^{\frac{1}{4}})^{|t-k|+\distance}\right)^2+4\sum^{T-1}_{t=0}\left(\sum^T_{k=0}(CD)^{\frac{1}{4}}\max(\rho^{\frac{1}{4}},\vartheta^{\frac{1}{4}})^{|t-k|+\distance}\right)^2\nonumber\\
        &\leq 4(CD)^{\frac{1}{2}}\left(\frac{1}{(1-\max(\rho^{\frac{1}{4}},\vartheta^{\frac{1}{4}}))^2}  + \left(\sum^{t}_{k=0}\max(\rho^{\frac{1}{4}},\vartheta^{\frac{1}{4}})^{|t-k|} + \sum^{T}_{k=t+1}\max(\rho^{\frac{1}{4}},\vartheta^{\frac{1}{4}})^{|t-k|}\right)^2\right)T\max(\rho^{\frac{1}{2}},\vartheta^{\frac{1}{2}})^{\distance}\nonumber \\ 
        &\leq\frac{4(CD)^{\frac{1}{2}}(1+(1+\max(\rho^{\frac{1}{4}},\vartheta^{\frac{1}{4}}))^2)T}{(1-\max(\rho^{\frac{1}{4}},\vartheta^{\frac{1}{4}}))^2}\cdot\max(\rho^{\frac{1}{2}},\vartheta^{\frac{1}{2}})^{\distance},\label{eq:bound 2}
    \end{align}
    where the second inequality is due to the following corollary:
    \begin{corollary}[Temporally-spatially decaying property]\label{lemma:temporal-spatial decay}
    Let the distributed closed-loop mappings $(\pxl_{t,k},\hpxl_{t,k},\pul_{t,k},\hpul_{t,k})$ and $(\px_{t,k},\hpx_{t,k},\pu_{t,k},\hpu_{t,k})$ denote optimal solutions to the problem \eqref{eq:k-decomp} corresponding to the $\kappa$-localized and centralized settings, respectively. Suppose the system $(A,B)$ is stabilizable. Under Assumption~\ref{assumption:subexponential}, 
    \begin{align}
        \left\|\begin{bmatrix}
               \px_{t,k}(j) -\pxl_{t,k}(j) & \hpx_{t,k}(j)-\hpxl_{t,k}(j) \\ \pu_{t,k}(j) - \pul_{t,k}(j) & \hpu_{t,k}(j)-\hpul_{t,k}(j)
           \end{bmatrix} \right\|\leq D_1\vartheta_1^{|t-k|+\distance}, \,\,\forall j\in[N]\nonumber
    \end{align}
    where $D_1\coloneqq\sqrt{2}(CD)^{1/4}$ and $\vartheta_1\coloneqq\max(\rho^{1/4},\vartheta^{1/4})$ with $\rho$ and $\vartheta$ defined in \Cref{theorem:temporal} and~\Cref{theorem:spatial}.
\end{corollary}
We defer the proof of \Cref{lemma:temporal-spatial decay} to Appendix \ref{proof:temporal-spatial_decay}. Putting \eqref{eq:bound 1} and \eqref{eq:bound 2} into \eqref{eq:bound 0}, we have
    \begin{align}\label{eq:summation_bound}
        \sum^{T-1}_{t=0}|\mathrm{x}_t^i(\PredSLS)-\mathrm{x}_t^i(\OPT)|^2\leq& 3C^2p^2(\distance)\cdot\eqref{eq:bound 1}+3W^2p^2(\distance)
        \cdot\eqref{eq:bound 2} \nonumber\\&+3TW^2\left(p(\mathrm{diam}({\mathcal{G}})) - p(\distance)\right)^2\cdot\frac{C^2(2+\rho)^2}{(1-\rho)^2}.
    \end{align}
    The action discrepancy $\sum^{T-1}_{t=0}|\mathrm{u}_t^i(\PredSLS)-\mathrm{u}_t^i(\OPT)|^2$ can be bounded using the same argument. Let $\mathrm{u}_t(\PredSLSE)$ and $\mathrm{x}_t(\PredSLSE)$ be the corresponding control action and state induced by $\PredSLSE$.
    To complete the proof of Theorem~\ref{theorem:regret}, Lemma \ref{lemma:regret bound 1} below connects the total cost with state and action discrepancies:
    \begin{lemma}
\label{lemma:regret bound 1}
Given the stabilizable system $A,B$ and cost matrices $Q,R$, the difference between the optimal cost  and the cost of $\PredSLS$ satisfies
    \begin{equation*}
    \begin{aligned} 
         &\cost(\PredSLS) - \cost(\OPT)
         = \sum^{T-1}_{t=0}\left\|V\left(\mathrm{u}_t(\PredSLS)-\mathrm{u}_t(\OPT)  \right) + B^\top PA\left(\mathrm{x}_t(\PredSLS)-\mathrm{x}_t(\OPT)  \right)\right\|_{V^{-1}}^2 ,
  \end{aligned}
\end{equation*}
where $V\coloneqq R+B^\top PB$; $\|\cdot\|^2_{V^{-1}}\coloneqq (\cdot)^{\top}V^{-1}(\cdot)$ denotes the generalized $\ell_2$-norm as a quadratic form and $P$ is the solution of DARE.
\end{lemma}
We defer the proof of \Cref{lemma:regret bound 1} to Appendix \ref{appendix:lemma_regret_bound_1}.
Lemma \ref{lemma:regret bound 1} directly implies that
\begin{align}
    &\cost(\PredSLS) - \cost(\OPT)\nonumber\\&\leq2\sum^{T-1}_{t=0}\left(\|\mathrm{u}_t(\PredSLS)-\mathrm{u}_t(\OPT)\|_V^2  \nonumber + \|\mathrm{x}_t(\PredSLS)-\mathrm{x}_t(\OPT)\|  _{A^\top PBV^{-1}B^\top PA}^2\right)\nonumber\\
    &\leq2\|V\|\sum^{T-1}_{t=0}\|\mathrm{u}_t(\PredSLS)-\mathrm{u}_t(\OPT)\|^2 + 2\|A\|^2\|B\|^2\|P\|^2\|V^{-1}\|\sum^{T-1}_{t=0}\|\mathrm{x}_t(\PredSLS)-\mathrm{x}_t(\OPT) \|^2\nonumber\\
    &\leq 2L(1+LL_P)\sum^{T-1}_{t=0}\sum_{i\in[N]}|\mathrm{u}_t^i(\PredSLS)-\mathrm{u}_t^i(\OPT)|^2  + 2L^4L_P^2\mu_R^{-1}\sum^{T-1}_{t=0}\sum_{i\in[N]}|\mathrm{x}_t^i(\PredSLS)-\mathrm{x}_t^i(\OPT) |^2, \nonumber\\
    &\leq \sum_{i\in[N]}\left(2L(1+LL_P)\sum^{T-1}_{t=0}|\mathrm{u}_t^i(\PredSLS)-\mathrm{u}_t^i(\OPT)|^2 + 2L^4L_P^2\mu_R^{-1}\sum^{T-1}_{t=0}|\mathrm{x}_t^i(\PredSLS)-\mathrm{x}_t^i(\OPT) |^2\right), \nonumber
\end{align}
where the third inequality uses the fact that $\mu_RI\preceq V \preceq  L(1+LL_P)I$. Substituting the sub-state and sub-action discrepancy bounds in \eqref{eq:summation_bound} into the preceding inequality, we obtain
\begin{align}
    \cost(\PredSLS) - \cost(\OPT)&\leq6TNL(1+LL_P+L^3L_P^2\mu_R^{-1})\left(p^2(\distance)\left(\frac{C(1+\rho)}{1-\rho}\right)^2\overline{\bm{\varepsilon}} \right.\nonumber\\ &\qquad+4p^2(\distance)\frac{W^2(CD)^{\frac{1}{2}}(1+(1+\max(\rho^{\frac{1}{4}},\vartheta^{\frac{1}{4}}))^2)}{(1-\max(\rho^{\frac{1}{4}},\vartheta^{\frac{1}{4}}))^2}\cdot\max(\rho^{\frac{1}{2}},\vartheta^{\frac{1}{2}})^{\distance}\nonumber\\&\qquad + \left(p(\mathrm{diam}({\mathcal{G}})) - p(\distance)\right)^2\cdot\frac{C^2(2+\rho)^2W^2}{(1-\rho)^2}\Bigg)\nonumber \\
    &= \mathcal{O}\left(p^2(\distance)\left(C_{1}\bm{{\varepsilon}}+C_{2} W^2\rho_0^{\distance}\right)+\left(p(\mathrm{diam}({\mathcal{G}})) - p(\distance)\right)^2C_{3}W^2\right).\nonumber
\end{align}
This completes the proof of \Cref{theorem:regret}.
\end{proof}

\subsection{Proof of Lemma \ref{lemma: discrepancy bound} \label{appendix:discrepancy bound}}
\begin{proof}

Unlike truncation and Ricatti-based controllers \cite{shin2023near,zhang2023optimal}, the predictive SLS closed-loop mappings \eqref{closedloop responses} allow for agent-wise control implementation, therefore this enables us to focus on the closed-loop behavior of each agent. This structural property implies that the performance of each agent directly depends on its local information structure. Consequently, it provides a clear pathway to bound the dynamic regret defined in~\eqref{eq:dynamic_regret} considering the impact of communication topology. 

We refer to the $i$th distributed version of the controller implementation \eqref{eq:policy}. Define $\PredSLSE$ as a fictitious version of $\PredSLS$ that is subject to the locality constraints \eqref{eq:locality constraint} but uses the true predictions. Let $\widetilde{w}_t^i(\PredSLS)$, $\widetilde{w}_t^i(\PredSLSE)$ and $\widehat{w}_t^i(\OPT)$ be the $i$th subsystem's internal disturbances implemented by $\PredSLS$, $\PredSLSE$ and $\OPT$ respectively. And by defining $\phi(i)$ as $i$th element of the optimal solution to the problem \eqref{eq:k-decomp} by removing \eqref{eq:locality constraint}, it follows:
    \begin{align}
        |\mathrm{x}_t^i(\PredSLS) - \mathrm{x}^i_t(\OPT)|&=|\mathrm{x}_t^i(\PredSLS) - \mathrm{x}^i_t(\PredSLSE)+\mathrm{x}_t^i(cf) - \mathrm{x}^i_t(\OPT)|\nonumber\\
        &\leq \underbrace{\left|\sum_{j\in\mathcal{N}^{\distance}_{\mathcal{G}}(i)}\sum^t_{k=0}\varphi^{\mathrm{x},j}_{t,k}(i)\cdot\left(\widetilde{\mathrm{w}}^j_{k}(\PredSLS) - \widetilde{\mathrm{w}}^j_{k}(\PredSLSE)\right)\right|}_{\text{(a)}}\nonumber\\
        &\qquad+\underbrace{\left|\sum_{j\in\mathcal{N}^{\distance}_{\mathcal{G}}(i)}\sum^t_{k=0}\left(\varphi^{\mathrm{x},j}_{t,k}(i)\cdot\widetilde{\mathrm{w}}^j_{k}(\PredSLSE) - \phi^{\mathrm{x},j}_{t,k}(i)\cdot\widetilde{\mathrm{w}}^j_{k}(\OPT)\right)\right|}_{\text{(b)}}\nonumber\\
        &\qquad+\underbrace{\left|\sum_{j\in\left([N]\backslash\mathcal{N}^{\distance}_{\mathcal{G}}(i)\right)}\sum^t_{k=0} \phi^{\mathrm{x},j}_{t,k}(i)\cdot\widetilde{\mathrm{w}}^j_{k}(\OPT)\right|}_{\text{(c)}}.\nonumber
        \end{align}
        where the first inequality use the distributed implementation in Appendix~\ref{sec:distributed implementation}. We now consider (a), (b) and (c) separately. Applying \Cref{theorem:spatial} and \Cref{assumption:subexponential},
        \begin{align}
        \text{(a)} &\leq\sum_{j\in\mathcal{N}^{\distance}_{\mathcal{G}}(i)}\left|\sum^T_{k=0}\widehat{\varphi}^{\mathrm{x},j}_{t,k}(i)\cdot\left(\widehat{\mathrm{w}}^j_{k-1}-\mathrm{w}^j_{k-1}\right)\right|\nonumber
        \\&\leq C\sum_{j\in\mathcal{N}^{\distance}_{\mathcal{G}}(i)}\sum^{T}_{k=0}\left|\widehat{\varphi}^{\mathrm{x},j}_{t,k}(i)\cdot\left(\widehat{\mathrm{w}}^j_{k-1}-\mathrm{w}^j_{k-1}\right)\right|\nonumber\\&\leq C\sum_{j\in\mathcal{N}^{\distance}_{\mathcal{G}}(i)}\sum^{T}_{k=0}\rho^{|t-k|}\left|\widehat{\mathrm{w}}^j_{k-1}-\mathrm{w}^j_{k-1}\right|\nonumber\\&\leq C(\distance+1)\sup_{d\in\{1,2,\dots,\kappa\}}g(d)\left(\sup_{j\in\mathcal{N}^{\distance}_{\mathcal{G}}(i)}\sum^{T}_{k=0}\rho^{|t-k|}\left|\widehat{\mathrm{w}}^j_{k-1}-\mathrm{w}^j_{k-1}\right|\right).\label{eq:(a)}\end{align}
    
        For the term (b), we have
    \begin{align}
       \text{(b)} &\leq \sum_{j\in\mathcal{N}^{\distance}_{\mathcal{G}}(i)}\left|\sum^t_{k=0}\left({\varphi}^{\mathrm{x},j}_{t,k}(i)- {\phi}^{\mathrm{x},j}_{t,k}(i)\right){\mathrm{w}}^j_{k-1}+\sum^T_{k=0}\left(\widehat{\varphi}^{\mathrm{x},j}_{t,k}(i)- \widehat{\phi}^{\mathrm{x},j}_{t,k}(i)\right){\mathrm{w}}^j_{k-1}\right|\nonumber\\&\leq W\sum_{j\in\mathcal{N}^{\distance}_{\mathcal{G}}(i)}\left(
        \sum^t_{k=0}\left|{\varphi}^{\mathrm{x},j}_{t,k}(i)- {\phi}^{\mathrm{x},j}_{t,k}(i)\right|+\sum^T_{k=0}\left|\widehat{\varphi}^{\mathrm{x},j}_{t,k}(i)- \widehat{\phi}^{\mathrm{x},j}_{t,k}(i)\right|\right)\nonumber\\&\leq W(\kappa+1)\sup_{d\in\{1,2,\dots,\kappa\}}g(d)\left(\sup_{j\in\mathcal{N}^{\distance}_{\mathcal{G}}(i)}
        \sum^t_{k=0}\left|{\varphi}^{\mathrm{x},j}_{t,k}(i)- {\phi}^{\mathrm{x},j}_{t,k}(i)\right|+\sum^T_{k=0}\left|\widehat{\varphi}^{\mathrm{x},j}_{t,k}(i)- \widehat{\phi}^{\mathrm{x},j}_{t,k}(i)\right|\right).\label{eq:(b)}
       \end{align}

       For the term (c), we have
       \begin{align}\text{(c)}&\leq\sum_{j\in\left([N]\backslash\mathcal{N}^{\distance}_{\mathcal{G}}(i)\right)}\left|\sum^t_{k=0}{\phi}^{\mathrm{x},j}_{t,k}(i)\cdot{\mathrm{w}}^j_{k-1}+\sum^T_{k=0} \widehat{\phi}^{\mathrm{x},j}_{t,k}(i)\cdot{\mathrm{w}}^j_{k-1}\right|\nonumber\\
        &{\leq}W\sum_{j\in\left([N]\backslash\mathcal{N}^{\distance}_{\mathcal{G}}(i)\right)}\left(
        \sum^t_{k=0}\left|{\varphi}^{\mathrm{x},j}_{t,k}(i)- {\phi}^{\mathrm{x},j}_{t,k}(i)\right|+\sum^T_{k=0}\left|\widehat{\varphi}^{\mathrm{x},j}_{t,k}(i)- \widehat{\phi}^{\mathrm{x},j}_{t,k}(i)\right|\right)\nonumber\\&\leq\frac{WC(2+\rho)}{1-\rho}\sum_{j\in\left([N]\backslash\mathcal{N}^{\distance}_{\mathcal{G}}(i)\right)}1\nonumber\\&\leq\frac{WC(2+\rho)}{1-\rho}\left((\mathrm{diam}({\mathcal{G}})+1)\sup_{d\in\{1,2,\dots\mathrm{diam}({\mathcal{G}})\}}g(d)-(\distance+1)\sup_{d\in\{1,2,\dots,\kappa\}}g(d)\right).\label{eq:(c)}
    \end{align}

Then, we return to bound $|\mathrm{x}_t^i(\PredSLS)-\mathrm{x}_t^i(\OPT)|$ as
\begin{align}
    |\mathrm{x}_t^i(\PredSLS)-\mathrm{x}_t^i(\OPT)|\leq \eqref{eq:(a)}+\eqref{eq:(b)}+\eqref{eq:(c)}. \label{eq:sub-state bound}
\end{align}

Similar to \eqref{eq:sub-state bound}, the sub-action bound is derived from the closed-loop implementation: 
   \begin{align}\nonumber
         &|\mathrm{u}_t^i(\PredSLS) - \mathrm{u}^i_t(\OPT)| =|\mathrm{u}_t^i(\PredSLS) - \mathrm{u}^i_t(\PredSLSE)+\mathrm{u}_t^i(\PredSLSE) - \mathrm{u}^i_t(\OPT)|\\
        &\leq \left|\sum_{j\in\mathcal{N}^{\distance}_{\mathcal{G}}(i)}\sum^t_{k=0}\varphi^{\mathrm{u},j}_{t,k}(i)\cdot\left((\widetilde{\mathrm{w}}^j_{k}-\Bar{\mathrm{w}}^j_{k})(\PredSLS) - (\widetilde{\mathrm{w}}^j_{k}-\Bar{\mathrm{w}}^j_{k})(\PredSLSE)\right)\right|\nonumber\\&\qquad+\left|\sum_{j\in\mathcal{N}^{\distance}_{\mathcal{G}}(i)}\sum^T_{k=0}\widehat{\varphi}^{\mathrm{u},j}_{t,k}(i)\cdot(\widehat{\mathrm{w}}^j_{k-1}-{\mathrm{w}}^j_{k-1})\right|\nonumber\\ &\qquad+\left|\sum_{j\in\mathcal{N}^{\distance}_{\mathcal{G}}(i)}\sum^t_{k=0}\left(\varphi^{\mathrm{u},j}_{t,k}(i)\cdot(\widetilde{\mathrm{w}}^j_{k}-\Bar{\mathrm{w}}^j_{k})(\PredSLSE) - \phi^{\mathrm{u},j}_{t,k}(i)\cdot(\widetilde{\mathrm{w}}^j_{k}-\Bar{\mathrm{w}}^j_{k})(\OPT)\right)\right|\nonumber\\&\qquad+\left|\sum_{j\in\mathcal{N}^{\distance}_{\mathcal{G}}(i)}\sum^T_{k=0}\left(\widehat{\varphi}^{\mathrm{u},j}_{t,k}(i)-\widehat{\phi}^{\mathrm{u},j}_{t,k}(i)\right){\mathrm{w}}^j_{k-1}\right|\nonumber\\
        &\qquad+\left|\sum_{j\in\left([N]\backslash\mathcal{N}^{\distance}_{\mathcal{G}}(i)\right)}\sum_{k=0}^t\phi^{\mathrm{u},j}_{t,k}(i)\cdot(\widetilde{\mathrm{w}}^j_{k}-\Bar{\mathrm{w}}^j_{k})(\OPT)\right|+\left|\sum_{j\in\left([N]\backslash\mathcal{N}^{\distance}_{\mathcal{G}}(i)\right)}\sum^T_{k=0}\widehat{\phi}^{\mathrm{u},j}_{t,k}(i)\cdot{\mathrm{w}}^j_{k-1}\right|\nonumber\\
        &\leq \left|\sum_{j\in\mathcal{N}^{\distance}_{\mathcal{G}}(i)}\sum^T_{k=0}\widehat{\varphi}^{\mathrm{u},j}_{t,k}(i)\cdot(\widehat{\mathrm{w}}^j_{k-1}-{\mathrm{w}}^j_{k-1})\right|\nonumber\\
        &\qquad + \left|\sum_{j\in\mathcal{N}^{\distance}_{\mathcal{G}}(i)}\sum^t_{k=0}\left(\left(\varphi^{\mathrm{u},j}_{t,k}(i) - \phi^{\mathrm{u},j}_{t,k}(i)\right){\mathrm{w}}^j_{k-1}\right)\right|+\left|\sum_{j\in\mathcal{N}^{\distance}_{\mathcal{G}}(i)}\sum^T_{k=0}\left(\widehat{\varphi}^{\mathrm{u},j}_{t,k}(i)-\widehat{\phi}^{\mathrm{u},j}_{t,k}(i)\right){\mathrm{w}}^j_{k-1}\right|\nonumber\\
        &\qquad+\left|\sum_{j\in\left([N]\backslash\mathcal{N}^{\distance}_{\mathcal{G}}(i)\right)}\sum_{k=0}^t\phi^{\mathrm{u},j}_{t,k}(i)\cdot{\mathrm{w}}^j_{k-1}\right|+\left|\sum_{j\in\left([N]\backslash\mathcal{N}^{\distance}_{\mathcal{G}}(i)\right)}\sum^T_{k=0}\widehat{\phi}^{\mathrm{u},j}_{t,k}(i)\cdot{\mathrm{w}}^j_{k-1}\right|\nonumber\\
        &\leq \eqref{eq:(a)}+\eqref{eq:(b)}+\eqref{eq:(c)},\nonumber
        \end{align}
where the second inequality is due to \Cref{lemma:w-w=w}. This completes the proof of Lemma \ref{lemma: discrepancy bound}. 
\end{proof}

\subsection{Proof of Corollary \ref{lemma:temporal-spatial decay}}\label{proof:temporal-spatial_decay}
\begin{proof}
    Applying Theorem \ref{theorem:temporal}, we have
    \begin{align}
        &\left\|\begin{bmatrix}
            \pxl_{t,k}(j)-\phi_{t,k}^{\mathrm{x},i}(j)&\hpxl_{t,k}(j)-\widehat{\phi}_{t,k}^{\mathrm{x},i}(j)\\\pul_{t,k}(j)-\phi_{t,k}^{\mathrm{u},i}(j)&\hpul_{t,k}(j)-\widehat{\phi}_{t,k}^{\mathrm{u},i}(j)
        \end{bmatrix}\right\|\leq\left\|\begin{bmatrix}
            \pxl_{t,k}(j)&\hpxl_{t,k}(j)\\\pul_{t,k}(j)&\hpul_{t,k}(j)
        \end{bmatrix}\right\| + \left\|\begin{bmatrix}
            \phi_{t,k}^{\mathrm{x},i}(j)&\widehat{\phi}_{t,k}^{\mathrm{x},i}(j)\\\phi_{t,k}^{\mathrm{u},i}(j)&\widehat{\phi}_{t,k}^{\mathrm{u},i}(j)
        \end{bmatrix}\right\|\nonumber
        \\&\leq\left(\left\|\begin{bmatrix}
            \pxl_{t,k}\\\pul_{t,k}
        \end{bmatrix}(j)\right\|^2 + \left\|\begin{bmatrix}
            \hpxl_{t,k}\\\hpul_{t,k}
        \end{bmatrix}(j)\right\|^2\right)^{\frac{1}{2}} + \left(\left\|\begin{bmatrix}
            {\phi}_{t,k}^{\mathrm{x},i}\\\widehat{\phi}_{t,k}^{\mathrm{u},i}
        \end{bmatrix}(j)\right\|^2 + \left\|\begin{bmatrix}
            \widehat{\phi}_{t,k}^{\mathrm{x},i}\\\widehat{\phi}_{t,k}^{\mathrm{u},i}
        \end{bmatrix}(j)\right\|^2\right)^{\frac{1}{2}}\nonumber\\
        &\leq\left(\left\|\begin{bmatrix}
            \pxl_{t,k}\\\pul_{t,k}
        \end{bmatrix}\right\|^2 + \left\|\begin{bmatrix}
            \hpxl_{t,k}\\\hpul_{t,k}
        \end{bmatrix}\right\|^2\right)^{\frac{1}{2}} + \left(\left\|\begin{bmatrix}
            {\phi}_{t,k}^{\mathrm{x},i}\\\widehat{\phi}_{t,k}^{\mathrm{u},i}
        \end{bmatrix}\right\|^2 + \left\|\begin{bmatrix}
            \widehat{\phi}_{t,k}^{\mathrm{x},i}\\\widehat{\phi}_{t,k}^{\mathrm{u},i}
        \end{bmatrix}\right\|^2\right)^{\frac{1}{2}}\leq 2C^{\frac{1}{2}}\rho^{\frac{|t-k|}{2}}.\label{eq:corollary bound 1}
    \end{align}
    Applying Theorem \ref{theorem:spatial}, we have
    \begin{align}
        \left\|\begin{bmatrix}\pxl_{t,k}(j)-\phi_{t,k}^{\mathrm{x},i}(j)&\hpxl_{t,k}(j)-\widehat{\phi}_{t,k}^{\mathrm{x},i}(j)\\\pul_{t,k}(j)-\phi_{t,k}^{\mathrm{u},i}(j)&\hpul_{t,k}(j)-\widehat{\phi}_{t,k}^{\mathrm{u},i}(j)
        \end{bmatrix}\right\|&\leq\sqrt{\sum^{T}_{t=0}\left\|\begin{bmatrix}\pxl_{t,k}(j)-\phi_{t,k}^{\mathrm{x},i}(j)&\hpxl_{t,k}(j)-\widehat{\phi}_{t,k}^{\mathrm{x},i}(j)\\\pul_{t,k}(j)-\phi_{t,k}^{\mathrm{u},i}(j)&\hpul_{t,k}(j)-\widehat{\phi}_{t,k}^{\mathrm{u},i}(j)
        \end{bmatrix}\right\|^2}\nonumber\\&{\leq}D^{\frac{1}{2}}\vartheta^{\frac{\distance}{2}}.\label{eq:corollary bound 2}
    \end{align}
    Combining \eqref{eq:corollary bound 1} and \eqref{eq:corollary bound 2}, we obtain
    \begin{align}
        \left\|\begin{bmatrix}\pxl_{t,k}(j)-\phi_{t,k}^{\mathrm{x},i}(j)&\hpxl_{t,k}(j)-\widehat{\phi}_{t,k}^{\mathrm{x},i}(j)\\\pul_{t,k}(j)-\phi_{t,k}^{\mathrm{u},i}(j)&\hpul_{t,k}(j)-\widehat{\phi}_{t,k}^{\mathrm{u},i}(j)
        \end{bmatrix}\right\|&\leq\sqrt{2C^{\frac{1}{2}}\rho^{\frac{|t-k|}{2}}D^{\frac{1}{2}}\vartheta^{\frac{\distance}{2}}}\nonumber\\
        & \leq\sqrt{2}(CD)^{\frac{1}{4}}\max(\rho^{\frac{1}{4}},\vartheta^{\frac{1}{4}})^{|t-k|+\kappa}.\nonumber
    \end{align}
    Let $D_1\coloneqq\sqrt{2}(CD)^{\frac{1}{4}}$ and $\vartheta_1\coloneqq\max(\rho^{\frac{1}{4}},\vartheta^{\frac{1}{4}})$. We complete the proof.
\end{proof}

\subsection{Proof of Lemma \ref{lemma:regret bound 1}\label{appendix:lemma_regret_bound_1}}

\begin{proof}
    Before proceeding to the proof, we state a useful lemma below.
    \begin{lemma}[Lemma 10 in \cite{yu2022competitive}, Lemma 3 in \cite{li2022robustness}]\label{lemma: referred regret bound 1}
        At each time $t\in[T]$, if a non-causal controller is by the following policy for any $\Xi_t\in\mathbb{R}^{n}$:
        \begin{align}
            \mathrm{u}_t=-(R+B^\top PB)^{-1}B^\top\left(PA\mathrm{x}_t+\sum^{T-1}_{k=t}\left(F^\top\right)^{k-t}P\mathrm{w}_k-\Xi_t\right),\nonumber
        \end{align}
        then the cost gap between an optimal controller $\OPT$ and the algorithm $\ALG$ induced by selecting control actions $(\mathrm{u}_0,\dots,\mathrm{u}_{T-1})$ equals to
        \begin{align} 
         \cost(\ALG) - \cost(\OPT) = \sum^{T-1}_{t=0}\Xi_t^\top Y\Xi_t^{\phantom{\top}}, \nonumber
         \end{align}
         where $Y\coloneqq B(R+B^\top PB)^{-1}B^{\top}$ and $F\coloneqq A-YPA$.
    \end{lemma}
    Recall the offline optimal policy illustrated in \eqref{eq:optimal policy}. The offline optimal action is given by
    \begin{align}
        \mathrm{u}_t(\OPT) = -(R+B^\top PB)^{-1}B^\top\left(PA\mathrm{x}_t(\OPT)+\sum^{T-1}_{k=t}\left(F^\top\right)^{k-t}P\mathrm{w}_k\right).\nonumber
    \end{align}
   Let $V\coloneqq R+B^\top PB$. Using Lemma \ref{lemma: referred regret bound 1}, it suffices to obtain the following
    \begin{align}
       &\mathrm{u}_t(\ALG)-\mathrm{u}_t(\OPT) =-V^{-1}B^\top PA(\mathrm{x}_t(\ALG)-\mathrm{x}_t(\OPT)) + V^{-1}B^\top\Xi_t\nonumber \\\Longleftrightarrow\quad & B^\top \Xi_t = V(\mathrm{u}_t(\ALG)-\mathrm{u}_t(\OPT))+B^\top PA(\mathrm{x}_t(\ALG)-\mathrm{x}_t(\OPT)).\nonumber
    \end{align}
    Then,
    \begin{align}
        \cost(\ALG) - \cost(\OPT) = \sum^{T-1}_{t=0}\left\|V\left(\mathrm{u}_t(\ALG)-\mathrm{u}_t(\OPT)  \right) + B^\top PA\left(\mathrm{x}_t(\ALG)-\mathrm{x}_t(\OPT)  \right)\right\|_{V^{-1}}^2.\nonumber
    \end{align}
    This finishes the proof of Lemma \ref{lemma:regret bound 1}.
\end{proof}
\newpage
\section{Additional Experiments}
\label{appendix:experiments}

This section extends the numerical experiments presented in the main body. 
We consider three additional graphs to define the communication topologies, 
as illustrated in sub-figures (b), (c), and (d) of~\Cref{fig:topological networks}. 
All other experimental settings remain consistent with those detailed in 
\Cref{sec:exp_performance}.

\begin{figure}[h]
    \centering
\includegraphics[width=1\linewidth]{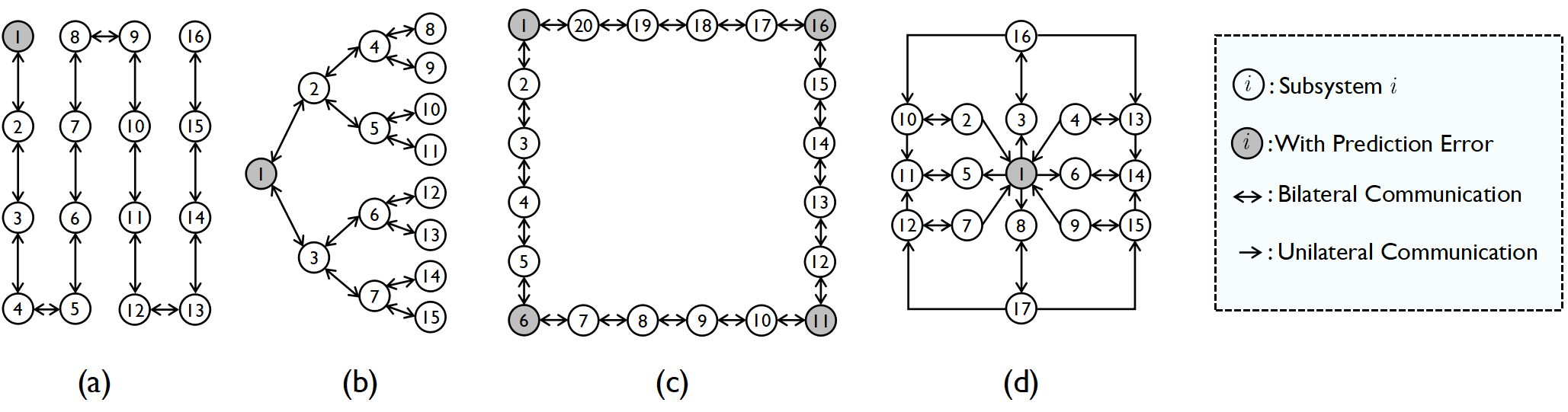}
    \caption{\textbf{Topological networks} tested in the experiment. These figures mark the total number of nodes and structural forms in different networks. We uniformly define the weight of each connection to be $0.5$.}
    \label{fig:topological networks}
\end{figure}

\subsection{Additional Numerical Evaluation}
We provide further experimental results that complement the results presented in \Cref{sec:case_studies}.
In \Cref{fig:result_comparison_tree}, \Cref{fig:result_comparison_cyclic}, and \Cref{fig:result_comparison_mesh}, we display the comparison results (as in \Cref{fig:regret}) on the tree, cyclic, and mesh topologies (\Cref{fig:topological networks}). The trajectory for \psls closely aligns the best with the centralized offline optimal trajectory upon $\kappa=\mathrm{diam}(\mathcal{G})$. In \Cref{fig:result_bound_tree}, \Cref{fig:result_bound_cyclic}, and \Cref{fig:result_bound_mesh}, for the analytic results (\Cref{theorem:regret}), we further demonstrate the trade-off discovery presented in the third paragraph of \Cref{sec:non-monotonic_perform}.

\newpage

\begin{figure}[H]
    \centering
    \includegraphics[width=1\linewidth]{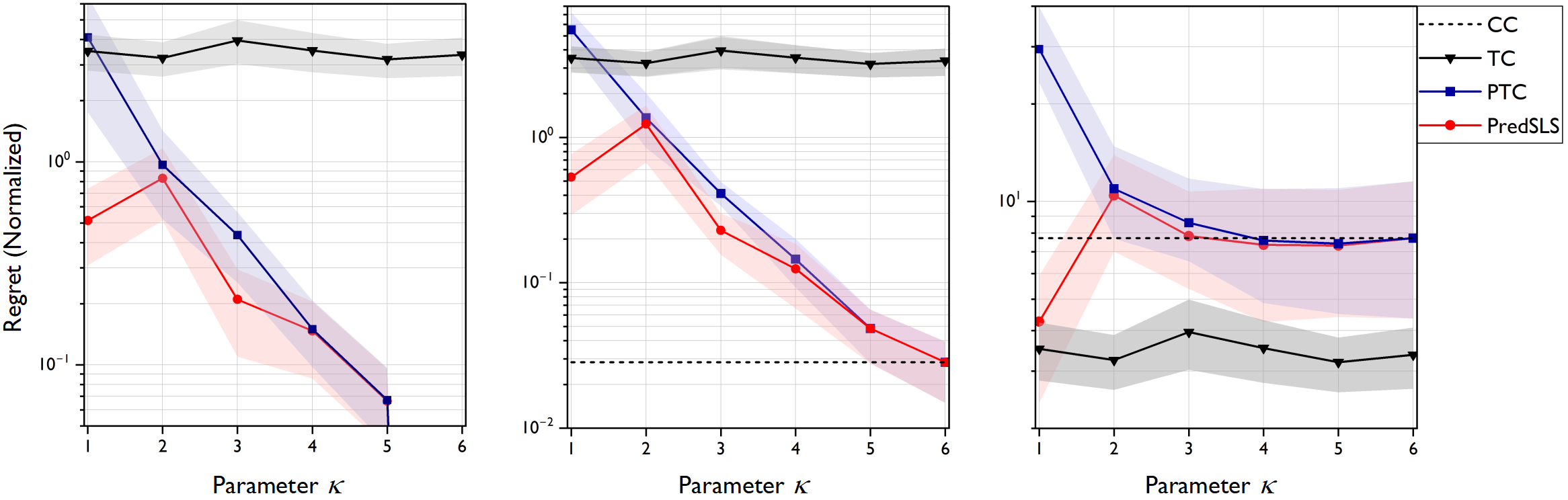}
    \caption{A comparison of different control methods on 100 experiments using normalized regret $\mathrm{DR}(\pi) / J^*$ in three cases. \textsc{Left}: the introduced example with $\|\mathrm{w}_t - \widehat{\mathrm{w}}_t\|=0$, \textsc{Mid}: $\|\mathrm{w}_t - \widehat{\mathrm{w}}_t\|=0.1$. \textsc{Right}: $\|\mathrm{w}_t - \widehat{\mathrm{w}}_t\|=1$. The topology of this experimental figure is the sub-figure (b) of \Cref{fig:topological networks}.}\label{fig:result_comparison_tree}
\end{figure}
\begin{figure}[H]
    \centering
    \includegraphics[width=1\linewidth]{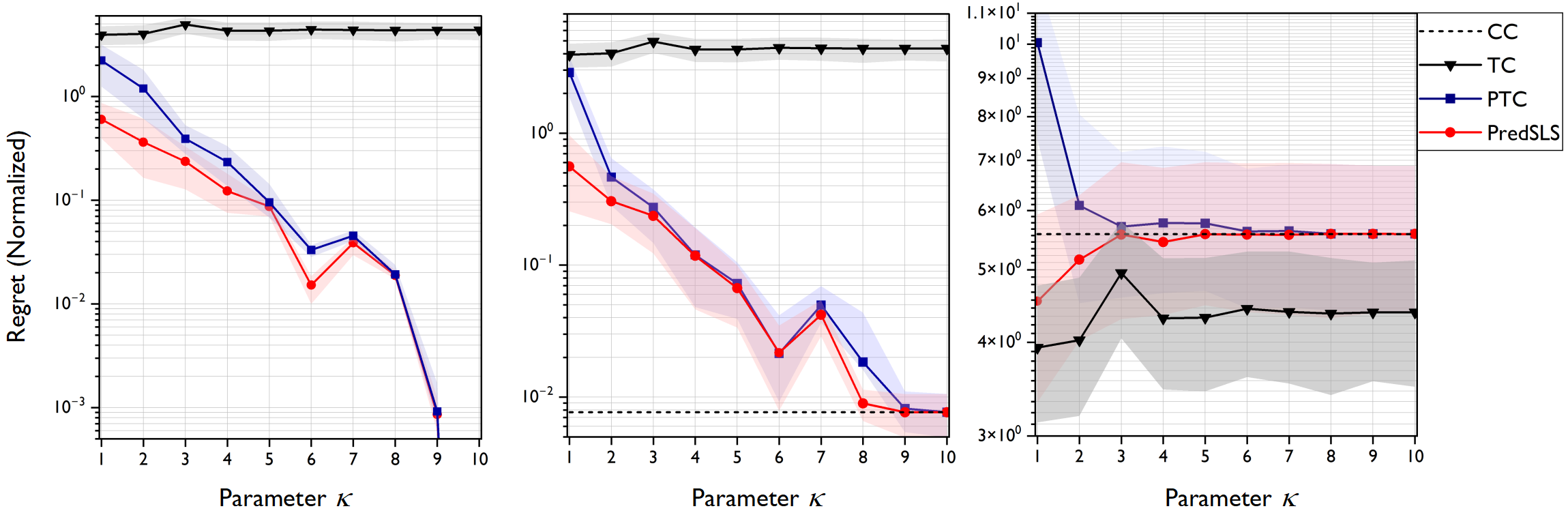}
    \caption{A comparison of different control methods on 100 experiments using normalized regret $\mathrm{DR}(\pi) / J^*$ in three cases. \textsc{Left}: the introduced example with $\mathbb{E}[\|\mathrm{w}_t - \widehat{\mathrm{w}}_t\|]=0$, \textsc{Mid}: $\mathbb{E}[\|\mathrm{w}_t - \widehat{\mathrm{w}}_t\|]=0.1$. \textsc{Right}: $\mathbb{E}[\|\mathrm{w}_t - \widehat{\mathrm{w}}_t\|]=1$. The topology of this experimental figure is the sub-figure (c) of \Cref{fig:topological networks}.}\label{fig:result_comparison_cyclic}
\end{figure}
\begin{figure}[H]
    \centering
    \includegraphics[width=1\linewidth]{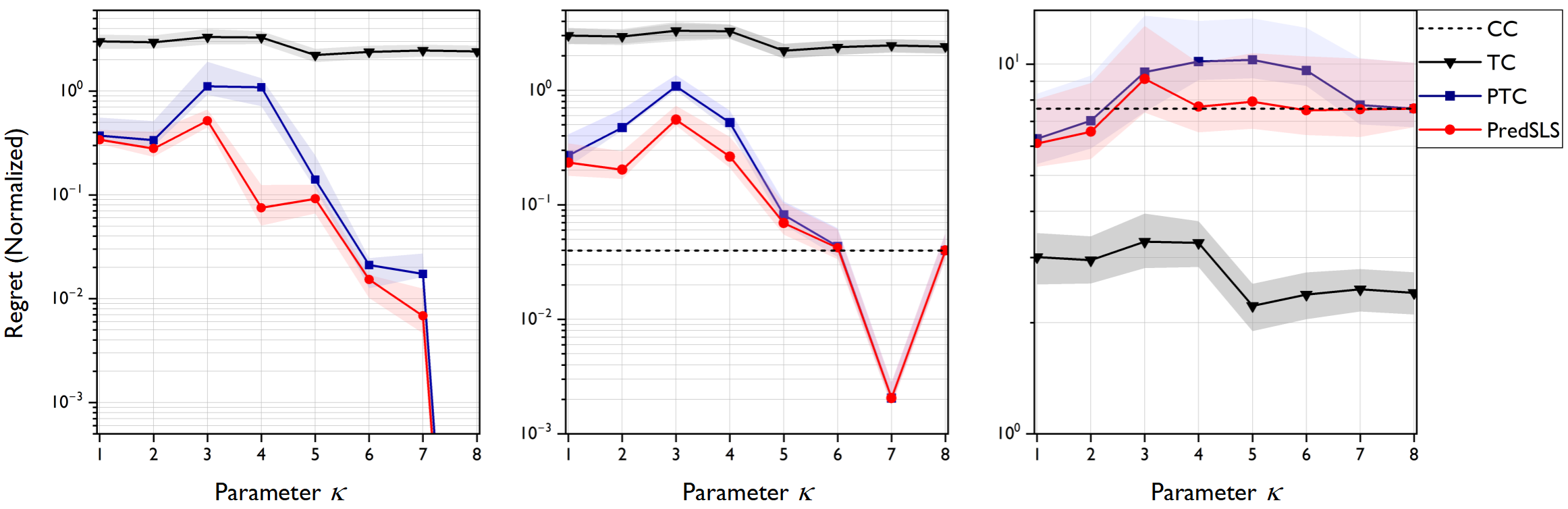}
    \caption{A comparison of different control methods on 100 experiments using normalized regret $\mathrm{DR}(\pi) / J^*$ in three cases. \textsc{Left}: the introduced example with $\|\mathrm{w}_t - \widehat{\mathrm{w}}_t\|=0$, \textsc{Mid}: $\|\mathrm{w}_t - \widehat{\mathrm{w}}_t\|=0.1$. \textsc{Right}: $\|\mathrm{w}_t - \widehat{\mathrm{w}}_t\|=1$. The topology of this experimental figure is the sub-figure (d) of \Cref{fig:topological networks}.}\label{fig:result_comparison_mesh}
\end{figure}

\newpage

\begin{figure}[H]
    \centering
\includegraphics[width=0.93\linewidth]{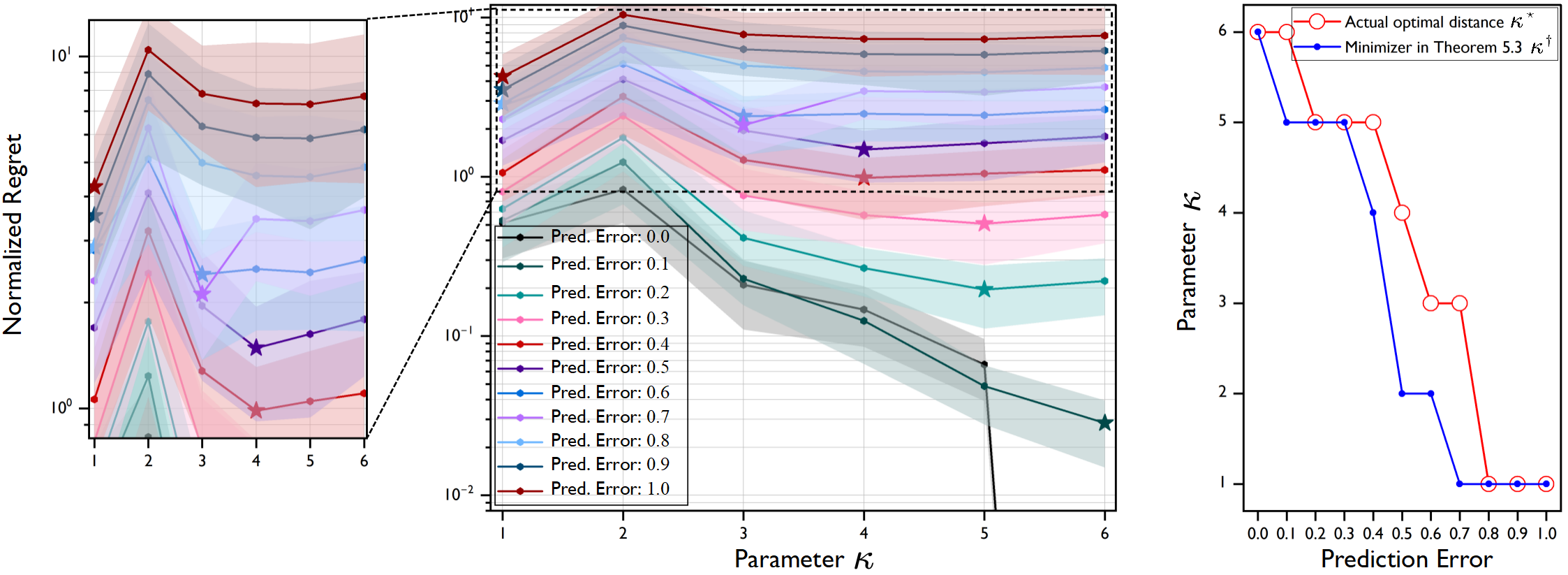}
    \caption{Performance of \psls for the tree-graph-induced (sub-figure (b) of \Cref{fig:topological networks}) LQR. \textbf{\textsc{Left}}: Algorithm performance (normalized regret) versus communication distance $\kappa$ with varying prediction errors; \textbf{\textsc{Right}}: Actual optimal communication distance  $\kappa^{\star}$ that minimizes the normalized regret, compared with the minimizer $\kappa^\star$ of the regret bound in \Cref{theorem:regret}.}\label{fig:result_bound_tree}
\end{figure}
\begin{figure}[H]
    \centering
    \includegraphics[width=0.93\linewidth]{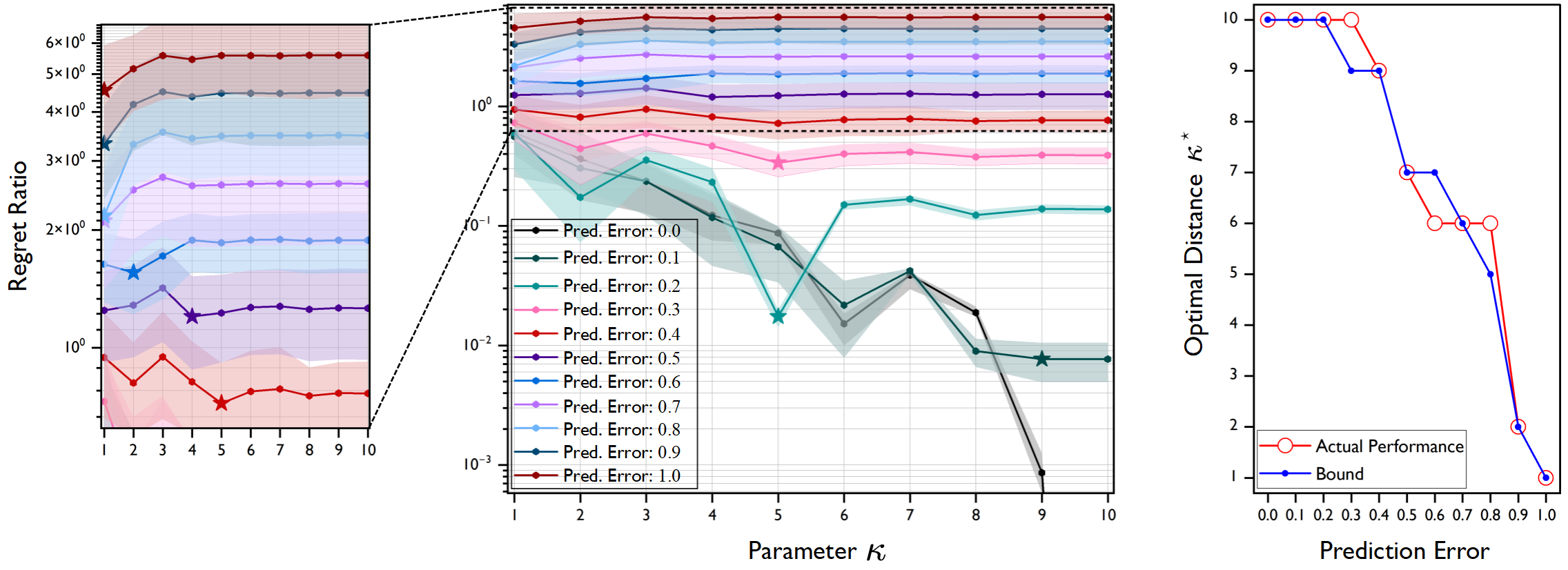}
    \caption{Performance of \psls for the cyclic-graph-induced (sub-figure (c) of \Cref{fig:topological networks}) LQR. \textbf{\textsc{Left}}: Algorithm performance (normalized regret) versus communication distance $\kappa$ with varying prediction errors; \textbf{\textsc{Right}}: Actual optimal communication distance  $\kappa^{\star}$ that minimizes the normalized regret, compared with the minimizer $\kappa^\star$ of the regret bound in \Cref{theorem:regret}.}\label{fig:result_bound_cyclic}
\end{figure}
\begin{figure}[H]
    \centering
    \includegraphics[width=0.93\linewidth]{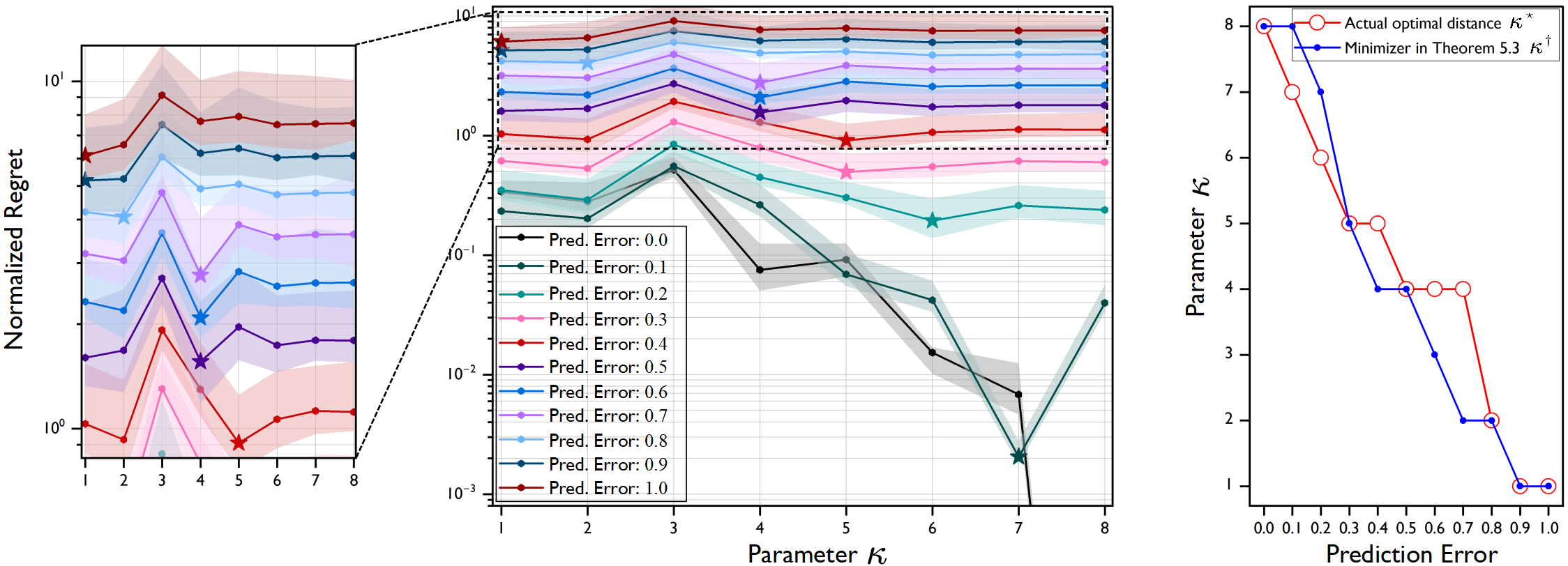}
    \caption{Performance of \psls for the mesh-graph-induced (sub-figure (d) of \Cref{fig:topological networks}) LQR. \textbf{\textsc{Left}}: Algorithm performance (normalized regret) versus communication distance $\kappa$ with varying prediction errors; \textbf{\textsc{Right}}: Actual optimal communication distance  $\kappa^{\star}$ that minimizes the normalized regret, compared with the minimizer $\kappa^\star$ of the regret bound in \Cref{theorem:regret}.}\label{fig:result_bound_mesh}
\end{figure}
    
\end{appendices}
\end{document}

%% file: preamble.tex
\usepackage[a4paper]{geometry}
\usepackage{subcaption}

\usepackage{amssymb}
\usepackage{fix-cm}
\usepackage{authblk}
\usepackage{booktabs} 
\usepackage{indentfirst}
\usepackage{times}
\usepackage{graphicx}
\graphicspath{{Figs/}}
\usepackage{comment}
\usepackage{caption}
\usepackage[utf8]{inputenc} 
\usepackage[T1]{fontenc}  
\usepackage{url}   
\usepackage{booktabs}       
\usepackage{nicefrac}     
\usepackage{microtype}     
\usepackage{makecell}
\usepackage{xcolor}
\usepackage[framemethod=tikz]{mdframed}
\usepackage{mathtools}
\usepackage{amsmath,amsfonts,amssymb,amscd,xspace}
\usepackage{mathrsfs}
\usepackage{amsthm}
\usepackage{hyperref}   
\usepackage{cleveref}
\usepackage{extarrows}
\usepackage{appendix}
\usepackage{tocbasic}

\usepackage[shortlabels]{enumitem}
\usepackage{bbm}
\usepackage{bm}
\usepackage{booktabs}
\usepackage[ruled]{algorithm2e}
\usepackage{algpseudocode}

\usepackage{algorithm}


\graphicspath{{Figs/}}

\usepackage{graphicx}
\providecommand{\keywords}[1]
{
  \small	
  \textbf{\textit{Keywords---}} #1
}

\newtheorem{assumption}{Assumption}
\newtheorem{theorem}{Theorem}
\numberwithin{theorem}{section}
\newtheorem{lemma}{Lemma}
\newtheorem{corollary}{Corollary}
\numberwithin{corollary}{section}

\newtheorem{definition}{Definition}
\newtheorem{proposition}{Proposition}

\newtheorem{remark}{Remark}

\newcommand{\psls}{{\texttt{PredSLS}}\xspace}
\newcommand{\dpsls}{{$\kappa$-$\texttt{PredSLS}(i)$}\xspace}
\newcommand{\decpsls}{{$\kappa$-$\texttt{PredSLS}{(i,k)}$}\xspace}
\newcommand{\causalk}{{$\texttt{Causal}{(i,k)}$}\xspace}
\newcommand{\mixedk}{{$\texttt{Noncau}{(i,k)}$}\xspace}

\newcommand{\PHI}{\mathbf{\Phi}}

\newcommand{\Phix}{{\Phi}^{\mathrm{x}}}
\newcommand{\hPhix}{\widehat{\Phi}^{\mathrm{x}}}
\newcommand{\Phiu}{{\Phi}^{\mathrm{u}}}
\newcommand{\hPhiu}{\widehat{\Phi}^{\mathrm{u}}}

\newcommand{\px}{\phi^{\mathrm{x},i}}

\newcommand{\pu}{\phi^{\mathrm{u},i}}

\newcommand{\hpx}{\widehat{\phi}^{\mathrm{x},i}}

\newcommand{\hpu}{\widehat{\phi}^{\mathrm{u},i}}

\newcommand{\pxl}{\varphi^{\mathrm{x},i}}

\newcommand{\pul}{\varphi^{\mathrm{u},i}}

\newcommand{\hpxl}{\widehat{\varphi}^{\mathrm{x},i}}

\newcommand{\hpul}{\widehat{\varphi}^{\mathrm{u},i}}

\newcommand{\distance}{\kappa}
\newcommand{\cost}{J}

\newcommand{\OPT}{\mathtt{OPT}}
\newcommand{\ALG}{\mathtt{ALG}}
\newcommand{\PredSLS}{\mathtt{PredSLS}}
\newcommand{\PredSLSE}{\mathtt{PredSLS}\star}

\newcommand{\ouralg}{{\texttt{PredSLS}}\xspace}